\documentclass[useAMS,usenatbib]{mn2e}
\usepackage{graphicx}

\newcommand\ub{\mbox{$U\!-\!B$}}
\newcommand\bv{\mbox{$B\!-\!V$}}
\newcommand\vk{\mbox{$V\!-\!K$}}
\newcommand\jk{\mbox{$J\!-\!K$}}
\newcommand{\mlv}{\mbox{$M/L_V$}}
\newcommand{\msun}{\mbox{$M_\odot$}}
\newcommand{\ha}{\hbox{H$\alpha$}}
\newcommand{\hb}{\hbox{H$\beta$}}

\newcommand{\mgii}{\hbox{Mg\,{\small II}}}
\newcommand{\caii}{\hbox{Ca\,{\small II}}}
\newcommand{\tauv}{\hbox{$\hat{\tau}_V$}}
\newcommand{\feh}{\hbox{\rm [Fe/H]}}
\newcommand{\sigv}{\hbox{$\sigma_V$}}
\newcommand{\chisq}{\hbox{$\chi^2$}}

\newcommand{\hii}{\hbox{H\,{\sc ii}}}
\newcommand{\mgfe}{\hbox{$\left[{\rm MgFe}\right]$}}
\newcommand{\mgfep}{\hbox{$\left[{\rm MgFe}\right]'$}}
\newcommand{\mgofe}{\hbox{$\left[{\rm Mg}_1{\rm Fe}\right]$}}
\newcommand{\mgtfe}{\hbox{$\left[{\rm Mg}_2{\rm Fe}\right]$}}
\newcommand{\kms}{\hbox{${\rm km}\,{\rm s}^{ -1}$}}

\title[Stellar population synthesis at the resolution of 2003]
{Stellar population synthesis at the resolution of 2003}
\author[G. Bruzual and S. Charlot]{G. Bruzual$^{1}$\thanks{E-mail:
bruzual@cida.ve (GBA); charlot@iap.fr (SC)} and S.
Charlot$^{2,3}$\footnotemark[1]\\
$^{1}$Centro de Investigaciones de Astronom{\'\i}a, AP 264, M\'erida 5101-A,
Venezuela\\
$^{2}$Max-Planck Institut f\"ur Astrophysik, Karl-Schwarzschild-Strasse 1, 
85748 Garching, Germany\\
$^{3}$Institut d'Astrophysique de Paris, CNRS, 98 bis Boulevard Arago, 75014 
Paris, France}
\begin{document}

\date{MNRAS, in press}

\pagerange{\pageref{firstpage}--\pageref{lastpage}} \pubyear{2002}

\maketitle

\label{firstpage}

\begin{abstract}
We present a new model for computing the spectral evolution of stellar
populations at ages between $1\times10^5\,$yr and $2\times10^{10}\,$yr
at a resolution of 3~{\AA} across the whole wavelength range from 3200~{\AA}
to 9500~{\AA} for a wide range of metallicities. These predictions are based
on a newly available library of observed stellar spectra. We also compute
the spectral evolution across a larger wavelength range, from 91~{\AA} to
160~$\mu$m, at lower resolution. The model incorporates recent progress in
stellar evolution theory and an observationally motivated prescription for 
thermally-pulsing stars on the asymptotic giant branch. The latter is supported
by observations of surface brightness fluctuations in nearby stellar 
populations. We show that this model reproduces well the observed optical and
near-infrared colour-magnitude diagrams of Galactic star clusters of various
ages and metallicities. Stochastic fluctuations in the numbers of stars in
different evolutionary phases can account for the full range of observed
integrated colours of star clusters in the Magellanic Clouds. The model
reproduces in detail typical galaxy spectra from the Early Data Release (EDR)
of the Sloan Digital Sky Survey (SDSS). We exemplify how this type of spectral
fit can constrain physical parameters such as the star formation history, 
metallicity and dust content of galaxies. Our model is the first to enable 
accurate studies of absorption-line strengths in galaxies containing stars over
the full range of ages. Using the highest-quality spectra of the SDSS EDR, we
show that this model can reproduce simultaneously the observed strengths of 
those Lick indices that do not depend strongly on element abundance ratios. The 
interpretation of such indices with our model should be particularly useful for
constraining the star formation histories and metallicities of galaxies.
\end{abstract}

\begin{keywords}
galaxies: formation -- galaxies: evolution -- galaxies: stellar content -- 
stars: evolution.
\end{keywords}

\section{Introduction}

The star formation history of galaxies is imprinted in their integrated light.
The first attempts to interpret the light emitted from galaxies in terms of
their stellar content relied on {\em trial and error} analyses (e.g., 
\citealt{1971ApJS...22..445S}; \citealt{1972A&A....20..361F};
\citealt{1976ApJ...206..370O}; \citealt{1976ApJ...210...33T};
\citealt{1977ApJS...35..397P}; \citealt{1985ApJ...296..340P}).
In this technique, one reproduces the integrated spectrum of a galaxy
with a linear combination of individual stellar spectra of various types taken
from a comprehensive library.  The technique was abandoned in the early 1980's
because the number of free parameters was too large to be constrained by 
typical galaxy spectra.  More recent models are based on the {\em evolutionary
population synthesis} technique (\citealt{1978ApJ...222...14T};
\citealt{1983ApJ...273..105B}; 
\citealt{1987A&A...173...23A}; \citealt{1987A&A...186....1G};
\citealt{1989ApJS...71..817B}; \citealt{1993ApJ...405..538B};
\citealt*{1994ApJS...94...63B}; \citealt{1994A&A...285..751F};
\citealt{1994ApJS...95..107W};
\citealt{1995ApJS...96....9L}; \citealt{1997A&A...326..950F}; 
\citealt{1998MNRAS.300..872M};
\citealt{1999ApJ...513..224V};
\citealt{2002A&A...392....1S}). In this approach, the main
adjustable parameters are the stellar initial mass function (IMF), the star
formation rate (SFR) and, in some cases, the rate of chemical enrichment. 
Assumptions about the time evolution of these parameters allow one to compute
the age-dependent distribution of stars in the Hertzsprung-Russell (HR) diagram,
from which the integrated spectral evolution of the stellar population can be
obtained. These models have become standard tools in the interpretation of
galaxy colours and spectra.

Despite important progress over the last decade, modern population synthesis
models still suffer from serious limitations. The largest {\em intrinsic}
uncertainties of the models arise from the poor understanding of some advanced
phases of stellar evolution, such as the supergiant phase and the 
asymptotic-giant-branch (AGB) phase (see \citealt{1996fstg.conf..275C}; 
\citealt*{1996ApJ...457..625C}; \citealt{2003ApJ...582..202Y}). Stars in these 
phases are very bright and have a strong influence on integrated-light 
properties. The limitations arising from these uncertainties in the
interpretation of galaxy spectra are further amplified by the fact that age, 
metallicity and dust all tend to affect spectra in similar ways. This is 
especially true at the low resolving power of most current population synthesis 
models, i.e., typically $\sim250$ at optical wavelengths (see however 
\citealt{1999ApJ...513..224V}; \citealt{2002ApJ...580..850S}). As a result, 
light-weighted ages and metallicities derived from integrated galaxy spectra tend
to be strongly degenerate (e.g., \citealt{1994ApJS...95..107W}). For old stellar
populations, this degeneracy may be broken by studying surface brightness 
fluctuations, that are more sensitive to the details of the stellar luminosity
function than ordinary integrated light (\citealt{2000ApJ...543..644L}; 
\citealt{2001MNRAS.320..193B}). This method, however, is mostly applicable to 
studies of nearby ellipticals and spiral bulges.

The general contention is that the age-metallicity degeneracy can be broken
by appealing to refined spectral diagnostics involving individual stellar 
absorption-line features (e.g., \citealt{1985AJ.....90.1927R};
\citealt{1995ApJ...446L..31J}; \citealt{1999ApJ...525..144V}). Several {\em 
spectral indices} of this kind have been defined at optical and near-infrared 
wavelengths (e.g., \citealt{1973ApJ...179..731F}; \citealt{1984AJ.....89.1238R};
\citealt*{1989MNRAS.239..325D}; \citealt{1994ApJS...94..687W}). In the widely 
used `Lick system', the strengths of 25 spectral indices were parametrized as 
functions of stellar effective temperature, gravity and metallicity using a
sample of 460 Galactic stars (\citealt{1984ApJ...287..586B}; 
\citealt{1993ApJS...86..153G}; \citealt{1994ApJS...94..687W}; 
\citealt{1997ApJS..111..377W}; \citealt{1998ApJS..116....1T}). This convenient 
parametrization allows one to compute integrated index strengths of model 
galaxies with any stellar population synthesis code. In practice, however, the\
applications are limited to studies of old stellar populations because of the 
lack of hot stars in the Lick stellar library. Also, the Lick indices were 
defined in spectra which were not flux-calibrated and whose resolution ($\sim
9$~{\AA} FWHM) is three times lower than achieved in modern spectroscopic galaxy
surveys, such as the {\it Sloan Digital Sky Survey} (SDSS; 
\citealt{2000AJ....120.1579Y}). Thus high-quality galaxy spectra must first be
{\em degraded} to the specific calibration and wavelength-dependent resolution 
of the Lick system for Lick index-strength analyses to be performed (Section~4.4).
Ideally, one requires a population synthesis model that can predict actual {\em
spectra} of galaxies at the resolution of modern surveys. The model of 
\citet{1999ApJ...513..224V} fulfills this requirement. However, it is limited to
two narrow wavelength regions, 3820--4500~{\AA} and 4780--5460~{\AA}.

In this paper, we present a new model for computing the spectral evolution
of stellar populations of different metallicities at ages between 
$1\times10^5\,$yr and $2\times10^{10}\,$yr at a resolution of 3~{\AA} FWHM 
across the whole wavelength range from 3200~{\AA} to 9500~{\AA} (corresponding to
a median resolving power $\lambda/\Delta \lambda \approx2000$). These predictions
are based on a new library of observed stellar spectra recently assembled by 
\citet{2003A&A...402..433L}. We also compute the spectral evolution across a larger
wavelength range, from 91~{\AA} to 160~$\mu$m, at lower spectral resolution. 
This model should be particularly useful for interpreting the spectra gathered
by modern spectroscopic surveys in terms of constraints on the star formation
histories and metallicities of galaxies.

The paper is organized as follows. In Section~2 below, we present the stellar 
evolution prescription and the stellar spectral library on which our model 
relies. We consider several alternatives for these ingredients. We adopt
an observationally motivated prescription for thermally-pulsing AGB stars, 
which is supported by observations of surface brightness fluctuations in 
nearby stellar populations (\citealt{2000ApJ...543..644L}; 
\citealt*{2002ApJ...564..216L}). We also briefly recall the principle of the
{\em isochrone synthesis} technique for computing the spectral evolution of 
stellar populations \citep{1991ApJ...367..126C}. In Section~3, we investigate
the influence of the main adjustable parameters of the model on photometric 
predictions and compare our results with previous work. Comparisons with 
observed colour-magnitude diagrams and integrated colours of star clusters of
various ages and metallicities are also presented in this section. In 
Section~4, we compute the spectral evolution of stellar populations and 
compare our model with observed galaxy spectra from the SDSS EDR
\citep{2002AJ....123..485S}. We compare in detail the predicted and observed 
strengths of several absorption-line indices and identify those indices that 
appear to be most promising for constraining the stellar content of galaxies.
We summarize our conclusions in Section~5, where we also suggest ways of 
including the effects of gas and dust in the interstellar medium on the stellar
radiation computed with our model. Readers interested mainly in the photometric
predictions of the model may skip directly to Section~3, while those interested
mainly in applications of the model to interpret galaxy spectra may skip 
directly to Section~4.

\section[]{The Model}

In this section, we present the main two ingredients of our population 
synthesis model: the stellar evolution prescription and the stellar spectral
library. We consider several alternatives for each of these. We also briefly
review the principle of the {\em isochrone synthesis} technique for computing
the spectral evolution of stellar populations.

\subsection{Stellar evolution prescription}

To account for current uncertainties in the stellar evolution theory, 
we consider three possible stellar evolution prescriptions in
our model (Table~\ref{tracks-options}). We first consider the library of
stellar evolutionary tracks computed by \citet{1993A&AS...97..851A}, 
\citet{1993A&AS..100..647B}, \citet{1994A&AS..104..365F}, 
\citet{1994A&AS..105...29F}, and \citet{1996A&AS..117..113G}.
This library encompasses a wide range of initial chemical compositions,
$Z=0.0001$, 0.0004, 0.004, 0.008, 0.02, 0.05, and 0.1 with $Y= 2.5Z + 0.23$ 
($Z_\odot =0.02$ and $Y_\odot= 0.28$) assumed. The range of initial masses is
$0.6 \leq m\leq 120\,M_{\sun}$ for all metallicities, except for $Z=0.0001$ 
($0.6 \leq m \leq 100\,M_{\sun}$) and $Z=0.1$ ($0.6 \leq m \leq 9\,M_{\sun}$).
The tracks were computed using the radiative opacities of 
\citet*{1992ApJ...397..717I}\footnote{The stellar evolutionary tracks for 
$Z=0.0001$, which were computed last, include slightly updated opacities and 
equation of state. According to \citet{1996A&AS..117..113G}, these updates 
do not compromise the consistency with the predictions at higher 
metallicities.} and include all phases of stellar evolution from the zero-age
main sequence to the beginning of the thermally 
pulsing regime of the asymptotic giant branch (TP-AGB; for low- and 
intermediate-mass stars) and core-carbon ignition (for massive stars). For 
solar composition, the models are normalized to the temperature, luminosity, 
and radius of the Sun at an age of 4.6~Gyr. The tracks include mild 
overshooting in the convective cores of stars more massive than $1.5\, 
M_{\odot}$, as suggested by observations of Galactic star clusters 
(\citealt{1993A&AS..100..647B}; \citealt*{1993A&AS...98..477M}; 
\citealt{1994ApJ...426..165D}). For stars with masses between 1.0 and 1.5$
\, M_\odot$, core overshooting is included with a reduced efficiency.
Overshooting is also included in the convective envelopes of low- and 
intermediate-mass stars, as suggested by observations of the red giant
branch and horizontal branch of star clusters in the Galactic halo
and the Large Magellanic Cloud (hereafter LMC; \citealt{1991A&A...244...95A}).
We refer to this set of tracks as the `Padova 1994 library'.

\begin{table}
  \caption{Different stellar evolution prescriptions.}
  \label{tracks-options}
  \begin{tabular}{@{}lcl@{}}
  \hline
Name & Metallicity range & Source \\
 \hline
Padova 1994 & 0.0001--0.10 & Alongi et al. (1993)\\
	    & 		   & Bressan et al. (1993)\\
	    & 		   & Fagotto et al. (1994a)\\
	    & 		   & Fagotto et al. (1994b)\\
	    & 		   & Girardi et al. (1996)\\
\\
Padova 2000 & 0.0004--0.03 & Girardi et al. {(2000)$^a$}\\
\\
Geneva	    & 	 0.02      & Schaller et al. (1992)\\
	    & 		   & Charbonnel et al. (1996)\\
	    & 		   & Charbonnel et al. (1999)\\
\hline
\end{tabular}

\medskip
$^a\,$Girardi et al. (2000) computed tracks only for low- and 
intermediate-mass stars. In the Padova 2000 library, these calculations
are supplemented with high-mass tracks from the Padova 1994 library,
as suggested by Girardi et al. (2002).
\end{table}

Recently, \citet{2000A&AS..141..371G} produced a new version of this 
library, in which the main novelties are a revised equation of state 
\citep{1990ApJ...350..300M} and new low-temperature opacities 
\citep{1994ApJ...437..879A}.  The revised library includes stars with masses
down to $m=0.15\,M_\odot$, but it does not contain stars more massive than $7\,
M_\odot$ (the new equation of state affects mainly the evolution of stars less
massive than $0.6\,M_\odot$). The chemical abundances also differ slightly
from those adopted in the 1994 release, $Z=0.0004$, 0.004, 0.008, 0.019, and
0.03, with $Y= 2.25Z + 0.23$ ($Z_\odot =0.019$ and $Y_\odot= 0.273$) 
assumed. Following the arguments of \citet{2002A&A...391..195G}, we combine the
new library of low- and intermediate-mass tracks with high-mass tracks from the
older Padova 1994 library to build an updated library encompassing a complete 
range of initial stellar masses. This can be achieved at all but the highest
metallicity ($Z=0.03$), for which there is no counterpart in the Padova 1994 
library ($Z=0.02$ and 0.05 available only). We refer to this set of tracks as
the `Padova 2000 library'.

The third stellar evolution prescription we consider, for the case of solar 
metallicity only, is the comprehensive library of tracks computed by 
\citet[ for $m\geq 2\, M_\odot$]{1992A&AS...96..269S},
\citet[ for $0.8\leq m< 2\,M_{\sun}$]{1996A&AS..115..339C} and 
\citet[ for $0.6\leq m< 0.8\,M_{\sun}$]{1999A&AS..135..405C}. The abundances 
are $X=0.68$, $Y=0.30$, and $Z=0.02$, and the opacities are from 
\citet[ for $m\geq2\,M_\odot$]{1992ApJS...79..507R} and 
\citet[ for $0.8\leq m/M_\odot< 2$]{1993ApJ...412..752I}. The tracks include 
all phases of stellar evolution from the zero-age main sequence to the 
beginning of the TP-AGB or core-carbon ignition, depending on the initial mass.
The models are normalized to the luminosity, temperature, and radius of the Sun
at an age of 4.6~Gyr. Mild overshooting is included in the convective cores of
stars more massive than $1.5\, M_{\odot}$. Differences with the 
solar-metallicity calculations of \citet{1993A&AS..100..647B} in the Padova 
1994 library include: the absence of overshooting in the convective cores of 
stars with masses between 1.0 and 1.5$\, M_\odot$ and in the convective 
envelopes of low- and intermediate-mass stars; the higher helium fraction; the
inclusion of mass loss along the red giant branch; the treatment of convection
during core-helium burning; and the internal mixing and mass loss of massive
stars.  The signatures of these differences in the stellar evolutionary tracks
have been investigated by \citet{1996ApJ...457..625C}. We refer to this 
alternative set of tracks for solar metallicity as the `Geneva library'.

We supplement the Padova and Geneva tracks of low- and intermediate-mass
stars beyond the early-AGB with TP-AGB and post-AGB evolutionary 
tracks.\footnote{The different stellar evolution prescriptions in the Padova 
and Geneva models lead to different upper mass limits for degenerate carbon 
ignition and hence AGB evolution. The limit is $M_{\rm up}\approx5\,M_\odot$ at
all metallicities in the Padova tracks and $M_{\rm up}\approx7\,M_\odot$ in the
Geneva tracks.} The TP-AGB phase is one of the most difficult evolutionary 
phases to model because of the combined effects of thermal pulses (i.e., helium
shell flashes), changes in surface abundance caused by heavy element dredge-up
(e.g., carbon) and important mass loss terminated by a superwind and the 
ejection of the stellar envelope (see the reviews by 
\citealt{1995MmSAI..66..627H} and \citealt{1996A&ARv...7...97H}). This phase
must be included in population synthesis models because the stochastic presence
of a few TP-AGB stars has a strong influence on the integrated colours of star
clusters (e.g., \citealt*{1990ApJ...352...96F}; \citealt{1997ApJ...479..764S};
see also Section~3.3.2 below). We appeal to recent models of TP-AGB stars which
have been calibrated using observations of stars in the Galaxy, the LMC and the
Small Magellanic Cloud (SMC). In particular, we adopt the effective 
temperatures, bolometric luminosities and lifetimes of TP-AGB stars from the
multi-metallicity models of 
\citet{1993ApJ...413..641V}.\footnote{\citet{1993ApJ...413..641V}
adopt slightly different stellar evolution parameters (e.g., helium fraction,
opacities, treatment of convection) from those used in the Padova models.
In the end, however, the duration of early-AGB evolution is similar to 
that in the Padova tracks. It is 10--25 per cent shorter for stars with initial
mass $m\la1\,M_\odot$ to 10--25 per cent longer for stars with $m=5\,M_\odot$
in the \citet{1993ApJ...413..641V} models for all the metallicities in
common with the Padova tracks ($Z=0.004$, 0.008, and 0.02). This similarity 
justifies the combination of the two sets of calculations.} These models, which
include predictions for both the optically-visible and the superwind phases, 
predict maximum TP-AGB luminosities in good agreement with those observed in 
Magellanic Cloud clusters. The models are for the metallicities $Z=0.001$, 
0.004, 0.008, and 0.016, which do not encompass all the metallicities in the 
Padova track library. For simplicity, we adopt the $Z=0.001$ prescription of 
\citet{1993ApJ...413..641V} at all metallicities $Z\le0.0004$ and their 
$Z=0.016$ prescription at all metallicities $Z\ge0.02$. 

Carbon dredge-up during TP-AGB evolution can lead to the transition from an 
oxygen-rich (M-type) to a carbon-rich (C-type) star (e.g., 
\citealt{1983ARA&A..21..271I}). Since C-type stars are much redder than M-type
stars and can dominate the integrated light of some star clusters, it is 
important to include them in the models. The minimum initial mass limit for a
carbon star to form increases with metallicity. This is supported 
observationally by the decrease in the ratio of C to M stars from the SMC, to
the LMC, to the Galactic bulge \citep*{1978Natur.271..638B}. While the 
formation of carbon stars is relatively well understood, no simple prescription 
is available to date that would allow us to describe accurately the 
transition from M to C stars over a wide range of initial masses and 
metallicities. \citet{1993A&A...267..410G} and \citet*{1995A&A...293..381G}
have computed models of TP-AGB stars, which reproduce the ratios of C to M 
stars observed in the LMC and the Galaxy. We use these models to define the 
transition from an M-type star to a C-type star in the TP-AGB evolutionary 
tracks of \citet{1993ApJ...413..641V}. We require that, for a given initial 
main-sequence mass, the relative durations of the two phases be the same as 
those in the models of \citet{1993A&A...267..410G} and 
\citet{1995A&A...293..381G}. Since these models do not extend to sub-Magellanic
($Z\la0.004$) nor super-solar ($Z>0.02$) metallicities, we apply fixed
relative durations of the M-type and C-type phases in the Padova tracks for
more extreme metallicities. As shown by \citet{2000ApJ...543..644L}, this 
simple but observationally motivated prescription for TP-AGB stars provides 
good agreement with the observed optical and near-infrared surface brightness 
fluctuations of (metal-poor) Galactic globular clusters and (more metal-rich) 
nearby elliptical galaxies.

For the post-AGB evolution, we adopt the evolutionary tracks of 
\citet{1994ApJS...92..125V}, whose calculations cover the range
of metallicities $0.001\leq Z\leq 0.016$. We use the \citet{1987A&A...188...74W}
relationship to compute the core mass of a star after ejection of the
planetary nebula (PN) at the tip of the AGB from its initial mass on the
main sequence (see \citealt{1990ARA&A..28..103W} for a review; and 
\citealt{1993ApJ...417..102M}). To each low- and intermediate-mass star in the
Padova and Geneva libraries, we then assign the post-AGB evolution computed by 
\citet{1994ApJS...92..125V} corresponding to the closest core mass and 
metallicity. These authors did not consider the evolution of stars with core
masses less than $0.569\, M_\odot$, corresponding to a main sequence progenitor
mass less than about $1.1\,M_\odot$. For lower-mass stars, we adopt the 
$0.546\, M_\odot$ post-AGB evolutionary track computed for the metallicity
$Z=0.021$ by \citet{1983ApJ...272..708S}, with an extension by 
\citet{1986A&A...154..125K}.  Since we will consider stellar population ages of
up to 20~Gyr, and the post-AGB calculations do not generally extend to this 
limit, we further supplement the tracks using white dwarf cooling models
by \citet{1987ApJ...315L..77W} at luminosities $L\la0.1L_\odot$. Following the
suggestion by Winget et al., we adopt their `pure carbon' models for masses
in the range $0.4 \leq m\leq1.0\,M_{\sun}$, in which the cooling times differ by
only 15 per cent from those in models including lighter elements. The 
prescription is thus naturally independent of the metallicity of the 
progenitor star. Specifically, we interpolate cooling ages for white dwarfs as
a function of luminosity at the masses corresponding to the 
\citet{1994ApJS...92..125V} and \citet{1983ApJ...272..708S} tracks. Since 
\citet{1987ApJ...315L..77W} do not tabulate the temperatures nor the radii of
their model white dwarfs, we assign effective temperatures as a function of 
luminosity using the slope of the white dwarf cooling sequence defined by the
calculations of \citet{1983ApJ...272..708S}, i.e., $\Delta \log T_{\rm eff}
\approx 0.23\Delta \log L$.

The resulting tracks in the Padova and Geneva libraries cover all phases
of evolution from zero-age main sequence to remnant stage for all stars 
more massive than $0.6\,M_\odot$ ($0.15\,M_\odot$ for the Padova 2000 
library). Since the main-sequence lifetime of a $0.6\,M_\odot$ star is 
nearly 80~Gyr, we supplement these libraries with multi-metallicity 
models of unevolving main-sequence stars in the mass range $0.09\leq m 
<0.6\,M_{\sun}$ \citep{1998A&A...337..403B}. These models provide smooth 
extensions of the Padova and Geneva calculations into the lower main sequence.
For the purpose of isochrone synthesis, all tracks must be resampled to a
system of evolutionary phases of equivalent physical significance 
\citep{1991ApJ...367..126C}. We define 311 such phases for low- and 
intermediate-mass stars and 260 for massive stars.

\subsection{Stellar spectral library and spectral calibration}

The second main ingredient of population synthesis models is the library
of individual stellar spectra used to describe the properties of stars at 
any position in the Hertzsprung-Russell diagram. We consider different
alternative stellar spectral libraries and different ways to calibrate them 
(see Tables~\ref{spectra-options} and \ref{calib-options}). We also refer
the reader to Table~\ref{x-accuracy} of Appendix~A for a qualitative assessment
of the spectral predictions of our model for simple stellar populations of 
various ages and metallicities computed using different spectral libraries.

\begin{table*}
 \centering
 \begin{minipage}{140mm}
  \caption{Different libraries of stellar spectra.}
  \label{spectra-options}
  \begin{tabular}{@{}llcccl@{}}
  \hline
Name & Type & Wavelength                 & Median          & Metallicity & 
Source \\
     &     &  range\footnote{The STELIB and Pickles
     libraries can be extended at shorter and longer wavelengths using the BaSeL
     library, as described in the text.} & resolving power &    range    & \\
 \hline
BaSeL & theoretical  & 91 {\AA} to 160 $\mu$m & 300 & $10^{-5}Z_{\sun}$ to
$10Z_{\sun}$ & Kurucz (1995, priv. comm.)\\
	    & & & & & Bessell et al. (1989)\\
	    & & & & & Bessell et al. (1991)\\
	    & & & & & Fluks et al. (1994)\\
	    & & & & & Allard \& Hauschildt (1995) \\
	    & & & & & Rauch (2002) \\
\\
STELIB & observational & 3200 {\AA} to 9500 {\AA} & 2000 & $-2.0<\feh<+0.50$
	    & Le~Borgne et al. (2003)\\
\\
Pickles & observational & 1205 {\AA} to 2.5 $\mu$m & 500 & $Z_{\sun}$
	    & Pickles (1998)\\
	    & & & & & Fanelli et al. (1992)\\
\hline
\end{tabular}
\end{minipage}
\end{table*}

\subsubsection{Multi-metallicity theoretical and semi-empirical libraries at
low spectral resolution}

Theoretical model atmospheres computed for wide ranges of stellar effective
temperatures, surface gravities and metallicities allow one to describe
the spectral energy distribution of any star in the HR diagram. 
\citet*{1997A&AS..125..229L} and \citet*{1998A&AS..130...65L} have compiled a
comprehensive library of model atmospheres for stars in the metallicity range
$10^{-5}Z_\odot\la Z\la 10\,Z_\odot$, encompassing all metallicities in the 
Padova track libraries (Section~2.1). The spectra cover the wavelength range 
from 91~\AA\ to 160~$\mu$m at resolving power $\lambda/\Delta \lambda\approx
200-500$. The library consists of Kurucz (1995, private communication to 
R.~Buser) spectra for the hotter stars (O--K), \citet{1989A&AS...77....1B}, 
\citet{1991A&AS...89..335B} and \citet{1994A&AS..105..311F} spectra for M 
giants, and \citet{1995ApJ...445..433A} spectra for M dwarfs.

There are three versions of this library. The first version contains the model 
spectra as originally published by their builders, only rebinned on to 
homogeneous scales of fundamental parameters (effective temperature, gravity,
metallicity) and wavelength. We refer to this library as the `BaSeL~1.0 
library'. In a second version of the library, \citet{1997A&AS..125..229L}
corrected the original model spectra for systematic deviations that become 
apparent when $UBVRIJHKL$ colour-temperature relations computed from the models
are compared to empirical calibrations. These semi-empirical blanketing 
corrections are especially important for M-star models, for which molecular 
opacity data are missing. The correction functions are expected to depend on 
the fundamental model parameters: temperature, gravity and metallicity. 
However, because of the lack of calibration standards at non-solar 
metallicities, \citet{1997A&AS..125..229L} applied the blanketing corrections
derived at solar metallicity to models of all metallicities. While uncertain,
this procedure ensures that the differentiation of spectral properties with 
respect to metallicity is at least the same as in the original library (and 
hence not worsened; see \citealt{1997A&AS..125..229L} for details).  This 
constitutes the `BaSeL~2.2 library'. Finally, \citet{2001wester} and 
\citet{2002A&A...381..524W} recently produced a new version of the library, in
which they derived semi-empirical corrections for model atmospheres at 
non-solar metallicities using metallicity-dependent $UBVRIJHKL$ colour 
calibrations. This new version is also free of some discontinuities affecting 
the colour-temperature relations of cool stars in the BaSeL~1.0 and BaSeL~2.2
libraries, which were linked to the assembly of model atmospheres from 
different sources. We refer to this as the `BaSeL~3.1' (WLBC99) library.

\begin{table}
  \caption{Different spectral calibrations.}
  \label{calib-options}
  \begin{tabular}{@{}lll@{}}
  \hline
Option & Calibration & Source \\
 \hline
BaSeL 1.0   & theoretical$^a$  & Lejeune et al. (1997) \\
            & & Lejeune et al. (1998) \\
\\
BaSeL 2.2   & semi-empirical$^b$ & Lejeune et al. (1997) \\
	    & & Lejeune et al. (1998) \\
\\
BaSeL 3.1   & semi-empirical$^c$ & Westera (2000) \\
	    & & Westera et al. (2002) \\
\hline
\end{tabular}

\medskip
$^a\,$Original calibration of model atmospheres included in the BaSeL library
(see Table~\ref{spectra-options}).\\
$^b\,$Empirical blanketing corrections derived at solar metallicity and applied
to models of all metallicities in the BaSeL library.\\
$^c\,$Metallicity-dependent blanketing corrections.\\
\end{table}

The BaSeL libraries encompass the range of stellar effective temperatures 
$2000 \le T_{\rm eff} \le 50,000\ $K. Some stars can reach temperatures 
outside this range during their evolution. In particular, in the stellar 
evolutionary tracks of Section~2.1, Wolf-Rayet stars and central stars of 
planetary nebulae can occasionally be hotter than 50,000~K. To describe the hot
radiation from these stars, we adopt the non-LTE model atmospheres of 
\citet{2002RMxAC..12..150R} for $Z = Z_\odot$ and $Z = 0.10\,Z_\odot$ that 
include metal-line blanketing from all elements from H to the Fe group (we
thank T.~Rauch for kindly providing us with these spectra). The models cover 
the temperature range $50,000 \le T_{\rm eff} \le 1,000,000\ $K at wavelengths
between 5 and 2000~{\AA} at a resolution of 0.1~{\AA}. We degrade these 
models to the BaSeL wavelength scale and extrapolate blackbody tails at 
wavelengths $\lambda>2000 \,${\AA}. We use the resulting spectra to describe
all the stars with $T_{\rm eff} \ge 50,000\ $K and $Z \ge 0.10\,Z_\odot$ in
the stellar evolutionary tracks.  For completeness, we approximate the 
spectra of stars hotter than 50,000~K at $Z=0.0004$ and $Z=0.0001$ by pure
blackbody spectra. Cool white dwarfs, when they reach temperatures cooler
than 2000~K, are also represented by pure blackbody spectra, irrespective 
of metallicity.

The BaSeL libraries do not include spectra for carbon stars nor for stars in 
the superwind phase at the tip of the TP-AGB. Our prescription for these stars
is common to all spectral libraries in Table~\ref{spectra-options} and is 
described in Section~2.2.4 below.

\subsubsection{Multi-metallicity observational library at higher spectral 
resolution}

To build models with higher spectral resolution than offered by the BaSeL 
libraries, one must appeal to observations of nearby stars. The difficulty
in this case is to sample the HR diagram in a uniform way. Recently, 
\citet{2003A&A...402..433L} have assembled a library of observed spectra of
stars in a wide range of metallicities, which they called `STELIB'. When 
building this library, Le~Borgne et al. took special care in optimizing the 
sampling of the fundamental stellar parameters across the HR diagram for the 
purpose of population synthesis modelling. The library contains 249 stellar 
spectra covering the wavelength range from 3200~{\AA} to 9500~{\AA} at a 
resolution of 3~{\AA} FWHM (corresponding to a median resolving power 
$\lambda/\Delta \lambda \approx2000$), with a sampling interval of 1~{\AA} and
a signal-to-noise ratio of typically 50 per pixel.\footnote{The STELIB spectra
were gathered from two different telescopes. At the 1~m Jacobus Kaptein Telescope
(La Palma), the instrumental setup gave a dispersion of 1.7~{\AA}/pixel and a
resolution of about 3~{\AA} FWHM. At the Siding Spring Observatory 2.3~m telescope,
the instrumental setup gave a dispersion of 1.1~{\AA}/pixel and the same 
resolution of 3~{\AA} FWHM. The two sets of spectra had to be resampled onto a
homogeneous wavelength scale for the purpose of population synthesis modelling. 
\citet{2003A&A...402..433L} adopted a uniform sampling interval of 1~{\AA}, a
`round' number close to the smallest of the two observational dispersions.} 
After correction for stellar radial velocities (Le~Borgne 2003, private 
communication), two narrow wavelength regions (6850--6950~{\AA} and 
7550--7725~{\AA}) had to be removed from the spectra because of contamination by
telluric features. For stars cooler than 7000~K, we replaced these segments in
the spectra with metallicity-dependent model atmospheres computed at 3~{\AA}
resolution using the SPECTRUM code \citep[ we thank C. Tremonti for kindly 
providing us with these computations based on the most recent Kurucz model
atmospheres]{1994AJ....107..742G}. For hotter stars, we replaced the segments
with spectra from the lower-resolution library of Pickles (see below), 
resampled to 1~{\AA}/pixel. These fixes are purely of cosmetic nature, and
we do not use the predictions of the population synthesis models in these two
narrow wavelength regions (we do not correct the STELIB spectra for the telluric
feature around 8950--9075~{\AA} that it is weaker than the other two features 
and falls in a noisier region of the spectra).

Most stars in the STELIB library were selected from the catalog of 
\citet{1992A&AS...95..273C}, which includes [Fe/H] determinations from 
high-resolution spectroscopic observations of stars in open and globular 
clusters in the Galaxy and of supergiant stars in the Magellanic Clouds. The
STELIB library contains stars with metallicities in the range $-2.0 <{\rm 
[Fe/H]} < +0.50$, spectral types from O5 to M9 and luminosity classes from
I to V. The coverage in spectral type is not uniform at all metallicities (see
Appendix A): hot ($T_{\rm eff}\ga 10,000\,$K) stars are under-represented at 
non-solar metallicities, and the library lacks very cool ($T_{\rm eff}
<3200\,$K) stars at all metallicities. These limitations are not
critical. The spectra of hot stars are not expected to depend strongly on 
metallicity because the opacities in these stars are dominated by electron 
scattering. Thus, the spectra of hot stars with solar metallicity should be 
representative of hot stars at all but the most extreme metallicities. Also, the
lack of cool M-dwarf stars has a negligible influence on model predictions, 
because these stars do not contribute significantly to the integrated light of
stellar populations (as found when adopting representative spectra 
for these stars; see Appendix A). For the coolest giant stars, we adopt in any 
case the prescription outlined in Section~2.2.4 below.

The main interest of the STELIB library is that it enables the interpretation
of integrated spectra of star clusters and galaxies taken at relatively high
resolution in the wavelength range 3200--9500~{\AA}. To allow for a consistent
modelling of spectral properties outside this range, we must extend the 
STELIB spectra at ultraviolet and infrared wavelengths using one of the 
spectral libraries described above. We consider three different types of 
extensions, corresponding to the three colour-temperature calibrations of the
BaSeL~1.0, 2.2 and 3.1 libraries (Section~2.2.1). To assign STELIB spectra to
stars on the evolutionary tracks, we therefore proceed as follows (the reader
is referred to Appendix A for more detail). We first distribute the stars
in several metallicity bins centered on the metallicities for which tracks are
available (Section~2.1).  Some stars with intermediate metallicities may be
included into two consecutive bins, while hot solar-metallicity stars are 
included in all bins.  We then select one of the three BaSeL libraries
to set the colour-temperature scale.\footnote{The effective 
temperatures published by \citet{2003A&A...402..433L} for the STELIB stars are
incomplete and were not derived in a homogeneous way. We prefer to rely on the
homogeneous colour-temperature scales of the BaSeL libraries.} For each
metallicity, we assign to each $\log g$--$\log T_{\rm eff}$ position in the HR
diagram the STELIB spectrum of the associated luminosity class that best matches
the BaSeL spectrum corresponding to these values of $\log g$ and $\log T_{\rm
eff}$ (here $\log g$ is the gravity).  We then extend the selected STELIB spectrum
blueward of 3200~{\AA} and redward of 9500~{\AA} with the ultraviolet and infrared
ends of the BaSeL spectrum. There are, therefore, three possible implementations
of the STELIB library in our model, which we refer to as the `STELIB/BaSeL~1.0',
the `STELIB/BaSeL~2.2' and the `STELIB/BaSeL~3.1' libraries. For solar 
metallicity, we can also use the Pickles library described below to extend the
STELIB/BaSeL~3.1 models blueward of 3200~{\AA} and redward of 9500~{\AA}
(Section~4.1 and Fig.~\ref{fig_ssp}).

\subsubsection{Solar-metallicity observational library with wider spectral
coverage}

\citet{1998PASP..110..863P} has assembled a library of 131 Galactic stars in 
wide ranges of spectral types (O5--M10) and luminosity classes (I--V) in three
metallicity groups (11 metal-weak, 12 metal-rich, and 108 solar-metallicity 
stars). The metal-weak and metal-rich stars sample very sparsely the HR diagram
and do not allow us to build accurate population synthesis models. We therefore
focus on solar-metallicity stars, for which the sampling is adequate. The 
interest of the \citet{1998PASP..110..863P} library is that it has a wider 
spectral coverage than the STELIB library at solar metallicity, despite the
coarser resolution. The spectra extend over the wavelength range from 1150~{\AA}
to 2.5~$\mu$m with a sampling interval of 5~{\AA}/pixel and a median resolving
power $\lambda/\Delta \lambda\approx 500$ (corresponding to the highest 
resolution at which spectra are available for all stars in the Pickles library).
The library does not include main-sequence and subgiant stars hotter than 
40,000~K, giant stars hotter than 32,000~K and supergiant stars hotter than
26,000~K and cooler than 4000~K. When needed, we select spectra for these stars
from the solar-metallicity BaSeL~3.1 library described above.

The quality of the spectra in the ultraviolet is of particular importance
for application to studies of distant galaxies. In the Pickles library, the
spectra at ultraviolet wavelengths are based on a limited number of {\it
International Ultraviolet Explorer} ({\it IUE}) observations for each stellar
type.  \citet{1992ApJS...82..197F} have compiled more refined average {\it IUE} 
spectra as a function of spectral type and luminosity class from a sample
of 218 stars. The sampling interval of these spectra is 1--1.2~{\AA} from 1205
to 1935~{\AA} and 2~{\AA} from 1935 to 3150~{\AA}. We replace the spectra of 
the Pickles library at wavelengths from 1205 to 3185~\AA\ by the type-averaged
spectra compiled by \citet{1992ApJS...82..197F}. For completeness, we extend 
the spectra further into the extreme ultraviolet using spectra from the 
BaSeL~3.1 library at wavelengths from 91 to 1195~{\AA}.

We also use BaSeL~3.1 spectra to extend the Pickles spectra into the
infrared at wavelengths from 2.5~$\mu$m to 160~$\mu$m. For cool M-giant stars,
we adopt a more refined prescription. The spectra of M0--M10 giant stars are 
the only non-observed ones in the Pickles library, as they are based on
the synthetic M-giant spectra computed by \citet{1994A&AS..105..311F}. To
extend these spectra into the infrared, we appeal to model atmospheres by 
\citet{1999schulthe}. These have a more refined definition of the strong 
infrared absorption features in cool stars than the BaSeL spectra (we thank 
M.~Schultheis for kindly providing us with these spectra). The Schultheis et
al. spectra cover the wavelength range from 5000~{\AA} to 10~$\mu$m and are 
available for 10 equally spaced stellar temperatures in the range $2600\le 
T_{\rm eff}\le4400$~K (the emission of these stars is negligible at wavelengths
less than 5000~{\AA}). The sampling interval increases from 2.5~{\AA} at the 
short wavelength end ($\lambda/\Delta\lambda \approx 2000$) to 400~{\AA} at the
long wavelength end ($\lambda/\Delta \lambda \approx250$). At wavelengths 
between 5000~{\AA} and 2.5~$\mu$m, the colours computed from these spectra
agree well with those computed from the Fluks et al. models in the Pickles 
library. We therefore extend the spectra of M-giant stars at wavelengths from 
2.5~$\mu$m to 10~$\mu$m in the Pickles library using the Schultheis et
al. spectra. For completeness, we extend the resulting M-star spectra further
into the infrared using spectra from the BaSeL~3.1 library at wavelengths from
10~$\mu$m to 160~$\mu$m. In what follows, we refer to this modified version of
the \citet{1998PASP..110..863P} library simply as the `Pickles library'.

\subsubsection{Carbon stars and stars in the superwind phase}

None of the libraries described above includes spectra for C-type stars
nor for stars in the superwind phase at the tip of the TP-AGB (Section~{2.1}).
We construct period-averaged spectra for these stars, based on models and 
observations of Galactic stars. We adopt these spectra to represent upper
TP-AGB stars of all metallicities in our model.

We construct period-averaged spectra for C-type TP-AGB stars as follows.
We use solar-metallicity model atmospheres for carbon stars with temperatures
in the range $2600 \la T_{\rm eff}\la 3400$~K from 
\citet[~we thank R.~Loidl for kindly providing us with these 
spectra]{2000ESASP.456..299H}. The spectra cover the wavelength range from
2500~{\AA} to 12.5~$\mu$m. The sampling interval increases from 2.5~\AA\ at 
the short-wavelength end ($\lambda/\Delta\lambda\approx1000$) to 800~\AA\ at
the long-wavelength end ($\lambda/\Delta\lambda\approx200$). The spectral 
features in the model spectra are in reasonable agreement with observations 
of carbon stars \citep*{2001A&A...371.1065L}. However, we find that the model
$UBVRIJHK$ broadband colours do not reproduce well the observations of 39 carbon
stars by \citet{1965ApJ...141..161M}. We therefore apply an empirical 
correction to the model spectra as follows. We first derive mean colour-colour
relations for C stars from the sample of \citet{1965ApJ...141..161M} by fitting
2nd~order polynomials to the relations defined by the data. We derive in the 
same way a mean relation between $K$-band bolometric correction BC$_K$ and \jk\
colour.  These mean relations are taken to represent period-averaged 
observations.  The temperature scale proposed by \citet{1965ApJ...141..161M}
for carbon stars does not appear to be robust (e.g., 
\citealt*{1996AJ....112..294D}). Thus, we prefer to adjust the calibration of
$T_{\rm eff}$ as a function of \jk\ colour in such a way that the reddest stars
observed by \citet{1965ApJ...141..161M} have roughly the temperature of the
coolest C-type stars in our model ($T_{\rm eff} \approx2600\,$K). Based on
these relations, we apply a smooth continuum correction to each 
\citet{2000ESASP.456..299H} spectrum in order to reproduce the mean observed
$UBVRIJHK$ colours at the corresponding $T_{\rm eff}$. The absolute scale of
the corrected spectrum is then set by the relation between BC$_K$ and \jk\ 
colour. 

We also construct spectra for stars in the superwind phase at the end of the
TP-AGB evolution. These stars may be of either M or C type, depending on
initial mass, metallicity and age. Their spectra are difficult to model 
because of the influence of expanding circumstellar shells of gas and dust.
To describe stars in the superwind phase, we therefore rely primarily on 
observations.  \citet{1996A&A...314..896L} and \citet{1997A&A...324.1059L}
have assembled broadband spectral energy distributions of 27 oxygen-rich and 23
carbon-rich stars with circumstellar dust shells. They also derived bolometric
luminosities from the known pulsation periods of all stars. For only some 
objects, however, is a complete set of optical-infrared observations available
(we thank T. Le~Bertre for kindly providing us with some unpublished $JHKL$ 
data). We extract a sample of 16 TP-AGB stars with complete $VRIJHKL$ 
information and partial $UB$ information (this includes 4 M-type and 12 C-type
stars). The colours of M-type and C-type stars follow similar relations to 
within the observed scatter for this small sample. We thus do not distinguish 
between M-type and C-type stars and use the sample as a whole to build spectra
for TP-AGB stars in the superwind phase in our model. As before, we fit mean 
colour-colour relations to the data and take these to represent period-averaged
observations. Most optical/infrared colours correlate well with bolometric
luminosity $\log L$. In addition, there is a tight correlation between 
$K$-band bolometric correction BC$_K$ and $\log L$. These relations are useful
because the effective temperatures of the stars are not known. In the 
evolutionary tracks, the bolometric luminosity of stars in the superwind phase
increases with initial mass at roughly constant $T_{\rm eff} \approx 
2800\,$K. By analogy with our approach above, we apply smooth continuum 
corrections to the 2800~K carbon-star model of \citet{2000ESASP.456..299H} to
generate 12 new spectra reproducing the observed colours of TP-AGB stars in
the superwind phase for different luminosities in the range $3.55\leq\log 
L/L_\odot\leq 4.65$.  

\subsection{Isochrone synthesis}

In this paper, we use the {\em isochrone synthesis} technique to compute the 
spectral evolution of stellar populations (\citealt{1991ApJ...367..126C}; 
\citealt{1993ApJ...405..538B}). This technique is based on the property that
stellar populations with any star formation history can be expanded in series
of instantaneous starbursts, conventionally named `simple stellar populations'
(SSPs). The spectral energy distribution at time $t$ of a stellar population
characterized by a star formation rate $\psi(t)$ and a metal-enrichment law 
$\zeta(t)$ can be written (e.g., \citealt{1980FCPh....5..287T})
\begin{equation}
F_\lambda(t) = \int_0^t\,\psi(t-t')\,S_\lambda\left[t',\zeta(t-t')\right]\,
dt'\,,
\label{convol}
\end{equation}
where $S_\lambda\left[t',\zeta(t-t')\right]$ is the power radiated per 
unit wavelength per unit initial mass by an SSP of age $t'$ and metallicity
$\zeta(t-t')$. The above expression assumes that the initial mass function
(IMF) is independent of time.

The function $S_\lambda\left[t',\zeta(t-t')\right]$ is the sum of the spectra
of stars defining the isochrone of an SSP of metallicity $\zeta(t-t')$ at
age $t'$. To compute $S_\lambda(t',Z_i)$ at a given metallicity $Z_i$ of
the stellar evolutionary tracks (Table~\ref{tracks-options}), we interpolate
the isochrone at age $t'$ from the tracks in the HR diagram. In practice, each 
evolutionary stage defined in the tracks is interpolated separately (Section 2.1).
The different evolutionary stages along the isochrone are populated by stars of 
different initial masses in proportions given by the IMF weight $\phi(m)$
[defined such that $\phi(m)dm$ is the number of stars born with masses between $m$
and $m+dm$]. We then use one of the spectral libraries described in Section~2.2 to
assign spectra to stars in the various evolutionary stages. The spectral energy
distribution of the SSP is obtained by summing the spectra of individual stars
along the isochrone.

The IMF is an adjustable parameter of the model. Except when otherwise 
indicated, we adopt in this paper the parametrization by \citet[ his table
1]{2003PASPchab} of the single-star IMF in the Galactic disc. This is
\begin{eqnarray}
\phi(\log m) 
\propto\cases{
\exp\left[ -{{(\log m - \log m_c)^2}\over{2\sigma^2}}\right]\,,
&for $m\leq 1M_\odot$\,,\cr
m^{-1.3}\,,
&for $m >   1M_\odot$\,,\cr}
\label{imf}
\end{eqnarray}
with $m_c=0.08M_\odot$ and $\sigma=0.69$  (the two expressions in 
equation~\ref{imf} are forced to coincide at $1M_\odot$). The spectral properties
obtained using the above IMF are very similar to those obtained using the 
\citet{2001MNRAS.322..231K} universal IMF (see Fig.~\ref{fig_imf} below). We 
adopt here the \citet{2003PASPchab} IMF because it is physically motivated and
provides a better fit to counts of low-mass stars and brown dwarfs in the 
Galactic disc (\citealt{2001ApJ...554.1274C}; \citealt{2002ApJ...567..304C};
\citealt{2003ApJ...586L.133C}). For reference, the 
\citet{1955ApJ...121..161S} IMF corresponds to $\phi(\log
m)\propto m^{-1.35}$, or equivalently $\phi(m)\propto m^{-2.35}$. Unless
otherwise specified, we adopt lower and upper IMF mass cutoffs $m_L = 
0.1~M_\odot$ and $m_U = 100~M_\odot$. As in \citet{1993ApJ...405..538B}, the
spectral energy distribution of a model SSP is normalized to a total mass of
$1~M_\odot$ in stars at age $t'=0$, and the spectra are computed at 221 
unequally spaced time steps from 0 to 20 Gyr. Each spectrum covers the 
wavelength range from 91 \AA\ to 160 $\mu$m, with a resolution that depends on
the spectral library employed.

\section{Photometric Evolution}

In the isochrone synthesis framework, the spectral evolution of simple stellar
populations (SSPs) is the most fundamental prediction of population synthesis
models. It determines the spectral evolution of stellar populations with any
history of star formation (Section 2.3). In this section, we examine the predictions
of our model for the {\em photometric} evolution of SSPs. This allows us to
illustrate the basic influence of the various adjustable parameters on model 
properties. We compare our results with previous work. We also compare the 
photometric properties of the model with observations of nearby star clusters.

In all applications in the remainder of this paper, we adopt a `standard' 
reference model computed using the Padova~1994 evolutionary tracks, the 
STELIB/BaSeL~3.1 spectral library and the IMF of equation~(\ref{imf}) truncated
at $0.1\,M_\odot$ and $100\,M_\odot$. We mention in Section~3.1 below the reason
for preferring the Padova~1994 tracks over the Padova~2000 tracks for the
standard model.

\begin{figure}
\includegraphics[width=90mm]{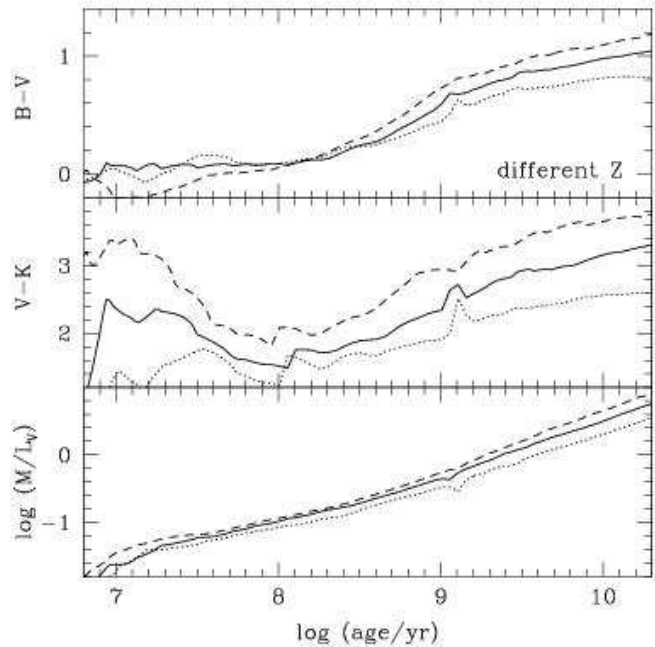}
\caption{Evolution of the \bv\ and \vk\ colours and stellar mass-to-light 
ratio \mlv\ of simple stellar populations for different metallicities, 
$Z=0.004$ (dotted line), $Z=Z_{\sun}=0.02$ (solid line) and $Z=0.05$ (dashed 
line), for the standard model of Section 3. All models have the 
\citet{2003PASPchab} IMF truncated at $0.1\,M_\odot$ and $100\,M_\odot$ (see
equation~\ref{imf}).}
\label{fig_z}
\end{figure}

\subsection{Influence of the adjustable parameters}

The spectral evolution of an SSP depends primarily on the assumed metallicity,
stellar evolution prescription, stellar spectral library and IMF. Here, we 
illustrate the influence of these adjustable parameters on the evolution of 
the \bv\ and \vk\ colours and stellar mass-to-visual light ratio \mlv\ of an
SSP. Optical and near-infrared colours reflect the relative contributions of 
hot and cool stars to the integrated light, while the stellar mass-to-light 
ratio reflects the absolute magnitude scale of the model. When computing \mlv,
we account for the mass lost by evolved stars to the interstellar medium in 
the form of winds, planetary nebulae and supernova ejecta.

Fig.~\ref{fig_z} shows the evolution of the \bv\ and \vk\ colours and \mlv\ for
three different metallicities, $Z=0.004$, $Z=Z_{\sun}=0.02$ and $Z=0.05$, for
our standard SSP model. The irregularities in the photometric evolution arise
both from the discrete sampling of initial stellar masses in the track library
and from `phase' transitions in stellar evolution. For example, the evolution
of low-mass stars through the helium flash causes a characteristic feature in
all properties in Fig.~\ref{fig_z} at ages near $10^9\,$yr. At fixed age, the
main effect of increasing metallicity is to redden the colours and increase
\mlv. The reason for this is that, at fixed initial stellar mass, lowering 
metallicity causes stars to evolve at higher effective temperatures and higher
luminosities (\citealt{1992A&AS...96..269S}; \citealt{1994A&AS..104..365F},
\citealt{2000A&AS..141..371G}). Another noticeable effect of varying $Z$ is to
change the relative numbers of red and blue supergiants. The evolution of the 
\bv\ colour at early ages in Fig.~\ref{fig_z} shows that the signature of red 
supergiants in the colour evolution of an SSP depends crucially on metallicity
(see also \citealt{1994A&A...284..749C}). We note that increasing metallicity at
fixed age has a similar effect as increasing age at fixed metallicity, which
leads to the well-known age-metallicity degeneracy. 

\begin{figure}
\includegraphics[width=90mm]{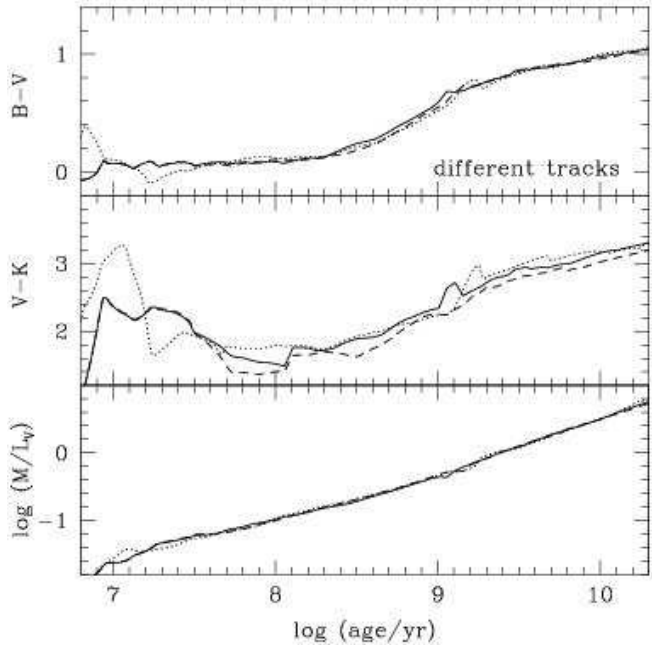}
\caption{Evolution of the \bv\ and \vk\ colours and stellar mass-to-light ratio
\mlv\ of simple stellar populations of solar metallicity computed using
the Geneva (dotted line), Padova 1994 (standard model; solid line) and Padova
2000 (dashed line) stellar evolution prescriptions and the STELIB/BaSeL~3.1 
spectral calibration. All models have the \citet{2003PASPchab} IMF truncated at
$0.1\,M_\odot$ and $100\,M_\odot$ (see equation~\ref{imf}).}
\label{fig_tracks}
\end{figure}

In Fig.~\ref{fig_tracks}, we illustrate the influence of the stellar evolution
prescription on the predicted photometric evolution of an SSP for fixed (solar)
metallicity and fixed (STELIB/BaSeL~3.1) spectral calibration. We show models
computed using the Padova 1994, the Geneva and the Padova 2000 track libraries
(Section~2.1). The largest difference between the Padova 1994 and Geneva 
prescriptions arises at early ages and results from the larger number of 
evolved, blue massive (Wolf-Rayet) stars in the Padova models than in the Geneva 
models (see fig.~2b of \citealt{1996fstg.conf..275C}). Also, since the minimum
mass for quiet helium ignition is lower in the Geneva model than in the 
Padova 1994 model ($1.9\,\msun$ versus $2.2\,\msun$), the photometric signature
of the helium flash occurs at slightly later ages in the Geneva model
in Fig.~\ref{fig_tracks}. Differences between the Padova 1994 and Padova 2000 
track libraries pertain only to stars less massive than $7\,\msun$, with turnoff 
ages greater than about $5 \times10^7 \,$yr (Section~2.1). In the Padova 2000
model, the finer resolution in initial stellar mass around $2.0\,\msun$ makes the
evolution through the helium flash much smoother than in the Padova 1994 model.
At late ages, the \vk\ colour is significantly bluer in the 
Padova 2000 model than in the Padova 1994 model. The reason for this is that 
the red giant branch is 50 to 200~K warmer (from bottom to tip) in the Padova 
2000 tracks than in the Padova 1994 tracks. As a result, the integrated \vk\ 
colour of a solar-metallicity SSP in the Padova 2000 model reaches values 
typical of old elliptical galaxies ($\vk\sim 3.0$--3.3 along the 
colour-magnitude relation; \citealt*{1992MNRAS.254..601B}) only at ages 
15--20~Gyr. Since this is older than currently favored estimates of the age of
the Universe, and since the giant-branch temperature in the Padova 2000 tracks
has not been tested against observational calibrations (e.g. 
\citealt*{1981ApJ...246..842F}), we have adopted here the Padova 1994 library
rather than the Padova 2000 library in our standard model (see above).\footnote{It
is intriguing that the Padova~2000 models, which include more recent input 
physics than the Padova~1994 models, tend to produce worse agreement with 
observed galaxy colours. The relatively high giant branch temperatures 
in the Padova~2000 models, though attributable to the adoption of new 
opacities, could be subject to significant coding uncertainties 
(L. Girardi 2002, private communication). This is supported by the fact that the 
implementation of the same input physics as used in the Padova~2000 models into a
different code produces giant branch temperatures in much better agreement with
those of the Padova 1994 models (A. Weiss 2002, private communication). We regard
the agreement between the \citet{2002A&A...391..195G} model and our standard model
at late ages in Fig.~\ref{fig_photocomp} as fortuitous, as the spectral calibration
adopted by \citet{2002A&A...391..195G} relies on purely theoretical model 
atmospheres, which do not reproduce well the colour-temperature relations of cool
stars (e.g., \citealt{1997A&AS..125..229L}).} 

\begin{figure}
\includegraphics[width=90mm]{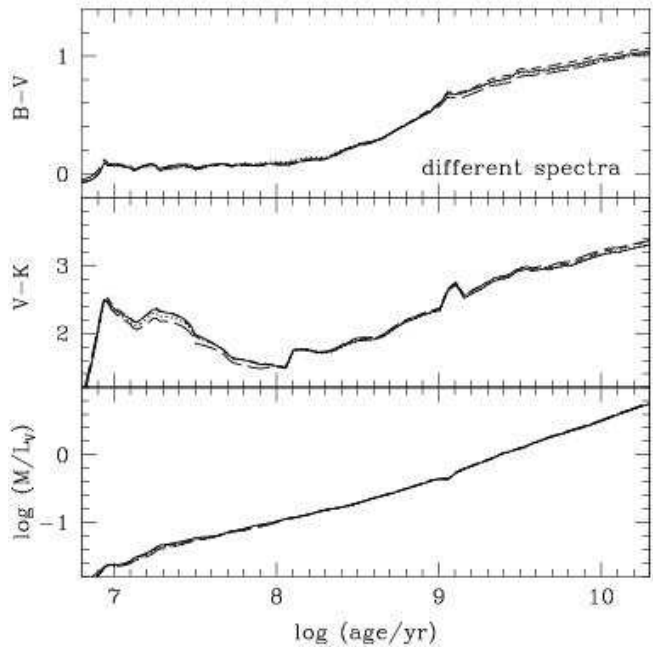}
\caption{Evolution of the \bv\ and \vk\ colours and stellar mass-to-light ratio
\mlv\ of simple stellar populations of solar metallicity computed using
the Padova 1994 stellar evolution prescription and the BaSeL~3.1 (dotted line),
STELIB/BaSeL~1.0 (short-dashed line), STELIB/BaSeL~3.1 (standard model; solid
line) and Pickles (long-dashed line) spectral calibrations. All models have the
\citet{2003PASPchab} IMF truncated at $0.1\,M_\odot$ and $100\,M_\odot$ (see 
equation~\ref{imf}).}
\label{fig_calib}
\end{figure}

We now consider the influence of the spectral calibration on the photometric
evolution of an SSP for fixed (solar) metallicity and fixed (Padova 1994) 
stellar evolution prescription. In Fig.~\ref{fig_calib}, we compare the results
obtained with four different spectral libraries: the STELIB/BaSeL~3.1 library
(standard model); the BaSeL~3.1 library; the STELIB/BaSeL~1.0 library; and the
Pickles library (recall that, at solar metallicity, the BaSeL~3.1 library is
identical to the BaSeL~2.2 library; Section~2.2.1). Fig.~\ref{fig_calib} shows 
that the differences between these spectral calibrations have only a weak 
influence on the predicted photometric evolution of an SSP. The good agreement
between the STELIB/BaSeL~3.1, the BaSeL~3.1 and the Pickles calibrations follows
in part from the consistent colour-temperature scale of the three libraries. 
Also the empirical corrections applied by \citet{1997A&AS..125..229L} and
\citet{2002A&A...381..524W} to the BaSeL~1.0 spectra, illustrated by the 
differences between the STELIB/BaSeL~3.1 and STELIB/BaSeL~1.0 models in 
Fig.~\ref{fig_calib}, imply changes of at most a few hundredths of a magnitude
in the evolution of the \bv\ and \vk\ colours. It is important to note that 
the spectral calibration has a stronger influence on observable quantities 
which are more sensitive than integrated colours to the details of the stellar
luminosity function, such as colour-magnitude diagrams (Section~3.3) and 
surface brightness fluctuations \citep{2000ApJ...543..644L}.  Fig.~8 of 
\citet{2000ApJ...543..644L} shows that, for example, the observed near-infrared
surface brightness fluctuations of nearby galaxies clearly favor the 
BaSeL~2.2/3.1 spectral calibration over the BaSeL~1.0 one.

\begin{figure}
\includegraphics[width=90mm]{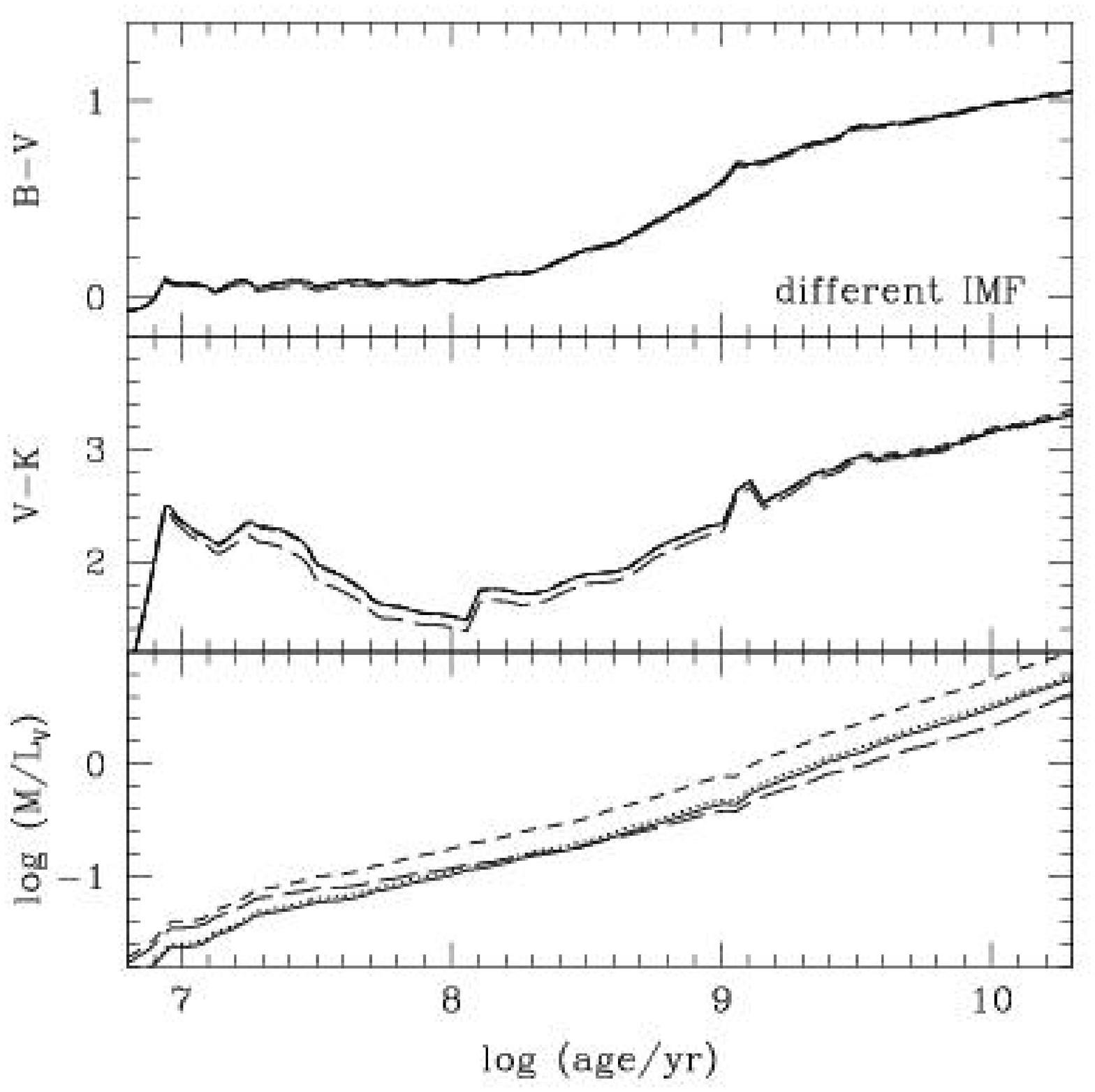}
\caption{Evolution of the \bv\ and \vk\ colours and stellar mass-to-light ratio
\mlv\ of simple stellar populations of solar metallicity computed using the 
Padova 1994 stellar evolution prescription and the STELIB/BaSeL~3.1 spectral
calibration, for different IMFs: \citet[ standard model; solid line; see
equation~\ref{imf}]{2003PASPchab}, \citet[ dotted line]{2001MNRAS.322..231K},
\citet[ short-dashed line]{1955ApJ...121..161S} and 
\citet[ long-dashed line]{1998simf.conf..201S}. All IMFs are truncated at 
$0.1\,M_\odot$ and $100\,M_\odot$.}
\label{fig_imf}
\end{figure}

It is also of interest to examine the influence of the IMF on the photometric
evolution of an SSP for fixed (solar) metallicity, fixed (Padova 1994) stellar
evolution prescription and fixed (STELIB/BaSeL~3.1) spectral calibration. 
Fig.~\ref{fig_imf} shows the evolution of the \bv\ and \vk\ colours and 
\mlv\ for four different IMFs: \citet*[ see equation~\ref{imf} 
above]{2003PASPchab}, \citet[ universal IMF]{2001MNRAS.322..231K}, \citet{1955ApJ...121..161S}
and \citet{1998simf.conf..201S}. In all cases, the IMF is truncated at 
$0.1\,M_\odot$ and $100\,M_\odot$. The evolution of the \bv\ colour does not
depend sensitively on the IMF, because the optical light is dominated at any
age by stars near the turnoff. The \vk\ colour is slightly more sensitive to 
the relative weights of stars of different masses along the isochrone, 
especially at ages less than about $10^9\,$yr, when the mass of the most 
evolved stars differs significantly from the turnoff mass. The \mlv\ ratio
is far more sensitive to the shape of the IMF, especially near the low-mass end
that determines the fraction of the total mass of the stellar population locked
into faint, slowly evolving stars. For reference, the fraction of mass returned
to the ISM by evolved stars at the age of $10\,$Gyr is 31, 44, 46, and 48 
per cent for the Salpeter, the Scalo, the Kroupa and the Chabrier IMFs, 
respectively.

\subsection{Comparison with previous work}

Most current population synthesis models rely on readily available computations
of stellar evolutionary tracks and stellar atmospheres, such as those mentioned
in Section~2.1 and Section~2.2.1 above. In general, however, publically 
available stellar evolutionary tracks do not include the uncertain evolution of
stars beyond the early-AGB phase. Also, the widely used model atmospheres of 
\citet*[ and other releases]{1992kurmod} do not include spectra of stars 
outside the temperature range $3500\,{\rm K}\le T_{\rm eff} \le 50,000\,$K. We
therefore expect differences between our model and previous work to originate
mainly from our observationally motivated prescription for TP-AGB stars, the 
spectral calibration of very hot and very cool (giant) stars and the adoption
of a new library of observed stellar spectra at various metallicities.

\begin{figure}
\includegraphics[width=90mm]{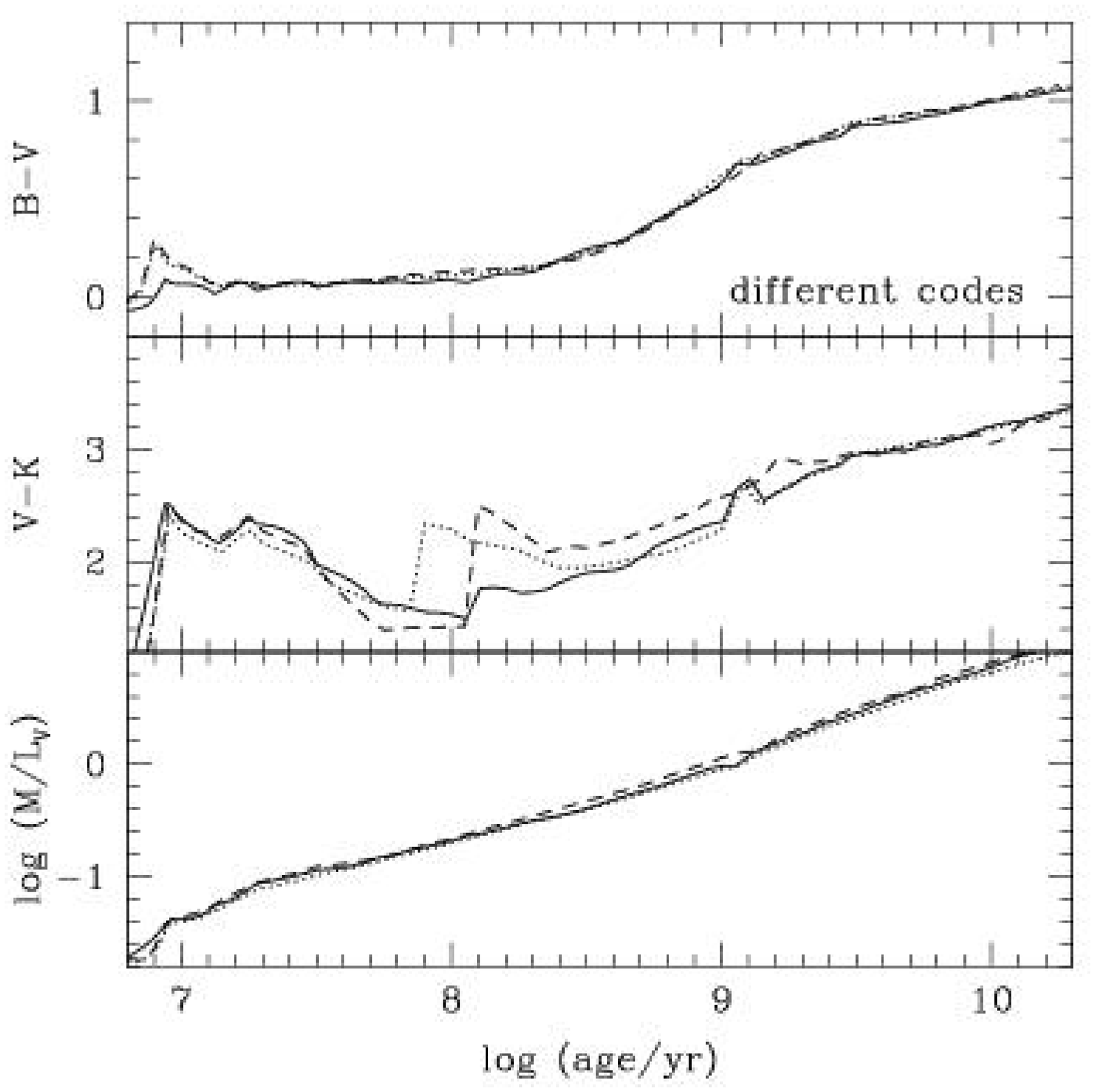}
\caption{Evolution of the \bv\ and \vk\ colours and stellar mass-to-light ratio
\mlv\ of simple stellar populations of solar metallicity computed using our
model (with the Padova 1994 stellar evolution prescription and the 
STELIB/BaSeL~2.2 spectral calibration; solid line), the 
\citet{1997A&A...326..950F} P{\small \'EGASE} version 2.0 model (dotted line)
and the \citet{2002A&A...391..195G} model (dashed line). All models have the
\citet{2001MNRAS.322..231K} {\em present-day} IMF truncated at
$0.01\,M_\odot$ and $100\,M_\odot$.}
\label{fig_photocomp}
\end{figure}

In Fig.~\ref{fig_photocomp}, we compare the evolution of the \bv\ and \vk\ 
colours and the mass-to-visual light ratio \mlv\ of a solar-metallicity SSP
predicted by our model with those predicted by two publically available 
population synthesis codes: the P{\small \'EGASE} model 
(\citealt{1997A&A...326..950F}; version 2.0) and the \citet{2002A&A...391..195G}
model. For practical reasons, we adopt in all models the same IMF as in 
\citet{2002A&A...391..195G}, i.e., the \citet{2001MNRAS.322..231K}
present-day IMF truncated at $0.01\,\msun$ and $100\, \msun$.\footnote{The 
present-day IMF in equation~(6) of \citet{2001MNRAS.322..231K} is much steeper
at masses between $0.08\,M_\odot$ and $1.0\,M_\odot$ than the universal 
Galactic-disc IMF proposed in his equation~(2). The universal IMF should be
better suited to studies of the past history of star formation in galaxies.} 
Also, for the purpose of this comparison, we compute our model using the Padova
1994 stellar evolution prescription and the STELIB/BaSeL~2.2 spectral 
calibration (which is identical to the BaSeL~3.1 calibration for solar 
metallicity; Section~2.2.1).

The P{\small \'EGASE} model shows good general agreement with our model in
Fig.~\ref{fig_photocomp}. There are marked discrepancies at ages around
$10^7\,$yr, where the P{\small \'EGASE} model is redder in \bv\ but bluer 
in \vk\ than our model, and at ages around $10^8\,$yr, where it is nearly a 
magnitude redder in \vk. General agreement is expected because the P{\small 
\'EGASE} model relies on the same Padova 1994 tracks as used in our model to
describe the evolution of stars up to the end of the early-AGB and on the same
BaSeL~2.2 spectral calibration. The discrepancy at early ages arises from a 
difference in the spectral calibration of stars hotter than 50,000~K. In the
P{\small \'EGASE} model, the spectra of these stars are taken from 
\citet{1987MNRAS.228..759C}, while in our model, they are taken from the more
recent computations of \citet{2002RMxAC..12..150R}. The discrepancy in the \vk\
colour at ages around $10^8\,$yr arises from a different prescription for 
TP-AGB evolution. \citet{1997A&A...326..950F} use `typical' TP-AGB luminosities
and evolutionary time-scales from \citet{1993A&A...267..410G}, while in our 
model, the evolution through this phase and its spectral calibration are more
refined (Section~2.1 and Section~2.2). 

The \citet{2002A&A...391..195G} model in Fig.~\ref{fig_photocomp} relies on the
Padova 2000 stellar evolutionary tracks and on model atmospheres by 
\citet*{1997A&A...324..432C} and \citet{1994A&AS..105..311F}. These model 
atmospheres do not include any empirical colour-temperature correction and are
akin to the Kurucz (1995, private communication to R.~Buser) and
\citet{1994A&AS..105..311F} spectra included in the BaSeL~1.0 library. It is
interesting to note that, when combined with these purely theoretical model 
atmospheres, the Padova 2000 evolutionary tracks, in which the giant branch is
relatively warm (Section~3.1), produce \bv\ and \vk\ colours in good agreement
with those predicted both by the P{\small \'EGASE} model and by our model at
late ages. At ages less than $10^7\,$yr and around $10^8\,$yr, the 
\citet{2002A&A...391..195G} model deviates from our model in a similar way as
the P{\small \'EGASE} model. The discrepancy at early ages is caused again by a
different treatment of stars hotter than 50,000~K, which 
\citet{2002A&A...391..195G} describe as simple blackbody spectra. The 
discrepancy at ages $10^8$--$10^9\,$yr follows primarily from the treatment of
TP-AGB evolution, which is based on a semi-analytic prescription by 
\citet{1998MNRAS.300..533G} in the \citet{2002A&A...391..195G} model. It is 
worth recalling that our prescription for TP-AGB evolution has been tested 
successfully against observed optical and near-infrared surface brightness 
fluctuations of nearby star clusters and galaxies (Section 2).

\subsection{Comparison with observations of star clusters}

\subsubsection{Colour-magnitude diagrams}

\begin{table*}
 \centering
 \begin{minipage}{140mm}
  \caption{Star cluster data.}
  \label{cluster-data}
  \begin{tabular}{@{}llcclllcl@{}}
  \hline
Cluster & Alias & $(m-M)_0$ & $E(\bv)$ & $\feh_{\rm obs}$ & 
	$Z_{\rm mod}$ & $\feh_{\rm mod}$ & $t_{\rm mod}$/Gyr
	& References\\
NGC6397 &             & 12.31 & 0.18 & $-1.94$ & 0.0004 & $-1.65$ & 14  
		& 1, 2, 3 \\
NGC6809 & M55         & 13.82 & 0.07 & $-1.80$ & 0.0004 & $-1.65$ & 13  
		& 4, 5 \\
NGC5139 & $\omega$Cen & 13.92 & 0.12 & $-1.62$ & 0.0004 & $-1.65$ & 13  
		& 4, 5 \\
NGC104  & 47Tuc       & 13.32 & 0.05 & $-0.71$ & 0.004  & $-0.64$ & 13  
		& 4, 6 \\
NGC6528 &             & 14.45 & 0.52 & $-0.35$ & 0.008  & $-0.33$ & 13  
		& 7, 8 \\
NGC6553 &             & 13.60 & 0.70 & $-0.35$ & 0.008  & $-0.33$ & 13  
		& 7, 8, 9 \\
NGC2682 & M67         &  9.50 & 0.06 & $+0.01$ & 0.02   & $+0.09$ &  4  
		& 10, 11, 12, 13, 14 \\
Hyades  &             &  3.40 & 0.00 & $+0.15$ & 0.02   & $+0.09$ & 0.7 
		& 15, 16, 17, 18 \\
\hline
\end{tabular}

\medskip
(1) \citet{1998ApJ...492L..37K};
(2) \citet{astro-ph/9910312};
(3) \citet{1997A&AS..121..455K};
(4) \citet{2000A&AS..144....5R};
(5) \citet{2000rosenb};
(6) \citet{1998AcA....48..439K};
(7) \citet{1997AJ....114.1531B};
(8) \citet{1995Natur.377..701O};
(9) \citet{1998A&A...331...70G};
(10) \citet{1964ApJ...140..130E};
(11) \citet{1991AJ....101..541G};
(12) \citet{1984AJ.....89..487J};
(13) \citet*{1993AJ....106..181M};
(14) \citet{1971ApJ...168..393R};
(15) \citet{1988ApJ...325..798M};
(16) \citet{1974ApJ...193..359U};
(17) \citet{1977AJ.....82..978U};
(18) \citet{2000mermil}.
\end{minipage}
\end{table*}

\begin{figure*}
\centering
\begin{minipage}{140mm}
\hspace{0cm}
\includegraphics[height=140mm,angle=270]{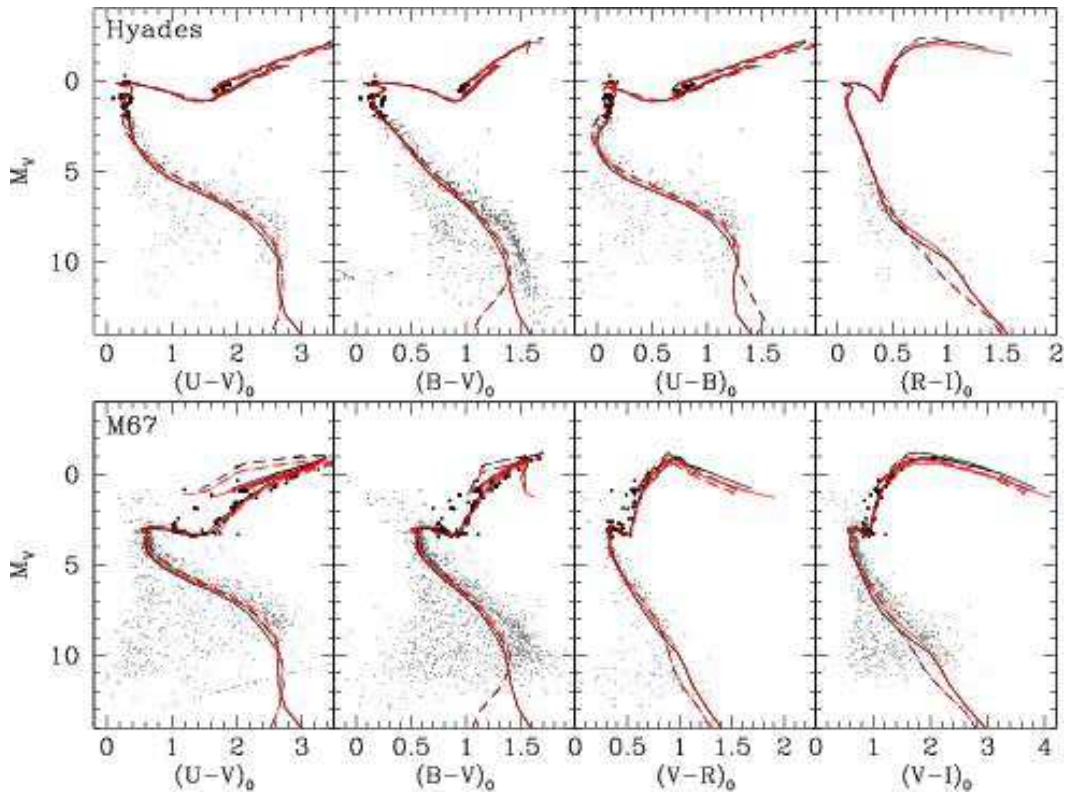}
\caption{Comparison of model isochrones with observed colour-magnitude 
diagrams of the Hyades and M67 Galactic open clusters in various photometric 
bands. For clarity, stars near and past the turnoff are plotted as large 
symbols. For each cluster, the adopted distance modulus and colour excess 
are listed in Table~\ref{cluster-data} along with the sources of the stellar
photometry. Each panel contains four isochrones: the red isochrones are computed
using the Padova 1994 tracks, while the black isochrones are computed using the
Padova 2000 tracks. In each case, the dashed and solid isochrones are computed
using the BaSeL~1.0 and BaSeL~3.1 spectral calibrations, respectively. All 
isochrones pertaining to a given cluster have fixed age and metallicity 
(see Table~\ref{cluster-data}).}
\label{fig_hym67}
\end{minipage}
\end{figure*}

\begin{figure*}
\centering
\begin{minipage}{140mm}
\hspace{0cm}
\includegraphics[height=140mm,angle=270]{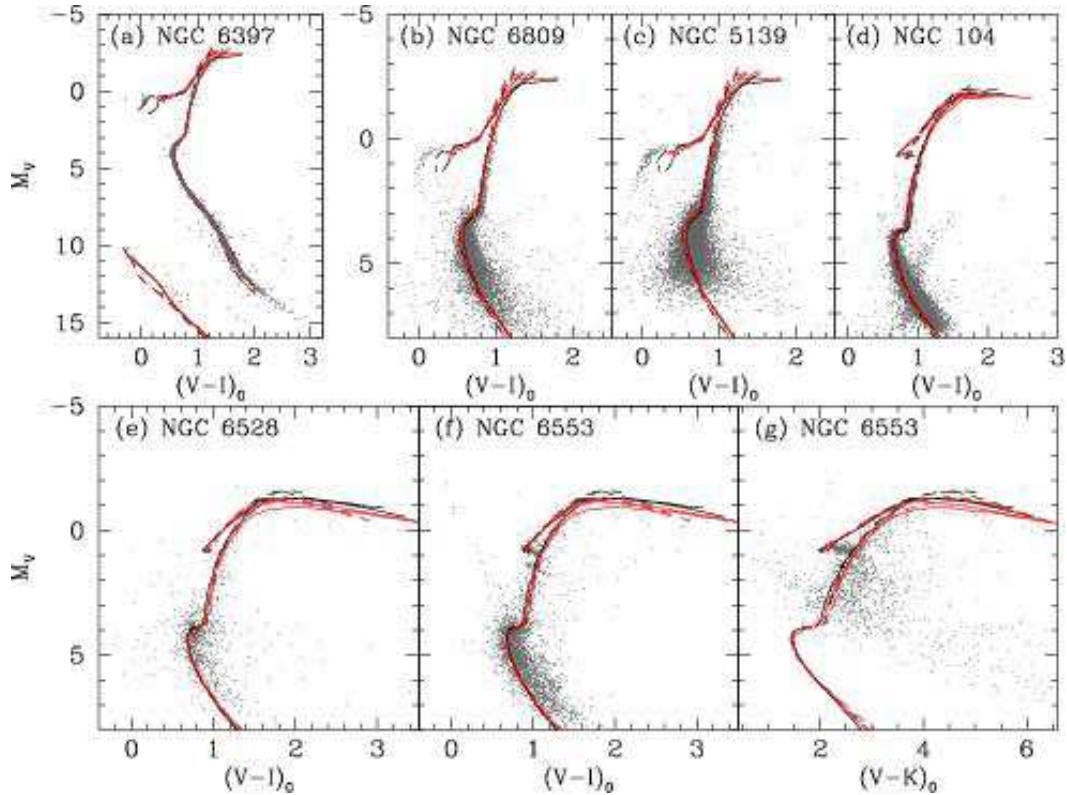}
\caption{Comparison of model isochrones with observed colour-magnitude diagrams
of six old Galactic globular clusters. For each cluster, the adopted distance
modulus and colour excess are listed in Table~\ref{cluster-data} along with
the sources of the stellar photometry. Each panel contains four isochrones: the
red isochrones are computed using the Padova 1994 tracks, while the black 
isochrones are computed using the Padova 2000 tracks. In each case, the dashed 
and solid isochrones are computed using the BaSeL~1.0 and BaSeL~3.1 
spectral calibrations, respectively. All isochrones pertaining to a given 
cluster have fixed age and metallicity (see Table~\ref{cluster-data}).}
\label{fig_globs}
\end{minipage}
\end{figure*}

To establish the reliability of our model, it is important to examine the
accuracy to which it can reproduce observed colour-(absolute) magnitude 
diagrams (CMDs) of star clusters of different ages and metallicities. 
Table~\ref{cluster-data} contains a list of star clusters for which extensive
data are available from the literature. The clusters are listed in order of
increasing \feh. For each cluster, we list the distance modulus $(m-M)_0$ and
the colour excess $E(\bv)$ from the same references as for the stellar 
photometry. The reddening-corrected CMDs of these clusters are presented in
Figs~\ref{fig_hym67} and \ref{fig_globs}, where we show the absolute $V$
magnitude as a function of various available optical and infrared colours.
Superimposed on the data in each frame are four isochrones. The red isochrones
are computed using the Padova 1994 tracks, while the black isochrones are
computed using the Padova 2000 tracks.  In each case, the dashed and solid
isochrones are computed using the BaSeL~1.0 and BaSeL~3.1 spectral 
calibrations, respectively. The isochrones were selected by adopting the
available model metallicity closest to the cluster \feh\ value and then 
choosing the age that provided the best agreement with the data. Age and 
metallicity are the same for all the isochrones for each cluster. Columns 6, 
7 and 8 of Table~\ref{cluster-data} list the metallicity $Z_{\rm mod}$, the 
corresponding $\feh_{\rm mod}$ and the age $t_{\rm mod}$ adopted for each 
cluster. The listed ages are in good agreement with previous determinations.

Fig.~\ref{fig_hym67} shows the CMDs of two Galactic open clusters of near-solar 
metallicity in various photometric bands: the young Hyades cluster and the 
intermediate-age M67 cluster. For clarity, stars near and past the turnoff are
plotted as large symbols. In the case of the Hyades, the 700~Myr 
Padova~1994/BaSeL~3.1 isochrone reproduces well the upper main sequence, the 
turnoff and the core-He burning phase in all $UBVIR$ bands. For M67, the 4~Gyr
Padova~1994/BaSeL~3.1 isochrone fits remarkably well the upper main sequence,
the subgiant branch, the red giant branch, the core-He burning clump and the
AGB in all bands. For both clusters, the models predict slightly bluer $UBV$ 
colours than observed on the lower main sequence. The offset is smaller in 
$R-I$ for the Hyades and in $V-R$ for M67, but the data are sparse in both
cases. It is worth noting that, for $M_V>10$, the BaSeL~3.1 spectral calibration
provides better agreement with the data than the BaSeL~1.0 calibration. 
Lower-main sequence stars, in any case, contribute negligibly to the integrated
light of a star cluster or a galaxy. At the age of the Hyades, the Padova 1994
and 2000 isochrones differ very little, as they rely on the same stellar 
evolution prescription for massive stars (Section~2.1). At the age of M67, the
Padova 2000 isochrone tends to predict stars bluer and brighter than the Padova
1994 isochrone near the tip of the red giant branch. As seen in Section~3.1 
above, this small but significant difference has a noticeable influence on 
integrated-light properties (see also below).

Fig.~\ref{fig_globs} shows the optical-infrared CMDs of six old Galactic 
globular clusters of different metallicities. NGC~6397 is the most metal-poor
cluster in our sample, with $\feh=-1.94$ (Table~\ref{cluster-data}). 
Fig.~\ref{fig_globs}(a) shows that models with $Z=0.0004$ ($\feh_{\rm mod}
\approx -1.65$) at an age of 14~Gyr provide excellent fits to the {\it Hubble 
Space Telescope} ({\it HST}) data for this cluster, all the way from the main
sequence, to the red giant branch, to the AGB and to the white-dwarf cooling
sequence. The models, however, do not fully reproduce the observed extension of
the blue horizontal branch (see also below). This problem persists even if
the age of the isochrones is increased. For this cluster, the Padova 
2000/BaSeL~3.1 isochrone appears to fit the {\em shape} of the horizontal 
branch and the main sequence near $M_V = 8$ marginally better than the Padova 
1994/BaSeL~3.1 isochrone. Ground-based data for the other two low-metallicity
clusters in our sample, NGC~6809 ($\feh= -1.80$) and NGC~5139 ($\feh=-1.62$), 
are also reproduced reasonably well by the $\feh_{\rm mod}=-1.65$ isochrones at
an age of 13~Gyr (Figs~\ref{fig_globs}b and \ref{fig_globs}c). As in the case of
NGC~6397, the models do not reproduce the full extension of the blue horizontal
branch. This suggests that this mismatch is not purely a metallicity effect and
that the evolution of these stars, or their spectral calibration, or both, may
have to be revised in the models. It is worth pointing out that the BaSeL~3.1 
spectral calibration provides a better fit of the upper red-giant stars than
the BaSeL~1.0 calibrations at these low metallicities.  

The CMD of the intermediate-metallicity cluster NGC~104 ($\feh= -0.71$) is well
reproduced by the Padova~1994/BaSeL~3.1 model with $\feh_{\rm mod}=-0.64$ at the
age of 13~Gyr (Fig.~\ref{fig_globs}d). This age should be regarded only as 
indicative, as the stars in NGC~104 are known to be overabundant in $\alpha$ 
elements relative to the solar composition, whereas the model has scaled-solar
abundances (see \citealt{2001ApJ...549..274V} for a more detailed analysis).
The NGC~6528 and NGC~6553 clusters of the Galactic bulge 
in Figs.~\ref{fig_globs}e--\ref{fig_globs}g are more metal-rich, 
with $\feh=-0.35$. The Padova~1994/BaSeL~3.1 model with $\feh_{\rm mod}= -0.41$ 
provides good fits to the CMDs of these clusters at the age of 13~Gyr. For both
clusters, the position of the core-He burning clump and the extension of the red
giant branch toward red $V-I$ and \vk\ colours are especially well accounted for.
As in the case of M67 (Fig.~\ref{fig_hym67}), the Padova 2000 isochrones tend to 
predict stars bluer and brighter than the Padova 1994 isochrones on the upper red
giant branch, providing a slightly worse fit to the observations.  Also, as is 
the case at other metallicities, the BaSeL~3.1 spectral calibration provides a
better fit of the upper red-giant stars than the BaSeL~1.0 calibration. 

Overall, Figs~\ref{fig_hym67} and \ref{fig_globs} show that our model provides
excellent fits to observed CMDs of star clusters of different ages and 
metallicities in a wide range of photometric bands. The data tend to favor the
combination of the Padova 1994 stellar evolution prescription with the BaSeL~3.1
spectral calibration. This justifies our adoption of this combination in our
standard model.

\subsubsection{Integrated colours}

\begin{figure*}
\centering
\begin{minipage}{140mm}
\includegraphics[width=140mm]{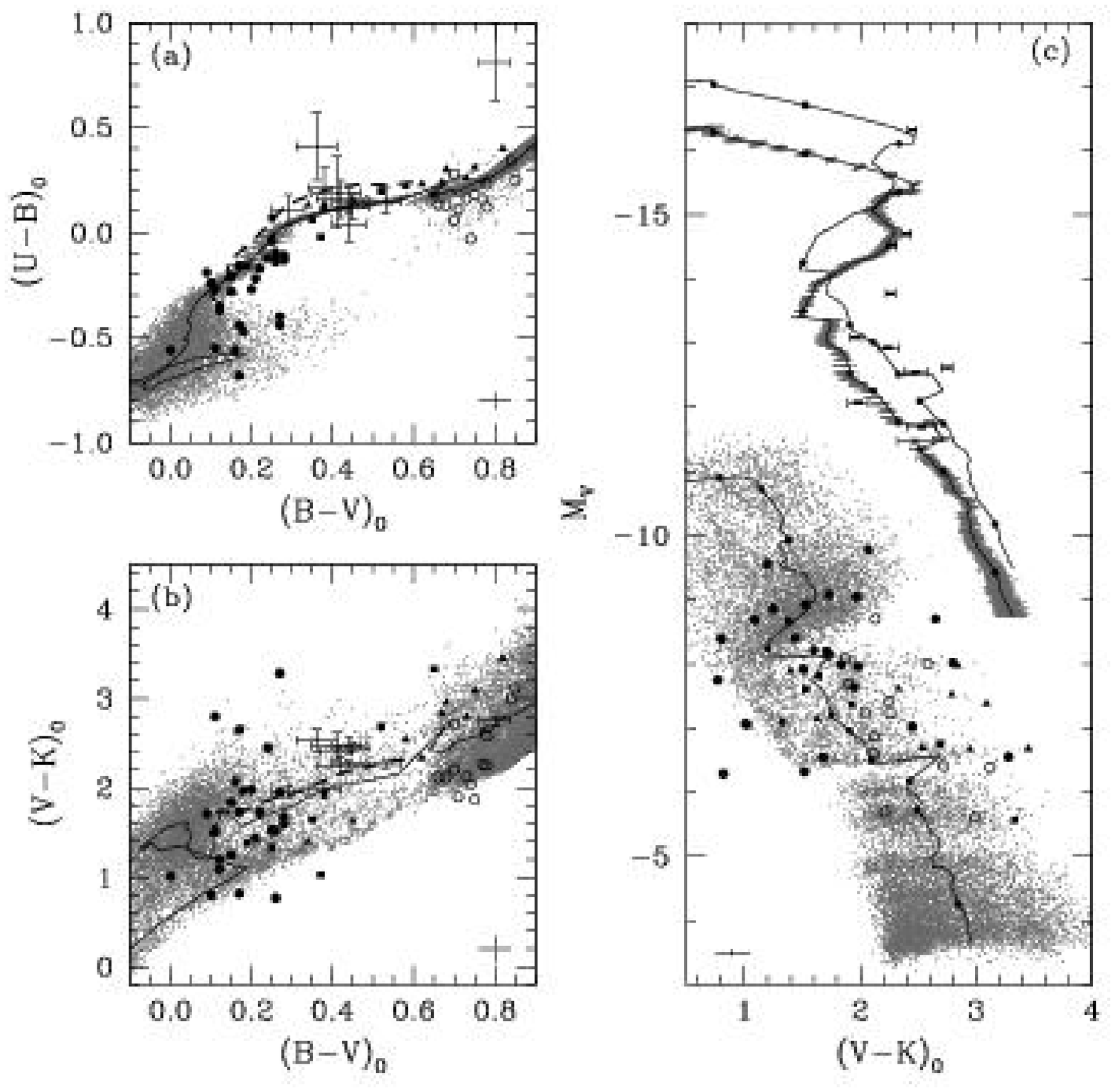}
\caption{(a) \ub\ versus \bv\ and (b) \vk\ versus \bv\ integrated
colours of star clusters. The different symbols represent LMC globular clusters
in various age ranges according to the SWB classification scheme
(classes I--III: filled circles; class IV: squares;  class V: triangles; 
classes VI--VII: open circles). The \ub\ and \bv\ colours are from 
\citet{1981A&AS...46...79V} and the \vk\ colours from 
\citet{1983ApJ...266..105P}. The points with error bars are young star clusters
in the merger remnant galaxy NGC~7252 (\citealt{1997AJ....114.2381M};
\citealt{2001A&A...370..176M}). The solid
line shows the evolution of the standard SSP model of Section~3 for the 
metallicity $Z = 0.4Z_\odot$ at ages from a few Myr to 13 Gyr. The small dots
show the results of 22,000 stochastic realizations of the integrated colours of
clusters of mass $2\times 10^4\, M_\odot$ at ages between $10^5$ and 13~Gyr,
for the same metallicity and IMF as for this SSP model. The heavy dashed line
shows the colours of the standard SSP model of Section~3 for the metallicity $Z
= Z_\odot$ at ages from 100 Myr to 1 Gyr. (c) Absolute magnitude $M_V$ versus
\vk\ colour.  The data are the same as in (a) and (b). Three models show the 
evolution of, from bottom to top, a $2\times 10^4\, M_\odot$ SSP with 
metallicity $Z=0.4 Z_\odot$, a $3 \times 10^6\,M_\odot$ SSP with metallicity
$Z=Z_\odot$ and a $6\times10^6\, M_\odot$ SSP with metallicity $Z=Z_\odot$. 
Small circles indicate the positions of the models at the ages 6, 7, 10, 100,
400 and 500 Myr and 1, 1.4, 2 and 10 Gyr (these marks can be used to roughly
date the clusters).  Stochastic realizations of integrated colours are shown
only for the two least massive models, as the predicted scatter is small for
the most massive one. Typical observational error bars are indicated at the 
bottom of each panel.}
\label{fig_simul}
\end{minipage}
\end{figure*}

We must also check that our model can reproduce the {\em integrated} colours of
star clusters of various ages and metallicities, which are sensitive to the 
{\em numbers} of stars populating different phases along the isochrones.
Figs~\ref{fig_simul}(a) and \ref{fig_simul}(b) show the integrated, 
reddening-corrected \ub, \bv\ and \vk\ colours of LMC clusters in various age
ranges, according to the classification scheme of 
\citet*[~hereafter SWB]{1980ApJ...239..803S}. Also shown as error bars are the
colours of young star clusters in the merger remnant galaxy NGC 7252 from
\citet{1997AJ....114.2381M} and \citet{2001A&A...370..176M}. The solid line shows
the evolution of our standard SSP model for $Z = 0.4Z_\odot$, at ages ranging 
from a few Myr at the blue end of the line to 13 Gyr at the red end of the line.
The scatter in cluster colours in Figs~\ref{fig_simul}(a) and \ref{fig_simul}(b)
is intrinsic (typical observational errors are indicated in each panel). It is 
largest in \vk\ colour (Fig.~\ref{fig_simul}b) but is also present, to a lesser 
extent, in \ub\ and \bv\ colours (Fig.~\ref{fig_simul}a). This scatter cannot be 
accounted for by metallicity variations. The age-metallicity degeneracy implies 
that the evolution of SSPs with various metallicities are similar to that of the 
$Z = 0.4Z_\odot$ model in these colour-colour diagrams. For reference, the heavy 
dashed line in Figs~\ref{fig_simul}(a) and \ref{fig_simul}(b) shows the colours
of the standard SSP model of Section~3 for the metallicity $Z = Z_\odot$ at 
ages from 100 Myr to 1 Gyr. The scatter in the observed integrated colours of
star clusters is most likely caused by stochastic fluctuations in the numbers
of stars populating different evolutionary stages. 

We illustrate this by generating random realizations of integrated cluster
colours using a Monte Carlo technique pioneered by \citet{1977A&A....54..243B}
(see also \citealt*{1988A&A...196...84C}; \citealt{1995A&A...298...87G}; 
\citealt{1997ApJ...479..764S}; \citealt{2002IAUS..207..616B};
\citealt{2001A&A...376..422C}; \citealt{2002A&A...381...51C}). For a given 
cluster age, we draw stars randomly from the IMF of equation~(\ref{imf}) and
place them in their evolutionary phase along the isochrone at that age, 
until a given cluster mass is reached. The small dots in Figs~\ref{fig_simul}(a)
and \ref{fig_simul}(b) show the results of 22,000 such realizations for clusters
of mass $2\times 10^4\, M_\odot$ and metallicity $Z = 0.4Z_\odot$, at ages 
between $10^5$ and 13~Gyr (see \citealt{2002IAUS..207..616B} for more detail). It
is clear from these figures that the models can account for the full observed
ranges of integrated cluster colours, including the scatter of nearly 2~mag in 
\vk\ colour. The reason for this is that the \vk\ colour is highly sensitive
to the small number of bright stars populating the upper giant branch. 
Fluctuations are smaller in the \ub\ and \bv\ colours, which are dominated by
the more numerous main-sequence stars. The predicted scatter would be smaller 
in all colours for clusters more massive than $2\times 10^4\,M_\odot$, as the 
number of stars in any evolutionary stage would then be larger 
(\citealt{2002IAUS..207..616B}; \citealt{2002A&A...381...51C}).

To further illustrate the relation between cluster mass and scatter in 
integrated colours, we plot in Fig.~\ref{fig_simul}(c) the absolute $V$ 
magnitude as a function of \vk\ colour for the same clusters as in 
Figs~\ref{fig_simul}(a) and \ref{fig_simul}(b). The three models shown 
correspond to the evolution of, from bottom to top, a $2\times 10^4\, M_\odot$
SSP with metallicity $Z=0.4 Z_\odot$, a $3 \times 10^6\,M_\odot$ SSP with 
metallicity $Z=Z_\odot$ and a $6\times10^6\, M_\odot$ SSP with metallicity 
$Z=Z_\odot$. We show stochastic realizations of integrated colours only for the
two least massive models, as the predicted scatter is small for the most 
massive one. As in Figs~\ref{fig_simul}(a) and \ref{fig_simul}(b), random 
realizations at various ages of $2\times 10^4\, M_\odot$ clusters with 
metallicity $Z=0.4Z_\odot$ can account for the full observed range of LMC 
cluster properties in this diagram. The NGC~7252 clusters are consistent with
being very young (100--800 Myr) and massive ($10^6-10^7\, M_\odot$) at solar 
metallicity, in agreement with the results of \citet{1998AJ....116.2206S}.

Our models, therefore, reproduce remarkably well the full observed ranges of
integrated colours and absolute magnitudes of star clusters or various ages and 
metallicities. It is worth pointing out that, because of the stochastic nature
of the integrated-light properties of star clusters, single clusters may not be 
taken as reference standards of simple stellar populations of specific age and 
metallicity.

\section{Spectral evolution}

We now turn to the predictions of our models for the {\em spectral} evolution of
stellar populations. In Section~4.1 below, we begin by describing the canonical 
evolution of the spectral energy distribution of a simple stellar population.
We also illustrate the influence of metallicity on the spectra. Then, in 
Section~4.2, we compare our model with observed galaxy spectra extracted from
the SDSS EDR. Section~4.3 presents a more detailed comparison of the predicted
and observed strengths of several absorption-line indices.

\subsection{Simple stellar population}

\begin{figure}
\includegraphics[width=90mm]{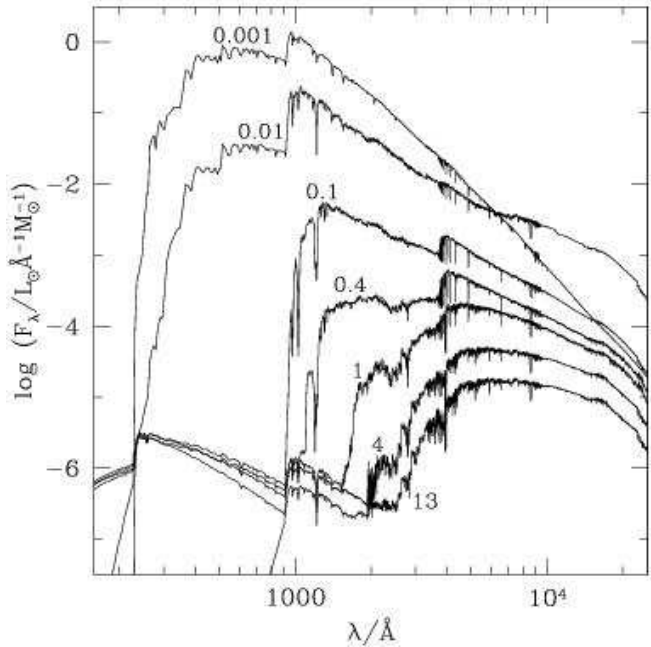}
\caption{Spectral evolution of the standard SSP model of Section~3 for the 
solar metallicity. The STELIB/BaSeL~3.1 spectra have been extended blueward of
3200~{\AA} and redward of 9500~{\AA} using the Pickles medium-resolution 
library. Ages are indicated next to the spectra (in Gyr).}
\label{fig_ssp}
\end{figure}

Fig.~\ref{fig_ssp} shows the spectral energy distribution of the standard SSP
model of Section~3 at various ages and for solar metallicity. As is possible
for this metallicity (Section~2.2.2), we have extended the STELIB/BaSeL~3.1 
library blueward of 3200~{\AA} and redward of 9500~{\AA} using the Pickles 
medium-resolution library (Section~2.2.3). In Fig.~\ref{fig_ssp}, therefore, 
the model includes libraries of observed stellar spectra across the whole 
wavelength range from 1205~{\AA} to 2.5~$\mu$m.

The spectral evolution of an SSP may be understood in terms of the evolution of
its stellar content. At $10^6$~yr, the spectrum in Fig.~\ref{fig_ssp} is 
entirely dominated by short-lived, young massive stars with strong ultraviolet
emission on the upper main sequence. Around $10^7$~yr, the most massive
stars leave the main sequence and evolve into red supergiants, causing
the ultraviolet light to decline and the near-infrared light to rise. From a 
few times $10^8$~yr to over $10^9$~yr, AGB stars maintain a high near-infrared
luminosity. The ultraviolet light continues to drop as the turnoff mass 
decreases on the main sequence. After a few gigayears, red giant stars account
for most of the near-infrared light. Then, the accumulation of low-mass, 
post-AGB stars causes the far-ultraviolet emission to rise until 13~Gyr. The 
most remarkable feature in Fig.~\ref{fig_ssp} is the nearly unevolving shape of
the optical to near-infrared spectrum at ages from 4 to 13~Gyr. The reason for
this is that low-mass stars evolve within a narrow temperature range all the way
from the main sequence to the end of the AGB.

The scale of Fig.~\ref{fig_ssp} is not optimal to fully appreciate the spectral
resolution of the model. However, some variations can be noticed in the 
strengths of prominent absorption lines. At ages between 0.1 and 1~Gyr, for 
example, there is a marked strengthening of all Balmer lines from \ha\ at 
6563~{\AA} to the Balmer continuum limit at 3646~{\AA}.  This characteristic
signature of a prominent population of late-B to early-F stars is, in fact, a 
standard diagnostic of recent bursts of star formation in galaxies (e.g., 
\citealt{1987MNRAS.229..423C}; \citealt{1999ApJ...518..576P}; 
\citealt{2003MNRAS.341...33K}). It is interesting to note how, over the same age
interval, the `Balmer break' (corresponding to the Balmer continuum limit) 
evolves into the `4000~{\AA} break' (arising from the prominence in cool stars
of a large number of metallic lines blueward of 4000~{\AA}). Other prominent 
absorption features in Fig.~\ref{fig_ssp} include the \mgii\ resonance doublet
near 2798~{\AA}, the \caii~H and K lines at 3933~{\AA} and 3968~{\AA}, and the
\caii\ triplet at 8498, 8542, and 8662~{\AA}. These features tend to strengthen
with age as they are stronger in late-type stars than in early-type stars, 
whose opacities are dominated by electron scattering. However, the strengths of
these features also depend on the abundances of the heavy elements that produce
them.

\begin{figure}
\includegraphics[width=90mm]{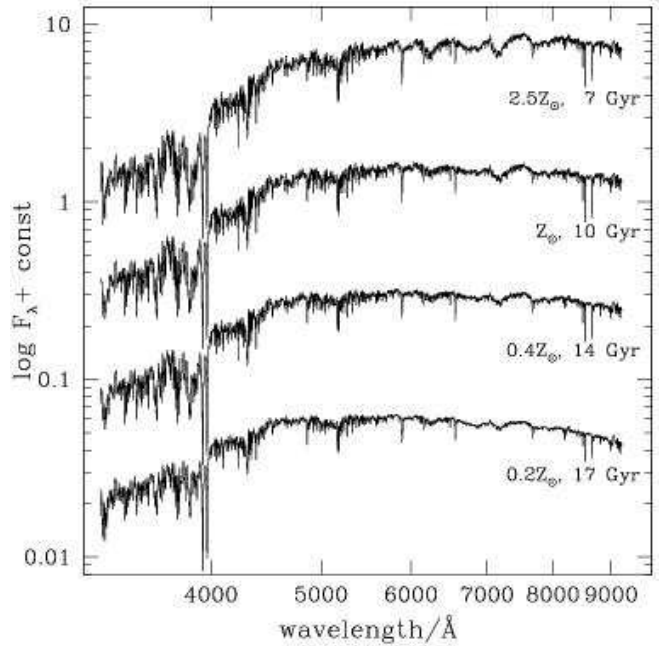}
\caption{Spectra of the standard SSP model of Section 3 at different ages for
different metallicities, as indicated. The prominent metallic features show a
clear strengthening from the most metal-poor to the most metal-rich models, 
even though the shape of the spectral continuum is roughly similar in all 
models.}
\label{fig_degen}
\end{figure}

Fig.~\ref{fig_ssp} also shows that the strengths of many absorption lines, in
contrast to the spectral continuum shape, continue to evolve significantly at
ages between 4 and 13~Gyr. Since the strengths of such features are expected to
react differently to age and metallicity, they can potentially help us resolve
the age-metallicity degeneracy that hampers the interpretation of galaxy spectra
(see Section~3.1; \citealt{1985AJ.....90.1927R}; \citealt{1994ApJS...95..107W};
\citealt{1999ApJ...513..224V}). This is illustrated by Fig.~\ref{fig_degen}, in
which we show the spectra of SSPs of different ages and metallicities, whose 
spectral continua have roughly similar shapes. The prominent metallic features
in these spectra, such as the \caii~H and K lines, the many Fe and Mg lines 
between 4500 and 5700~{\AA}, and several TiO, H$_2$O and O$_2$ molecular 
absorption features redward of 6000~{\AA}, show a clear strengthening from the
most metal-poor to the most metal-rich stellar populations. As we shall see in
Sections~4.2 and 4.3 below, the analysis of these features in observed 
galaxy spectra provide useful constraints on the metallicities, and in turn on
the ages, of the stellar populations that dominate the emission.

\subsection{Interpretation of galaxy spectra}

\begin{figure*}
\centering
\begin{minipage}{140mm}
\includegraphics[width=140mm]{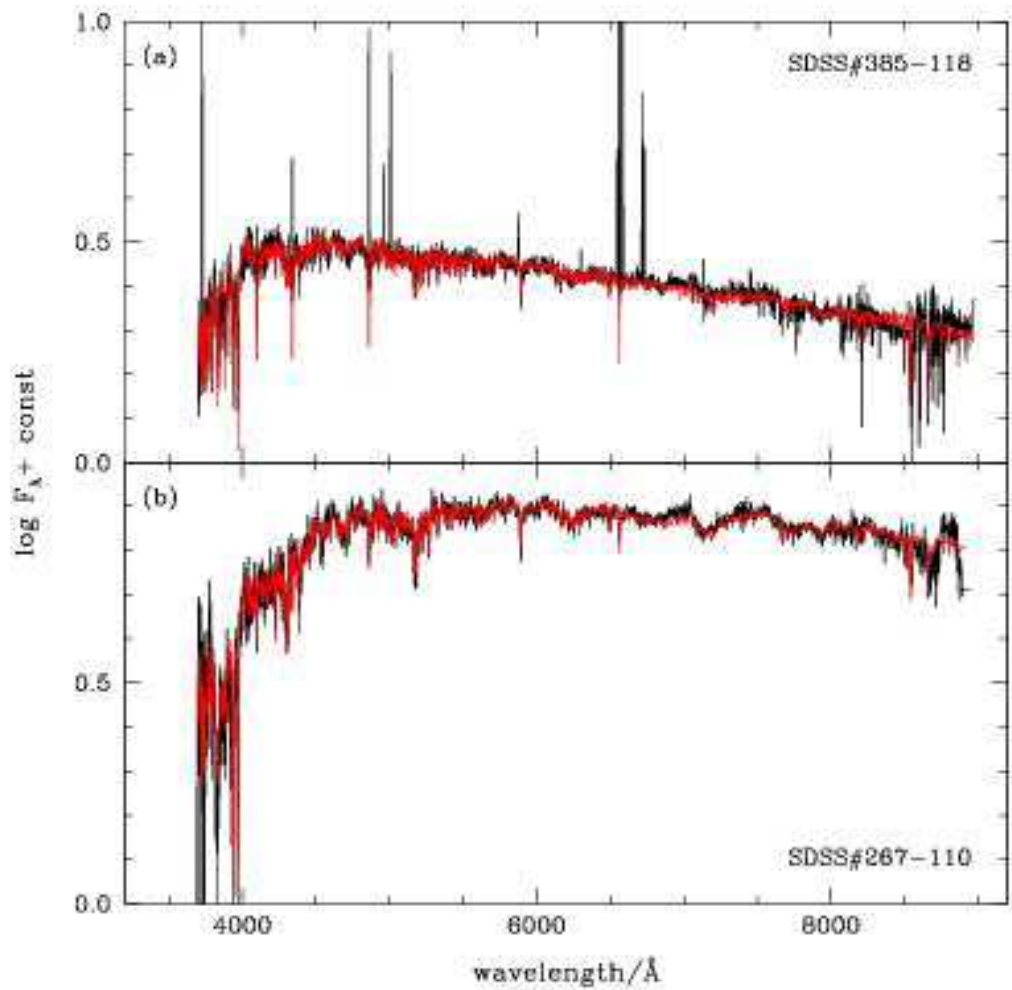}
\caption{Model fits (red spectra) of two galaxies extracted from the SDSS 
Early Data Release (black spectra). The fits were derived using the optimized
data compression algorithm of \citet*{2000MNRAS.317..965H}, as described in the
text. The emission lines of SDSS 385--118 were removed to perform the fit.}
\label{fig_fit}
\end{minipage}
\end{figure*}

We now exemplify how our model can be used to interpret observed galaxy 
spectra. The observational sample we consider is the Early Data Release 
of the Sloan Digital Sky Survey (\citealt{2002AJ....123..485S}; see Section~1).
This survey will obtain $u$, $g$, $r$, $i$, and $z$ photometry of almost a 
quarter of the sky and spectra of at least 700,000 objects. The `main galaxy
sample' of the EDR includes the spectra of 32,949 galaxies with $r$-band 
Petrosian magnitudes brighter than 17.77 after correction for foreground 
Galactic extinction \citep{2002AJ....124.1810S}. The spectra are flux- and 
wavelength-calibrated, with 4096 pixels from 3800~{\AA} to 9200~{\AA} at resolving
power $\lambda/\Delta\lambda\approx1800$.  This is similar to the resolution of
our model in the wavelength range from 3200~{\AA} to 9500~{\AA}. The SDSS spectra
are acquired using 3-arcsecond diameter fibres that are positioned as close as 
possible to the centres of the target galaxies. For the purpose of first
illustration, we select SDSS spectra of two representative galaxies of 
different types according to their 4000~{\AA} discontinuities. We adopt here 
the 4000~{\AA} discontinuity index defined by \citet{1999ApJ...527...54B} as 
the ratio of the average flux density $F_\nu$ in the narrow bands 
3850--3950~{\AA} and 4000--4100~{\AA}. The original definition of this index by
\citet{1983ApJ...273..105B} uses wider bands (3750--3950~{\AA} and 
4050--4250~{\AA}), and hence, it is more sensitive to reddening effects. We 
select two spectra with median signal-to-noise ratios per pixel larger than 30
and with discontinuity indices near opposite ends of the sample distribution,
$D_n(4000)=1.26$ (SDSS 385--118) and $D_n(4000)=1.92$ (SDSS 267--110). The
galaxies have measured line-of-sight velocity dispersions $\sigv\approx70~\kms$
and $130~\kms$, respectively. The spectra are corrected for foreground Galactic
extinction using the reddening maps of \citet*{1998ApJ...500..525S} and the 
extinction curve of \citet{1999PASP..111...63F}.

To interpret these spectra with our model, we use MOPED, the optimized data 
compression algorithm of \citet*{2000MNRAS.317..965H}. In this approach,
galaxy spectra are compressed into a reduced number of linear combinations 
connected to physical parameters such as age, star formation history, 
metallicity and dust content. The linear combinations contain as much 
information about the parameters as the original spectra. There are several
advantages to this method. First, it allows one to explore a wide range of
star formation histories, chemical enrichment histories and dust contents
by choosing appropriate parametrizations \citep*{2001MNRAS.327..849R}.
Second, it allows one to estimate the errors on derived physical parameters. 
And third, it is extremely fast and hence efficient to interpret large numbers
of galaxy spectra. Our model has already been combined with the MOPED 
algorithm to interpret SDSS EDR spectra (Mathis et al., in preparation).
The results presented for the two galaxies considered here are based on a 
decomposition of the star formation history into six episodes of constant star
formation in the age bins 0.0--0.01, 0.01--0.1, 0.1--1.0, 1.0--2.5, 2.5--5 and
5--13~Gyr. The metallicity in each bin can be one of $Z=0.4Z_{\sun}$, $Z_{\sun}$
or $2.5Z_{\sun}$. The attenuation by dust is parametrized using the simple 
two-component model of \citet*[ see Section~5 below]{2000ApJ...539..718C}. The
effective attenuation optical depth affecting stars younger than 0.01~Gyr can
be $\tauv= 0.0$, 0.1, 0.5, 1, 1.5, 2 or 3, while that affecting older stars is
$\mu\tauv$, with $\mu =0.0$, 0.1, 0.3, 0.5 or 1 (see equation~\ref{taueff} 
below; we are grateful to H.~Mathis for providing us with the results of these
fits). 

\begin{figure*}
\centering
\begin{minipage}{140mm}
\includegraphics[width=140mm]{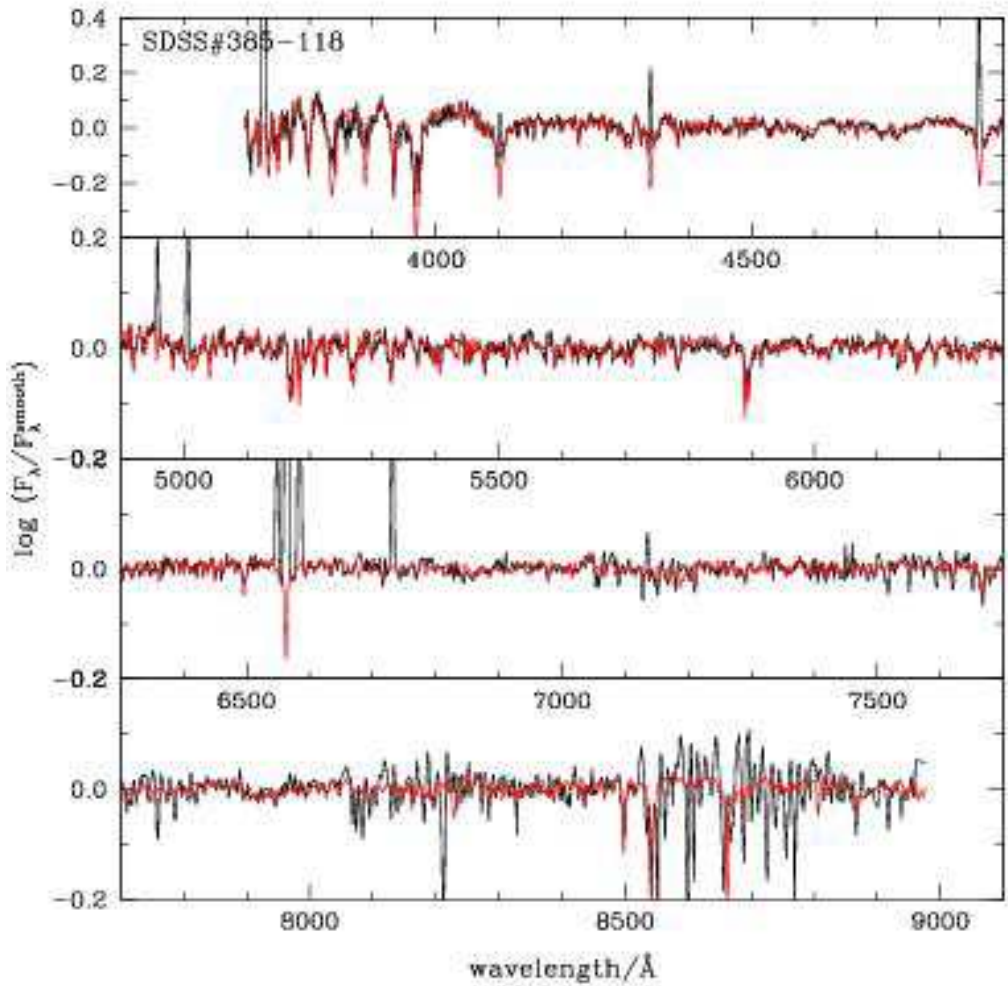}
\caption{Detailed comparison of `high-pass' spectra for the same model (in red)
of SDSS galaxy 385--118 (in black) as in Fig.~\ref{fig_fit}(a). The high-pass
spectra were obtained by smoothing the original spectra using a top-hat
function of width 200~{\AA} and then dividing the original spectra by the
smoothed spectra.}
\label{fig_sf}
\end{minipage}
\end{figure*}

\begin{figure*}
\centering
\begin{minipage}{140mm}
\includegraphics[width=140mm]{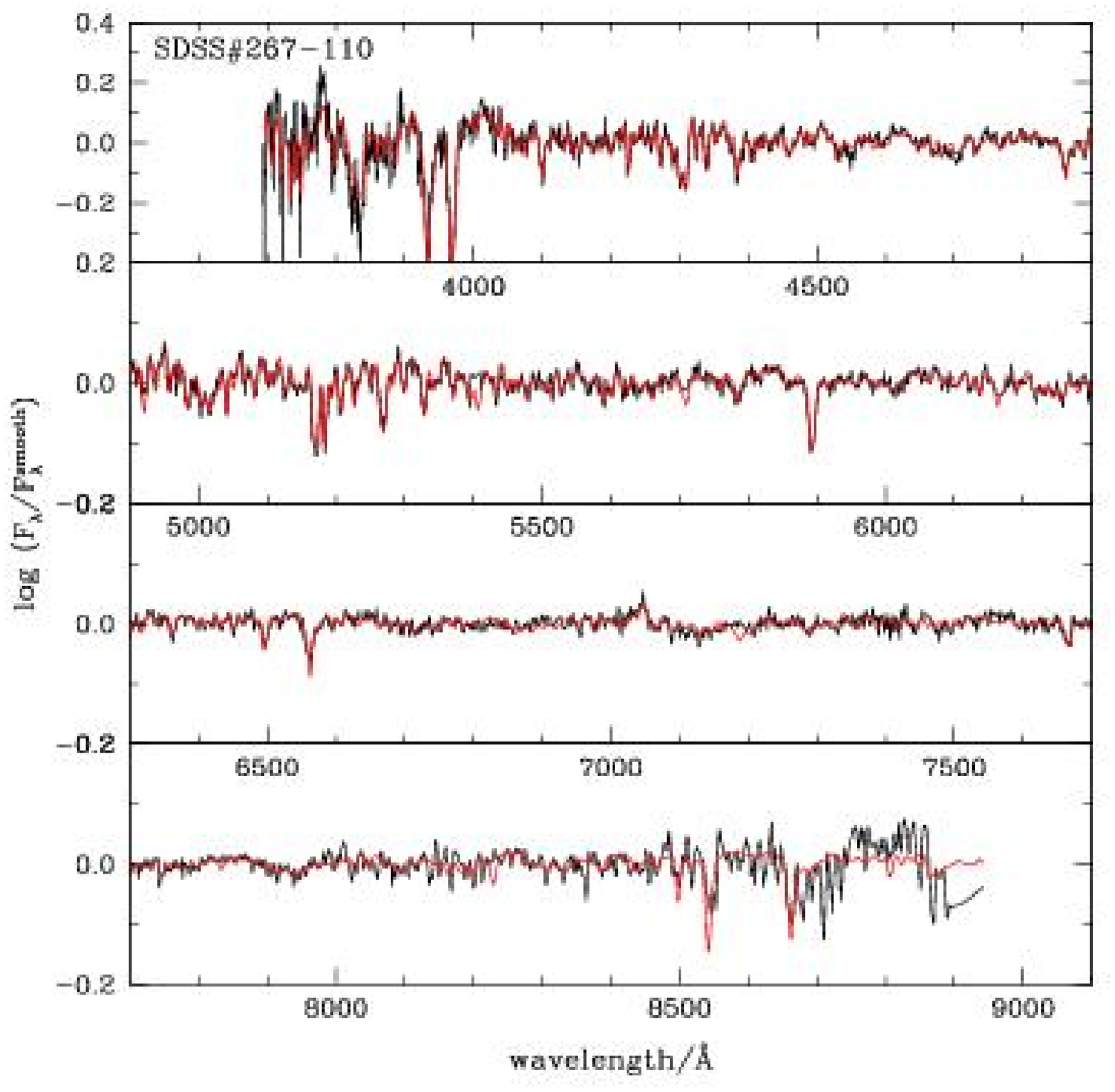}
\caption{Detailed comparison of `high-pass' spectra for the same model (in red)
of SDSS galaxy 267--110 (in black) as in Fig.~\ref{fig_fit}(b). The high-pass
spectra were obtained by smoothing the original spectra using a top-hat
function of width 200~{\AA} and then dividing the original spectra by the
smoothed spectra.}
\label{fig_el}
\end{minipage}
\end{figure*}

Fig.~\ref{fig_fit} shows the resulting spectral fits of SDSS 385--118 and SDSS
267--110.  Figs~\ref{fig_sf} and \ref{fig_el} show details of the `high-pass' 
spectra of the fitted models and observed galaxies (note that we display the 
emission lines of SDSS 385--118 in Figs~\ref{fig_fit}a and \ref{fig_sf}, even
though these were removed to perform the fit). The high-pass spectra were 
obtained by smoothing the original spectra using a top-hat function of width
200~{\AA} and then dividing the original spectra by the smoothed spectra (see
\citealt{2002ApJ...569..582B}).  The model reproduces the main stellar 
absorption features of both galaxies extremely well. In particular, in the 
spectrum of SDSS 385--118, the absorption wings of Balmer lines are well fitted
up to high orders in the series. With such an accuracy, the model can be used 
reliably to measure the contamination of Balmer emission lines by underlying 
stellar absorption in galaxies (see \citealt{CTPhD2003}).  This is especially 
important, for example, to constrain attenuation by dust using the \ha/\hb\ 
ratio. The spectrum of SDSS 267--110 in Fig.~\ref{fig_el} shows no obvious 
emission lines and exhibits strong stellar absorption features characteristic 
of old stellar populations. Among the most recognizable features, the \caii~H 
and K lines, the $G$ band near 4300~{\AA}, the magnesium features near 
5100~{\AA} and 5200~{\AA}, the iron features between 5270~{\AA} and 5800~{\AA},
the NaD feature near 5900~{\AA}, and the TiO bands near 6000~{\AA} and 
6200~{\AA} are all well reproduced by the model. In Section~4.3 below, we compare
in a more quantitative way the strengths of these features in our model with 
those in the SDSS EDR spectra.

It is of interest to mention the physical parameters of the model fits in
Figs~\ref{fig_fit}--\ref{fig_el}. For SDSS 385--118, the algorithm assigns 91
per cent of the total stellar mass of $\sim10^{9} M_{\sun}$ to stars with 
metallicity $Z=0.4Z_{\sun}$ formed between 2.5 and 13~Gyr ago, and the 
remainder to stars of the same metallicity formed in the last Gyr or so. The 
galaxy is best fitted with $\mu\tauv=0.5$.  For SDSS 267--110, 50 per cent of
the total stellar mass of $\sim10^{10} M_{\sun}$ is attributed to stars
formed 5--13~Gyr ago and the remainder to stars formed 2.5--5~Gyr ago, all
with solar metallicity. The dust attenuation optical depth is found to be 
negligible, with $\tauv=0$. The total stellar masses quoted here do
not include aperture corrections for the light missed by the 3-arcsec diameter
fibres. The errors in the derived mass {\em fractions} in our various bins are
relatively modest, of the order of 20 per cent, for these spectra with high 
signal-to-noise ratios (Mathis et al., in preparation).

The examples described above illustrate how our model can be used to interpret
observed high-resolution spectra of galaxies at wavelength from 3200~{\AA} to 
9500~{\AA} in terms of physical parameters such as age, star formation history,
chemical enrichment history and dust content.

\subsection{Spectral indices}

It is important to establish in a more quantitative way the ability of our 
model to reproduce the strengths of prominent stellar absorption features
in observed galaxy spectra. The atomic and molecular features that are most
commonly measured in the visible spectra of galaxies are those defined in the
extended Lick system (\citealt{1994ApJS...94..687W};
\citealt{1997ApJS..111..377W}; \citealt{1998ApJS..116....1T};
see Section~1). This system includes a total of 25 {\em spectral indices}
that were defined and calibrated in the spectra of 460 Galactic stars covering
the wavelength range from 4000~{\AA} to 6400~{\AA} at a resolution of 
$\sim9$~{\AA} FWHM.  In the Lick system, an index is defined in terms of a central
`feature bandpass' bracketed by two `pseudo-continuum bandpasses'. By convention,
atomic indices are expressed in angstroms of equivalent width, while molecular 
indices are expressed in magnitudes.

\begin{figure*}
\centering
\begin{minipage}{140mm}
\includegraphics[width=140mm]{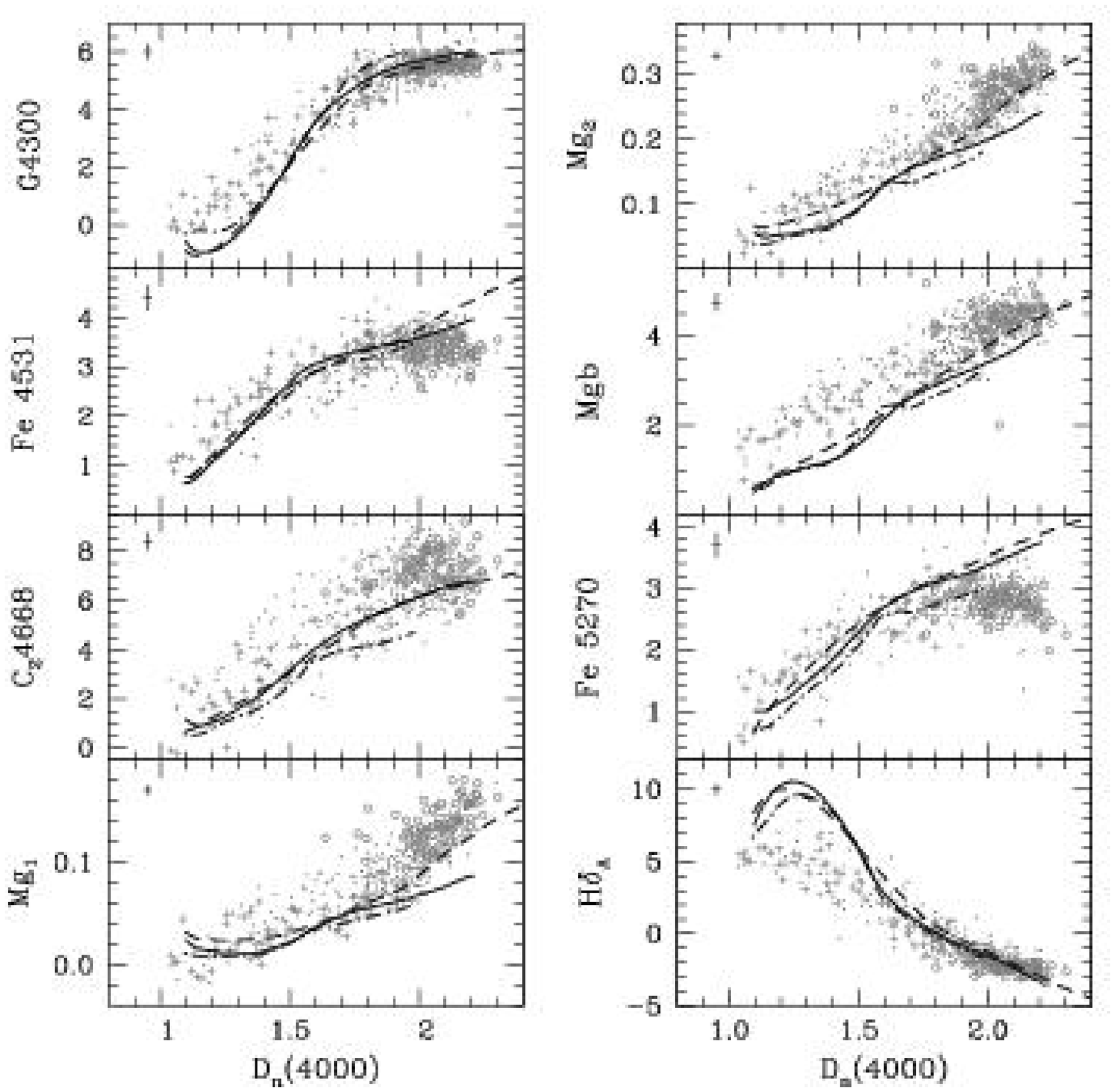}
\caption{Strengths of G4300, Fe4531, C$_2$4668, Mg$_1$, Mg$_2$, Mgb, Fe5270
and H$\delta_A$ as a function of $D_n(4000)$ for 543 galaxies with S/N$_{\rm
med} \ge 40$ in the `main galaxy sample' of the SDSS EDR (the median
observational error bars are indicated in the upper left corner of each panel).
Different symbols correspond to different velocity dispersions (crosses: 
$\sigv \leq  100~\kms$; dots: $100<\sigv \leq 250~\kms$; open circles: $\sigv>
250~\kms$). The lines show the evolution of the standard SSP model of Section
3 for the metallicities $Z=0.008$ (dot-and-dashed line), $Z=0.02$ (solid line)
and $Z=0.05$ (dashed line) at ages from $5\times10^7\,$yr to 15 Gyr. The
models have 3~{\AA} FWHM spectral resolution, corresponding to a nominal stellar 
velocity dispersion $\sigv\approx70~\kms$ at 5500~{\AA}.}
\label{fig_idxmet}
\end{minipage}
\end{figure*}

\begin{figure*}
\centering
\begin{minipage}{140mm}
\includegraphics[width=140mm]{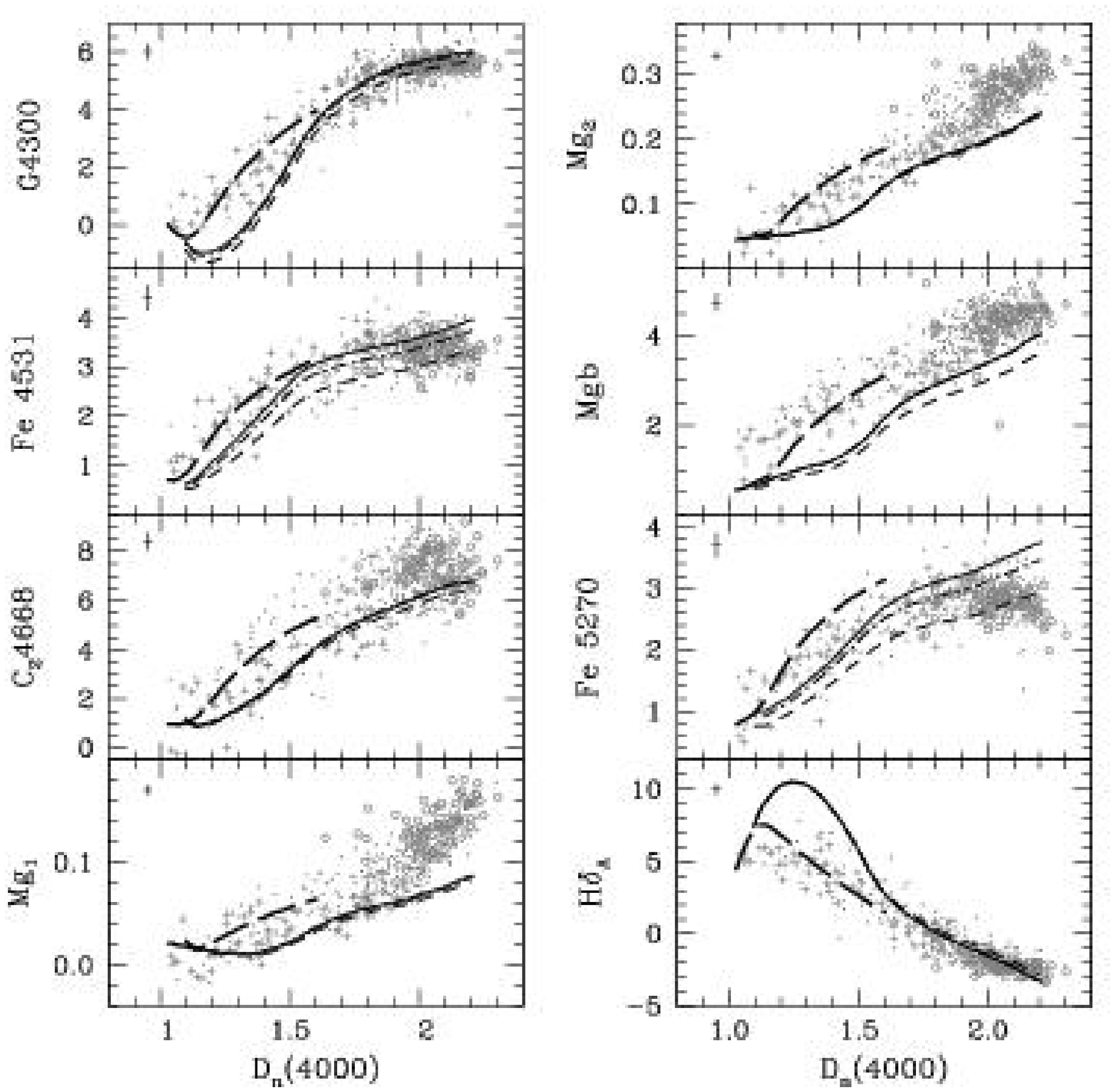}
\caption{Strengths of G4300, Fe4531, C$_2$4668, Mg$_1$, Mg$_2$, Mgb, Fe5270
and H$\delta_A$ as a function of $D_n(4000)$ for the same SDSS galaxies as in
Fig.~\ref{fig_idxmet}. The lines show the evolution of the standard SSP model
of Section 3 for solar metallicity and for stellar velocity dispersions 
$\sigv=70~\kms$ (nominal model velocity dispersion,
solid line), 150~\kms\ (dot-and-dashed line) and 300~\kms\
(dashed line) at ages from $5\times10^7\,$yr to 15 Gyr. The long-dashed line
shows the evolution of a model with continuous star formation with a law 
$\psi(t) \propto \exp\left[ -t/({\rm 4\,Gyr)}\right]$, for $Z=Z_\odot$ and 
$\sigv=70~\kms$.}
\label{fig_idxvel}
\end{minipage}
\end{figure*}

We compare the Lick index strengths predicted by our model with those measured
in SDSS EDR spectra with high signal-to-noise ratios. We do not use here the 
index strengths included in the SDSS EDR. Instead, we re-measure index strengths
in the SDSS spectra in the same way as in the model spectra (we are grateful to
J.~Brinchmann for providing us with the results of these measurements). Our
approach differs in two ways from previous absorption-line studies of galaxies.
First, in previous studies, Lick indices were generally modelled by
parametrizing index strengths as functions of stellar effective temperature,
gravity and metallicity (see Section~1). In our model, the index strengths are
measured directly from the spectra using the bandpass definitions of 
\citet{1997ApJS..111..377W} and \citet{1998ApJS..116....1T}. Second, we 
consider all types of galaxies in our analysis, including star-forming 
galaxies, while all previous analyses focused on passively evolving stellar 
populations older than about 1~Gyr. The inclusion of star-forming galaxies
requires that emission lines be removed from the SDSS spectra before measuring
the indices. This is achieved following the careful procedure outlined by 
\citet[ see also \citealt{2003MNRAS.341...33K}]{CTPhD2003} that is based on accurate
fits of the emission line-free regions of the spectra with model spectra 
broadened to the observed stellar velocity dispersion. The subtraction of 
emission lines is important mainly for Balmer-line indices and a few 
metallic-line indices (e.g., G4300, Fe5015). We include the errors resulting
from line subtraction into the observational errors of these indices.

Fig.~\ref{fig_idxmet} shows the strengths of eight indices measured in this way
in 543 spectra with high median signal-to-noise ratio per pixel, 
S/N$_{\rm med} \ge 40$, in the `main galaxy sample' of the SDSS EDR. These 
indices, whose strengths are plotted as a function of $D_n(4000)$, were chosen
for illustration purposes because they are produced by different elements
(see below). The typical observational errors, indicated in the upper left
corner of each panel, are very small for this sample of high-quality spectra. 
Different symbols in Fig.~\ref{fig_idxmet} correspond to different velocity
dispersions (crosses: $\sigv \leq 100~\kms$; dots: $100<\sigv \leq 250~\kms$;
open circles: $\sigv> 250~\kms$).  As expected, early-type galaxies 
characterized by large $D_n(4000)$ strengths tend to have large velocity 
dispersions (see \citealt{2003MNRAS.341...33K}). However, we cannot make 
statistical inferences about the properties of SDSS galaxies based on this sample
alone because of the selection by signal-to-noise ratio applied to test
the model. Superimposed on the data in Fig.~\ref{fig_idxmet}
are three models showing the evolution of the index strengths of SSPs with 
different metallicities, $Z=0.4 Z_\odot$, $Z_\odot$ and $2.5 Z_\odot$, at ages 
between $5\times10^7\,$yr and 15~Gyr. The models have 3~{\AA} FWHM spectral 
resolution, corresponding to a nominal stellar velocity dispersion $\sigv
\approx70~\kms$ at 5500~{\AA}. As a complement to Fig.~\ref{fig_idxmet}, 
Fig.~\ref{fig_idxvel} shows the evolution in the same diagrams of $Z=Z_\odot$ 
models broadened to velocity dispersions $\sigv= 70$, 150 and 300~\kms. We also 
show in Fig.~\ref{fig_idxvel} the evolution of a model with continuous star 
formation with a law $\psi(t) \propto \exp\left[ -t/({\rm 4\,Gyr)}\right]$, for
$Z=Z_\odot$ and $\sigv=70~\kms$.

The models in Figs~\ref{fig_idxmet} and \ref{fig_idxvel} summarize the influence
of metallicity, velocity dispersion and star formation history on the strengths
of Lick indices. Fig.~\ref{fig_idxmet} shows that, as found in many previous 
studies of Lick indices, `metallic-line indices' such as Fe4531, C$_2$4668, 
Mg$_1$, Mg$_2$, Mg$b$ and Fe5270 react sensitively to changes in metallicity
in old stellar populations [corresponding to the largest $D_n(4000)$ values],
while `Balmer-line indices' such as H$\delta_A$ do not, as they are controlled
mainly by the temperature of the main-sequence turnoff (see references in 
Section~1). We note that $D_n(4000)$ is also sensitive to metallicity, as it is
produced by the accumulation of a large number of metallic lines just blueward
of 4000~{\AA}. Stellar velocity dispersion is another parameter affecting the 
strengths of Lick indices (e.g., \citealt{1998ApJS..116....1T}; 
\citealt{2001Ap&SS.276..839V}). As Fig.~\ref{fig_idxvel} shows, $\sigv$ greatly
influences the strengths of indices whose definitions involve narrow 
pseudo-continuum bandpasses, such as Fe4531 and Fe5270. Fig.~\ref{fig_idxvel} 
further shows that a model with continuous star formation can account for the 
observed strengths of most indices in galaxies with low velocity dispersions. 
This is remarkable, as our model provides the first opportunity to study the 
index strengths of star-forming galaxies in this way. Old SSP models do not seem
to reproduce well the observed strengths of indices such as C$_2$4668, Mg$_1$,
Mg$_2$ and Mg$b$ in galaxies with high velocity dispersions. This is not 
surprising, as massive galaxies show departures in the relative abundances of 
different heavy elements from the Galactic stars on which the models are based
(e.g., \citealt*{1992ApJ...398...69W}).

We wish to investigate in more detail the ability of our model to reproduce
simultaneously the strengths of different indices in individual galaxy spectra.
To carry out such an analysis, we need a library of models encompassing a
full range of physically plausible star formation histories. We generate a 
library of Monte Carlo realizations of different star formation histories 
similar to that with which \citet{2003MNRAS.341...33K} interpreted the $D_n(4000)$
and H$\delta_A$ index strengths of a complete sample of 120,000 SDSS galaxies
using our model. In this library, each star formation history consists of two
parts: (1) an underlying continuous model parametrized by a formation time
$t_{\rm form}$ and a star formation time scale parameter $\gamma>0$, such that
galaxies form stars according to the law $\psi(t) \propto\exp\left[-\gamma t
({\rm\,Gyr)} \right]$ from time $t_{\rm form}$ to the present. The time 
$t_{\rm form}$ is taken to be distributed uniformly from the Big Bang to 
1.5~Gyr before the present day and $\gamma$ over the interval 0 to 1; (2)
random bursts are superimposed on these continuous models. Bursts occur with
equal probability at all times after $t_{\rm form}$. They are parametrized in
terms of the ratio between the mass of stars formed in the burst and the total
mass formed by the continuous model from time $t_{\rm form}$ to the present. 
This ratio is taken to be distributed logarithmically between 0.03 and 4.0. 
During a burst, stars form at a constant rate for a time distributed uniformly
in the range $3\times10^7$--$3\times10^8 \,$yr. The burst probability is set so
that 50 per cent of the galaxies in the library have experienced a burst over
the past 2~Gyr (see \citealt{2003MNRAS.341...33K} for more detail). We distribute
our models uniformly in metallicity from 0.25 to 2 times solar (all stars in a
given model have the same metallicity) and uniformly in velocity dispersion from
70~\kms\ (the nominal resolution of the models at 5500~{\AA}) to 350~\kms. Our
final library consists of 150,000 different star formation histories.

\begin{figure*}
\centering
\begin{minipage}{140mm}
\includegraphics[width=140mm]{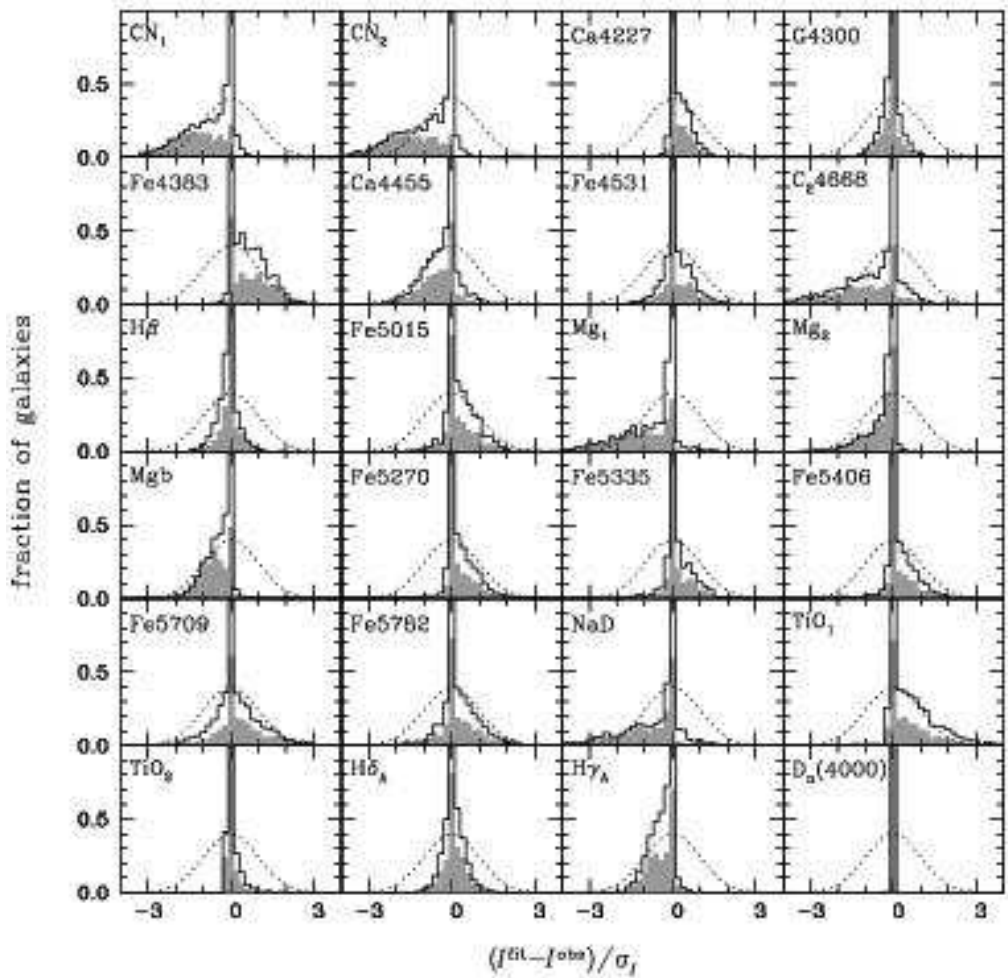}
\caption{Fits of the strengths of individual indices in combination with 
$D_n(4000)$ in the spectra of 2010 galaxies with S/N$_{\rm med} \ge 30$ in the
`main galaxy sample' of the SDSS EDR. The stellar velocity dispersion of the
models are required to be within 15~\kms\ of the observed ones. Each panel 
shows the distribution of the fitted index strength $I^{\rm fit}$ minus the 
observed one $I^{\rm obs}$, divided by the associated error $\sigma_I$ 
(equation~\ref{sigi}). For reference, a dotted line in each 
panel indicates a Gaussian distribution with unit standard deviation. The 
shaded histograms show the contributions to the total distributions by galaxies
with $\sigv>180~\kms$, corresponding roughly to the median stellar velocity 
dispersion of the sample.}
\label{fig_fitindiv}
\end{minipage}
\end{figure*}

We use this library to evaluate the ability of our model to reproduce the 
strengths of various indices in observed galaxy spectra. To start with, we 
examine the accuracy to which the model can reproduce the strength of any {\em
individual} index at the same time as the 4000~{\AA} discontinuity in SDSS
spectra. We also require consistency with the observed stellar velocity
dispersion. We enlarge here our observational sample to 2010 spectra with 
S/N$_{\rm med} \ge 30$ in the `main galaxy sample' of the SDSS EDR. For each
individual Lick index, we first select for each SDSS spectrum the models in 
the library whose stellar velocity dispersions are within 15~\kms\ of the 
observed one. We then select among these models the one that reproduces the
observed $D_n(4000)$ and selected index strength with the lowest \chisq. We 
report in Fig.~\ref{fig_fitindiv}, for 24 indices, the distribution of the 
index strength $I^{\rm fit}$ in the best-fitting model minus that $I^{\rm obs}$
in the observed spectrum, divided by the associated error $\sigma_I$ 
(equation~\ref{sigi} below), for the 2010 galaxies in our 
sample.  For reference, a dotted line in each panel indicates a Gaussian 
distribution with unit standard deviation. The shaded histograms in 
Fig.~\ref{fig_fitindiv} show the contributions to the total distributions by
galaxies with $\sigv>180~\kms$, corresponding roughly to the median stellar
velocity dispersion of our SDSS sample.

The error $\sigma_I$ associated to each index fit in this analysis includes 
both the observational error ${\sigma_I}^{\rm obs}$ and a theoretical error 
${\sigma_I}^{\rm mod}$ reflecting the uncertainties in the model spectral 
calibration. We therefore write
\begin{equation}
\sigma_I=\left[({\sigma_I}^{\rm obs})^2+
               ({\sigma_I}^{\rm mod})^2\right]^{1/2}\,.
\label{sigi}
\end{equation}
Table~\ref{idx-errors} lists, for each index, the median observational error
$\overline{\sigma_I}^{\rm obs}$ for the galaxies in our sample. For many
indices, this is much smaller than the observed 1\%--99\% percentile range 
$\Delta_I$ of index strengths in the sample (also listed), indicating that 
variations in index strength are highly significant. We adopt a representative,
fixed theoretical error ${\sigma_I}^{\rm mod}$ for each index. This is taken to
be half the maximum difference in the strength of the index between SSP models
calibrated using the STELIB/BaSeL~1.0 and STELIB/BaSeL~3.1 spectral libraries
over wide ranges of ages (1--13~Gyr) and metallicities (0.004--0.05). 
Table~\ref{idx-errors} shows that the theoretical error ${\sigma_I}^{\rm mod}$ 
obtained in this way is comparable to the median observational error
$\overline{\sigma_I}^{\rm obs}$ for most indices. Also listed in 
Table~\ref{idx-errors} is the quantity $\Delta_I/\overline{\sigma_I}$ indicating
the `resolving power' of each index. Here $\overline{\sigma_I}$ is the median 
error from equation~(\ref{sigi}) for the galaxies in our sample.

\begin{table}
  \caption{Observed 1\%--99\% percentile range $\Delta_I$ and median 
observational error $\overline{\sigma_I}^{\rm obs}$ for 28 spectral features 
in 2010 spectra with S/N$_{\rm med} \ge 30$ in the `main galaxy sample' of the
SDSS EDR. Also listed for each index is the theoretical error ${\sigma_I}^{\rm
mod}$ reflecting the uncertainties in the model spectral calibration. The 
quantity $\Delta_I/\overline{\sigma_I}$, where $\overline{\sigma_I}$ is the
median of $\sigma_I=[({\sigma_I}^{\rm obs})^2+ ({\sigma_I}^{\rm mod})^2]^{1/2}$
for the sample, indicates the `resolving power' of each index. The
values for atomic indices are expressed in angstroms of equivalent width, while
those for molecular indices (indicated by a star) are expressed in magnitudes.
An exception is $D_n(4000)$, whose value is the ratio of the average flux
densities in two narrow bands (Section 4.2).
}
  \label{idx-errors}
  \begin{tabular}{@{}lcllr@{}}
  \hline
Feature & 1\%--99\% 
        & $\overline{\sigma_I}^{\rm obs}$ 
	& ${\sigma_I}^{\rm mod}$%
	& $\Delta_I/\overline{\sigma_I}$ \\
        & range $\Delta_I$                                        &       &     \\
 \hline
$^{\star}$CN$_1$        & $-$0.137\ldots0.126\,\,\,\, &   0.013   & 0.023 &  10 \\
$^{\star}$CN$_2$        & $-$0.089\ldots0.174\,\,\,\, &   0.013   & 0.022 &  10 \\
\hskip1.5mm Ca4227      &     0.15\ldots1.87          &   0.22    & 0.22  &  6  \\
\hskip1.5mm G4300       &     0.06\ldots6.36          &   0.34    & 0.52  &  10 \\
\hskip1.5mm Fe4383      &     0.97\ldots6.27          &   0.36    & 0.40  &  10 \\
\hskip1.5mm Ca4455      &     0.05\ldots1.99          &   0.24    & 0.11  &  7  \\
\hskip1.5mm Fe4531      &     1.13\ldots4.18          &   0.32    & 0.25  &  8  \\
\hskip1.5mm C$_2$4668   &  $-$0.01\ldots8.62\,\,\,\,  &   0.41    & 0.39  &  15 \\
\hskip1.5mm H$\beta$    &     1.28\ldots4.73          &   0.24    & 0.39  &  7  \\
\hskip1.5mm Fe5015      &     1.40\ldots6.55          &   0.39    & 0.47  &  8  \\
$^{\star}$Mg$_1$        & $-$0.004\ldots0.164\,\,\,\, &   0.006   & 0.010 &  14 \\
$^{\star}$Mg$_2$        &    0.050\ldots0.322         &   0.008   & 0.019 &  13 \\
\hskip1.5mm Mgb         &     1.32\ldots5.13          &   0.24    & 0.59  &  6  \\
\hskip1.5mm Fe5270      &     0.81\ldots3.65          &   0.25    & 0.15  &  10 \\
\hskip1.5mm Fe5335      &     0.93\ldots3.48          &   0.26    & 0.15  &  9  \\
\hskip1.5mm Fe5046      &     0.50\ldots2.40          &   0.22    & 0.10  &  8  \\
\hskip1.5mm Fe5709      &     0.11\ldots1.40          &   0.18    & 0.10  &  6  \\
\hskip1.5mm Fe5782      &     0.17\ldots1.28          &   0.15    & 0.08  &  7  \\
\hskip1.5mm NaD         &     0.77\ldots5.45          &   0.17    & 0.33  &  12 \\
$^{\star}$TiO$_1$       & $-$0.014\ldots0.072\,\,\,\, &   0.005   & 0.006 &  11 \\
$^{\star}$TiO$_2$       &    0.025\ldots0.105         &   0.004   & 0.015 &  5  \\
\hskip1.5mm H$\delta_A$ &  $-$3.20\ldots6.09\,\,\,\,  &   0.49    & 0.93  &  9  \\
\hskip1.5mm H$\gamma_A$ &  $-$6.90\ldots4.69\,\,\,\,  &   0.45    & 1.01  &  11 \\
\hskip1.5mm $D_n(4000)$ &     1.12\ldots2.23          &   0.02    & 0.08  &  13 \\
\hskip1.5mm \mgfe       &     1.13\ldots3.91          &   0.15    & 0.18  &  12 \\
\hskip1.5mm \mgfep      &     1.12\ldots3.95          &   0.16    & 0.18  &  12 \\
$^{\star}$\mgofe        &    0.197\ldots0.483         &   0.011   & 0.014 &  16 \\
$^{\star}$\mgtfe        &    0.226\ldots0.576         &   0.011   & 0.014 &  19 \\
\hline
\end{tabular}

\end{table}

We can identify in Fig.~\ref{fig_fitindiv} those Lick indices that our model
fails to reproduce well when compared to high-quality galaxy spectra: CN$_1$,
CN$_2$, C$_2$4668, Mg$_1$ and NaD. The distributions of $\left(I^{\rm fit}-
I^{\rm obs} \right)/ \sigma_I$ for these indices all show significant tails 
relative to a Gaussian distribution. These departures are most likely caused by
differences in element abundance ratios between SDSS galaxies and the Galactic 
stars used to build our model. Several recent studies have addressed the influence
of changes in element abundance ratios on the strengths of Lick indices 
(\citealt{1995AJ....110.3035T}; \citealt{1998A&A...333..419T};
\citealt{2000AJ....119.1645T}; \citealt{2001ApJ...549..274V};
\citealt{2002MNRAS.333..517P}; \citealt*{2003MNRAS.339..897T}). These studies were
motivated by the observational evidence that the abundance ratios of 
$\alpha$-elements to iron are enhanced in massive early-type galaxies relative to
the solar composition. The discrepancies found in these studies between models 
with scaled-solar abundances and observations of Lick indices in nearby star 
clusters and galaxies are similar to those identified above in 
Fig.~\ref{fig_fitindiv}.  Enhanced abundances of Mg, C and N have been invoked to
account for the departures of models from observations of CN$_1$, CN$_2$, 
C$_2$4668 and Mg$_1$. Some indices like NaD could also be contaminated by 
interstellar absorption. As Fig.~\ref{fig_fitindiv} shows, the discrepancies 
pertaining to these indices tend to arise in galaxies with large velocity
dispersions.

\begin{figure*}
\centering
\begin{minipage}{140mm}
\includegraphics[width=140mm]{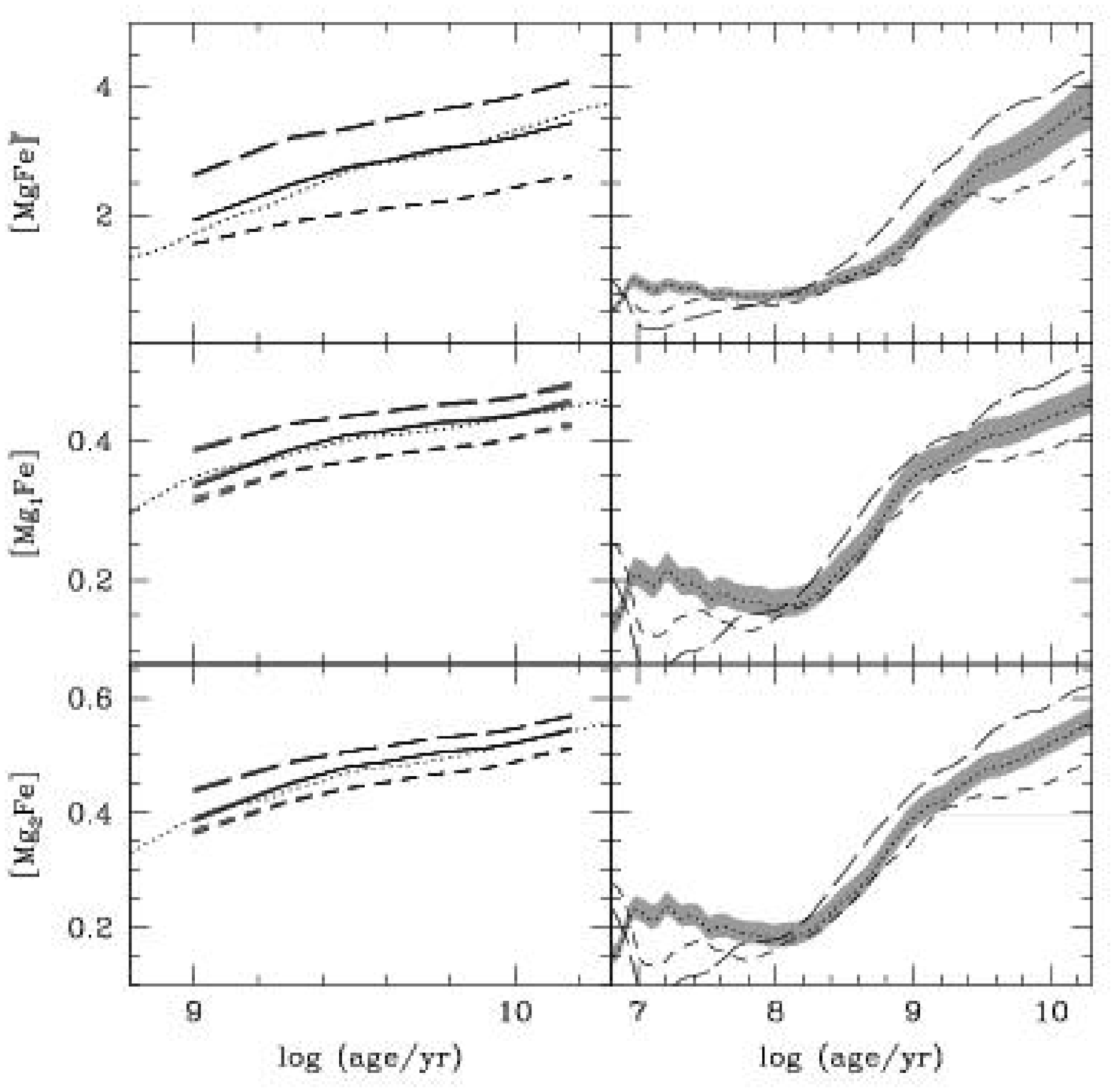}
\caption{{\it Left}: Evolution of the strengths of \mgfep, \mgofe\ and \mgtfe\
according to the model with variable element abundance ratios of 
\citet{2003MNRAS.339..897T} for the metallicities $Z=0.5Z_\odot$ (short-dashed
lines), $Z_\odot$ (solid lines) and $2.2 Z_\odot$ (long-dashed lines). For
each metallicity, three lines corresponding to [$\alpha$/Fe]=0.0, 0.3 and
0.5 are shown (hardly distinguishable). The dotted line shows the evolution of
an SSP with solar metallicity computed using the Padova 1994 stellar evolution
prescription, the STELIB/BaSeL~3.1 spectral calibration and a 
\citet{1955ApJ...121..161S} IMF (0.1--100~\msun), for a stellar 
velocity dispersion $\sigv=200~\kms$. {\it Right}: Evolution of the strengths
of \mgfep, \mgofe\ and \mgtfe\ in SSPs computed using the Padova 1994 stellar 
evolution prescription, the STELIB/BaSeL~3.1 spectral calibration and a 
\citet{1955ApJ...121..161S} IMF (0.1--100~\msun), for the metallicities 
$Z=0.4Z_{\sun}$ (short-dashed line), $Z_{\sun}$ (dotted line) and $2.5Z_{\sun}$
(long-dashed line) and for $\sigv=200~\kms$. The shaded area around the 
solar-metallicity model indicates the range spanned by models with stellar 
velocity dispersions $70\leq\sigv \leq300~\kms$.}
\label{fig_mgfe}
\end{minipage}
\end{figure*}

Our main goal here is to identify those indices that {\em can} be fitted
by our model in observed galaxy spectra. Among the indices the model can recover
within the errors at the same time as $D_n(4000)$ in Fig.~\ref{fig_fitindiv},
we expect a subset to also be reproducible simultaneously. In particular, since
the Balmer-line indices H$\beta$, H$\gamma_A$ and H$\delta_A$ are not expected
to depend sensitively on metallicity, they should also be reproducible in 
combination with metallic-line indices. We further expect our model to be able
to reproduce metallic-line indices that do not depend sensitively on changes in 
$\alpha$-element to iron abundance ratios. The model with variable element 
abundance ratios of \citet{2003MNRAS.339..897T} is useful for identifying 
`composite' indices that are sensitive to metallicity but not to $\alpha$/Fe. 
The original \mgfe\ index of \citet{1993PhDT.........7G} and the new \mgfep\ 
index proposed by \citet{2003MNRAS.339..897T}, which is even less sensitive to 
$\alpha$/Fe, are both well calibrated in our model (see Fig.~\ref{fig_fitall} 
below). After some experimentation, we identified two other indices with 
similarly weak dependence on $\alpha$/Fe,
\begin{equation}
\mgofe = 0.6{\rm Mg}_1+0.4\log({\rm Fe4531}+{\rm Fe5015})\,,
\label{mg1fe}
\end{equation}
\begin{equation}
\mgtfe = 0.6{\rm Mg}_2+0.4\log({\rm Fe4531}+{\rm Fe5015})\,.
\label{mg2fe}
\end{equation}
Fig.~\ref{fig_mgfe} illustrates the evolution of \mgfep, \mgofe\ and \mgtfe\ for
SSPs with different metallicities, $\alpha$/Fe abundance ratios and stellar 
velocity dispersions. The left-hand panels show the predictions of the 
\citet{2003MNRAS.339..897T} model for the metallicities $Z= 0.5Z_\odot$, $Z_\odot$
and $2.2 Z_\odot$ and for [$\alpha$/Fe]= 0.0, 0.3 and 0.5, for a 
\citet{1955ApJ...121..161S} IMF truncated at 0.1 and 100~\msun. The right-hand 
panels show the predictions of our model for the same IMF for the metallicities 
$Z=0.4Z_\odot$, $Z_\odot$ and $2.5Z_\odot$ and for stellar velocity dispersions 
$70\leq\sigv \leq300\,\kms$. It is clear from this figure that \mgfep, \mgofe\ and
\mgtfe\ are affected sensitively by changes in $Z$ and \sigv\ but not by changes
in $\alpha$/Fe. We note that, since Mg$_1$ and Mg$_2$ are defined over broader 
bandpasses than the Mgb index involved in the definition of \mgfep\ (471~{\AA}
versus 64~{\AA}), \mgofe\ and \mgtfe\ may be slightly more sensitive to
flux-calibration uncertainties and attenuation by dust in observed galaxy spectra. 

Based on these arguments, we use our library of models with different star 
formation histories to fit {\em simultaneously} the observed strengths of 
H$\beta$, H$\gamma_A$, H$\delta_A$, \mgfep, \mgofe, \mgtfe\ and
$D_n(4000)$ in the 2010 SDSS spectra with S/N$_{\rm med}>30$ in our sample. As
before, for each SDSS spectrum, we select the best-fitting model in the library
among those with stellar velocity dispersions within 15~\kms\ of the 
observed one. Fig.~\ref{fig_fitall} shows the resulting distributions of 
$\left(I^{\rm fit}-I^{\rm obs} \right)/ \sigma_I$ for all galaxies in the 
sample, for the same 24 indices as in Fig.~\ref{fig_fitindiv} and for \mgfe,
\mgfep, \mgofe\ and \mgtfe. The highlighted frames indicate the seven indices 
used to constrain the fits. Fig.~\ref{fig_fitall} demonstrates that our model
can account simultaneously for the observed strengths of H$\beta$, H$\gamma_A$, 
H$\delta_A$, \mgfep, \mgofe, \mgtfe\ and $D_n(4000)$ in high-quality galaxy
spectra. The strengths of these indices are always recovered within the
errors. In addition, the model recovers reasonably well the strengths of 
several other indices which were not used to constrain the fits, such as 
G4300, Ca4455, Fe4531, Fe5015, Fe5270, Fe5335, Fe5709 and Fe5782. As expected
from Fig.~\ref{fig_fitindiv}, indices like CN$_1$, CN$_2$, C$_2$4668, Mg$_1$
and NaD cannot be fitted accurately because of their strong dependence on 
element abundance ratios. The fact that Mgb appears to be better reproduced than
Mg$_1$ and Mg$_2$ in Fig.~\ref{fig_fitall} is a consequence of the larger 
relative error on this index (see Table~\ref{idx-errors}). Interestingly, 
TiO$_1$ and TiO$_2$ that could be fitted individually in 
Fig.~\ref{fig_fitindiv} do not appear to be well reproducible in combination
with the other indices in Fig.~\ref{fig_fitall}.

\begin{figure*}
\centering
\begin{minipage}{140mm}
\includegraphics[width=140mm]{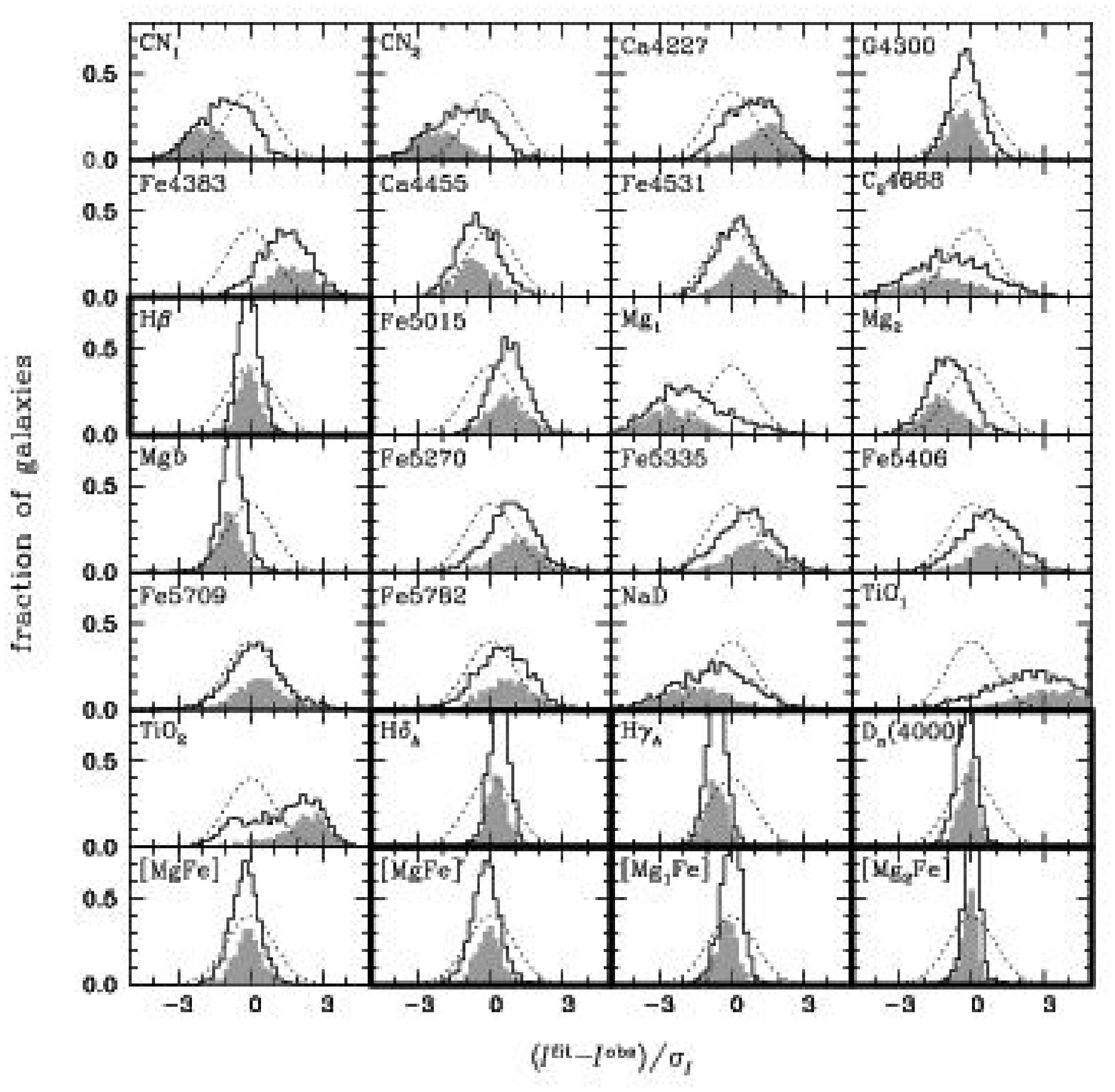}
\caption{Simultaneous fit of the strengths of several indices in the spectra 
of 2010 galaxies with S/N$_{\rm med} \ge 30$ in the `main galaxy sample' of the
SDSS EDR. The highlighted frames indicate the seven indices used to constrain 
the fits. The stellar velocity dispersion of the models are required to be 
within 15~\kms\ of the observed ones. Each panel shows the distribution of the
fitted index strength $I^{\rm fit}$ minus the observed one $I^{\rm obs}$, 
divided by the associated error $\sigma_I$ (equation~\ref{sigi}).
For reference, a dotted line in each panel indicates a Gaussian distribution 
with unit standard deviation. The shaded histograms show the contributions to
the total distributions by galaxies with $\sigv>180~\kms$, corresponding 
roughly to the median stellar velocity dispersion of the sample.}
\label{fig_fitall}
\end{minipage}
\end{figure*}

The agreement between model and observations for many indices in 
Fig.~\ref{fig_fitall} is all the more remarkable in that the measurement errors
for this sample of high-quality SDSS spectra are very small. As 
Table~\ref{idx-errors} shows, for indices such as H$\gamma_A$, H$\delta_A$, 
$D_n(4000)$, \mgfep, \mgofe\ and \mgtfe, the typical error $\overline{\sigma_I}$ 
amounts to only about 5--10 per cent of the total range spanned by the index over
the full sample. Our model, therefore, represents an ideal tool for interpreting 
the {\em distribution} of these indices in complete samples of galaxies in 
terms of the parameters describing the stellar populations. Such analyses may
include all types of galaxies, from star-forming galaxies to passively evolving,
early-type galaxies, for which stellar velocity dispersions should first be 
determined. 

\subsection{Comparison with previous work}

In most previous studies of absorption-line features in galaxy spectra,
index strengths were computed in the Lick/image dissector scanner (IDS)
system (see for example \citealt{1994ApJS...94..687W}). To compare our results
with these previous studies, we must therefore transform our predictions to the
Lick/IDS system. This amounts to computing index strengths as if they were 
measured in spectra which are not flux-calibrated and whose resolution ranges
from 8~{\AA} to 10~{\AA} FWHM, depending on wavelength. We follow the procedure 
outlined in the appendix of \citet{1997ApJS..111..377W} and calibrate the 
transformation from STELIB to Lick/IDS spectra using 31 stars in common between
the two libraries. We first broaden the STELIB spectra of these stars to the 
wavelength-dependent Lick/IDS resolution. Then, we identify the median 
offsets between the index strengths measured in the broadened STELIB spectra 
and those measured from the non-fluxed Lick/IDS spectra for the 31 stars. The
resulting median offsets are listed in Table~\ref{idx-shifts} for 25 spectral 
features. To compute predictions in the Lick/IDS system, therefore, we first 
broaden our model galaxy spectra to the Lick/IDS resolution and then subtract
the median offsets of Table~\ref{idx-shifts} from the index strengths measured 
in the broadened spectra.

\begin{table}
  \caption{Median offsets in the strengths of 25 spectral features between
the STELIB spectra broadened to the wavelength-dependent Lick/IDS resolution
and the non-fluxed Lick/IDS spectra for the 31 stars in common between the 
two libraries. A positive offset denotes that the index measured from
STELIB spectra is larger than that measured from Lick/IDS spectra. The values
for atomic indices are expressed in angstroms of equivalent width, while those
for molecular indices (indicated by a star) are expressed in magnitudes.
}
  \label{idx-shifts}
  \begin{tabular}{@{}lc@{}}
  \hline
Feature & Median Offset \\
        & (STELIB minus Lick/IDS) \\
 \hline
$^{\star}$CN$_1$       &  $-$0.011      \\
$^{\star}$CN$_2$       &  $-$0.001      \\
\hskip1.5mm Ca4227     &  $-$0.02\,\,\, \\
\hskip1.5mm G4300      & \,\,0.05       \\
\hskip1.5mm Fe4383     & \,\,0.52       \\
\hskip1.5mm Ca4455     &  $-$0.13\,\,\, \\
\hskip1.5mm Fe4531     &  $-$0.12\,\,\, \\
\hskip1.5mm C$_2$4668  & \,\,0.43       \\
\hskip1.5mm H$\beta$   & \,\,0.13       \\
\hskip1.5mm Fe5015     & \,\,0.47       \\
$^{\star}$Mg$_1$       &  $-$0.020      \\
$^{\star}$Mg$_2$       &  $-$0.018      \\
\hskip1.5mm Mgb        & \,\,0.03       \\
\hskip1.5mm Fe5270     & \,\,0.17       \\
\hskip1.5mm Fe5335     & \,\,0.07       \\
\hskip1.5mm Fe5046     & \,\,0.20       \\
\hskip1.5mm Fe5709     & \,\,0.03       \\
\hskip1.5mm Fe5782     & \,\,0.04       \\
\hskip1.5mm NaD        & \,\,0.01       \\
$^{\star}$TiO$_1$      & \,\,\,\,\,0.003\\
$^{\star}$TiO$_2$      & \,\,\,\,\,0.004\\
\hskip1.5mm H$\delta_A$& \,\,0.83       \\
\hskip1.5mm H$\gamma_A$&  $-$0.89\,\,\, \\
\hskip1.5mm H$\delta_F$& \,\,0.20       \\
\hskip1.5mm H$\gamma_F$&  $-$0.29\,\,\, \\
\hline
\end{tabular}

\end{table}

\begin{figure*}
\centering
\begin{minipage}{140mm}
\includegraphics[width=140mm]{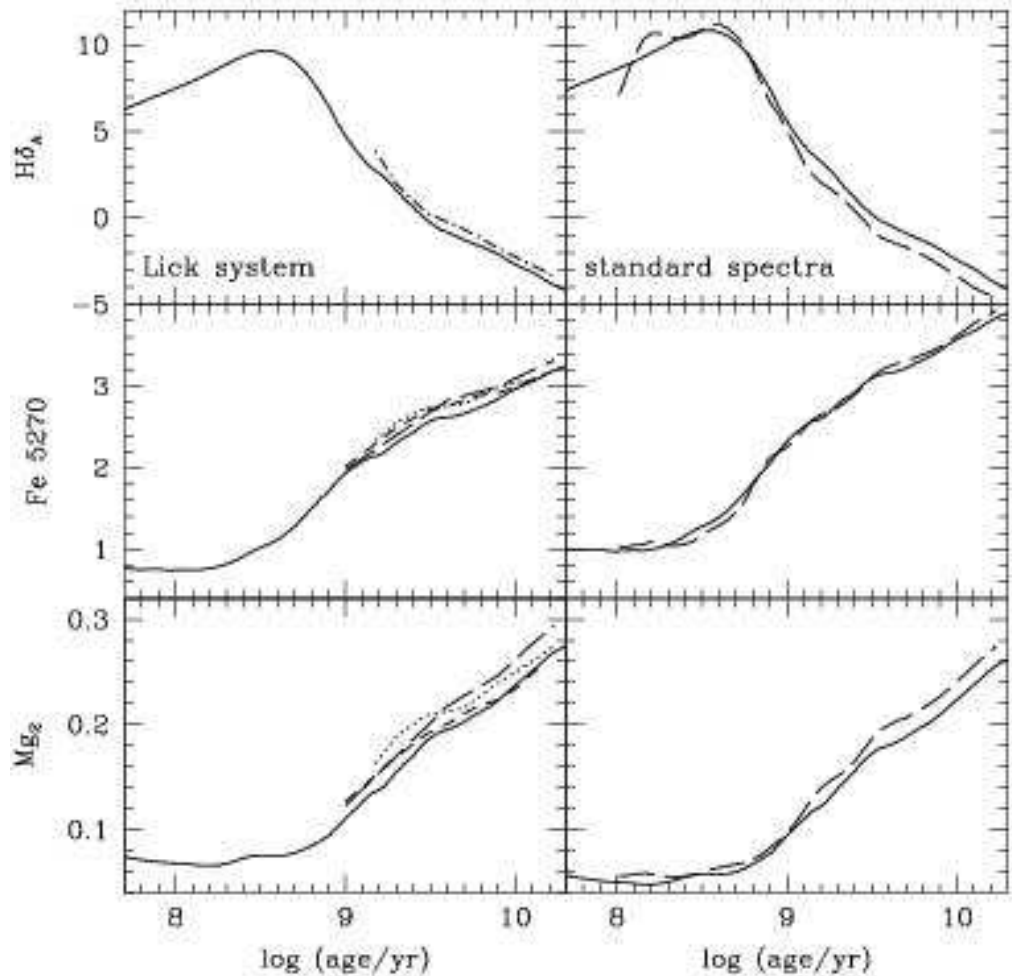}
\caption{Evolution of the strengths of H$\delta_A$, Fe5270 and Mg$_2$ of an
SSP with solar composition ($Z=Z_\odot$, [$\alpha$/Fe]=0) and a 
\citet{1955ApJ...121..161S} IMF according to different models. {\it Left}: 
Predictions computed in the Lick/IDS system. In each panel, the solid line 
shows the standard model of Section 3 transformed to the Lick/IDS system, as
described in the text. Also shown, where available, are the models of 
\citet[ dotted line]{1994ApJS...95..107W},
\citet[ long-dashed line]{1996ApJS..106..307V},
\citet[ dot-and-dashed line]{1997ApJS..111..377W}
and \citet[ short-dashed line]{2003MNRAS.339..897T}. {\it Right}: Index
strengths measured in the same way in the spectra of the standard model of 
Section~3 (solid line) and in the spectra of the \citet{1999ApJ...513..224V}
model broadened to a resolution of 3~{\AA} FWHM (long-dashed line).}
\label{fig_idxcomp}
\end{minipage}
\end{figure*}

In Fig.~\ref{fig_idxcomp}, we compare different models for the evolution of the 
H$\delta_A$, Fe5270 and Mg$_2$ index strengths of a simple stellar population 
with solar composition ($Z=Z_\odot$, [$\alpha$/Fe] =0) and a 
\citet{1955ApJ...121..161S} IMF. The left-hand panels show model predictions
computed in the Lick/IDS system. The solid line shows our standard 
model transformed to the Lick/IDS system, as described above. Also shown, where
available, are the models of \citet[ dotted line]{1994ApJS...95..107W},
\citet[ long-dashed line]{1996ApJS..106..307V}, 
\citet[ dot-and-dashed line]{1997ApJS..111..377W}
and \citet[ short-dashed line]{2003MNRAS.339..897T}. These models differ from
ours in that they rely on the implementation of the `fitting formulae' of 
\citet{1994ApJS...94..687W} and \citet{1997ApJS..111..377W} -- who parametrized
index strengths as functions of stellar effective temperature, gravity and 
metallicity (see Section 1 above) -- into different population synthesis codes.
The predictions are restricted to stellar populations older than 1~Gyr. As
Fig.~\ref{fig_idxcomp} shows, the strengths of H$\delta_A$, Fe5270 and Mg$_2$
predicted by our model when transformed to the Lick/IDS system are consistent
with the predictions from these previous models to within the typical theoretical
errors listed in Table~\ref{idx-errors}.

\begin{figure*}
\centering
\begin{minipage}{140mm}
\includegraphics[width=140mm]{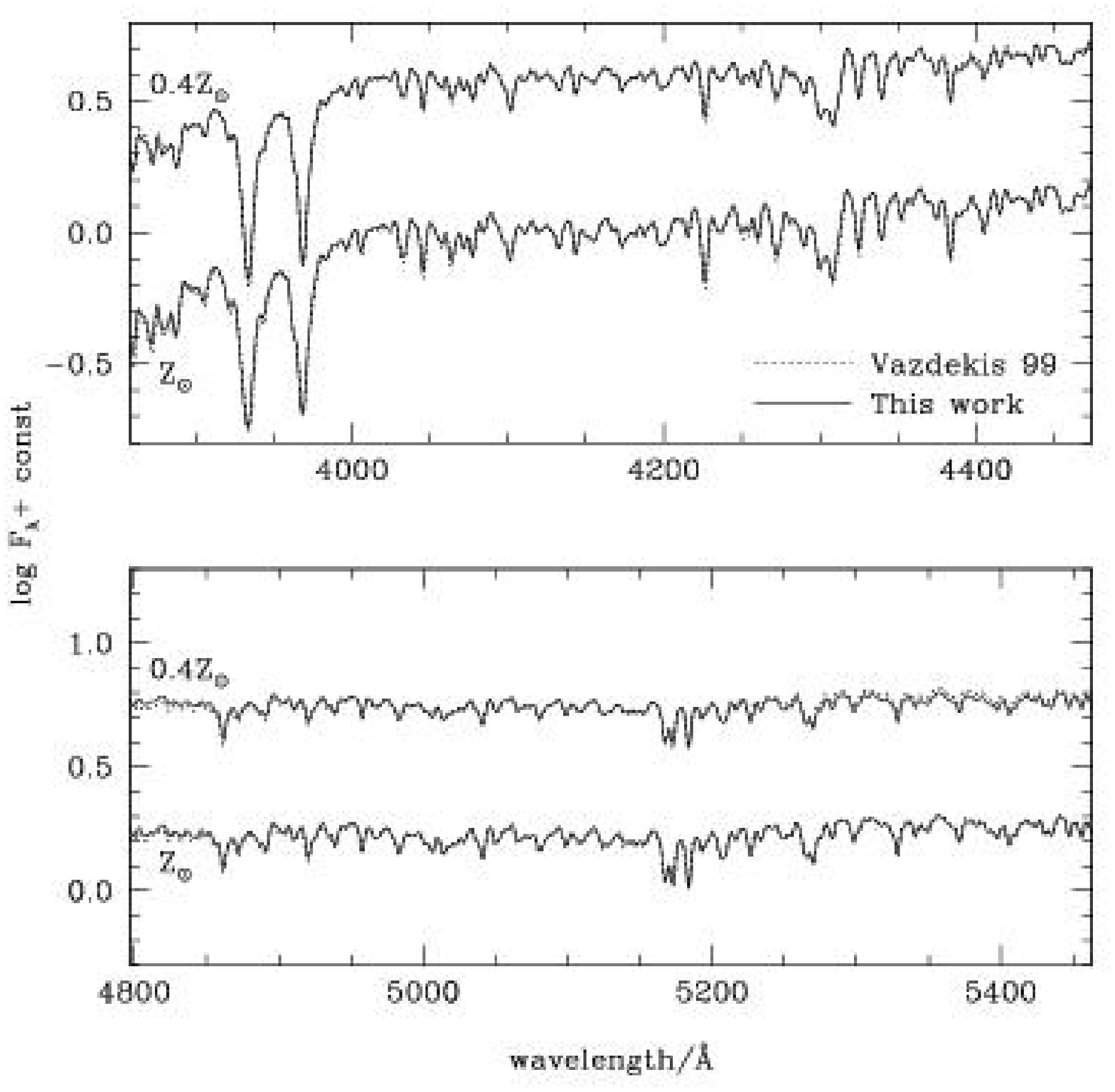}
\caption{Comparison of the standard model of Section~3 (solid line) with the 
\citet{1999ApJ...513..224V} model broadened to a resolution of 3~{\AA} FWHM 
(dotted line), for 10~Gyr-old SSPs with scaled-solar abundances and a
\citet{1955ApJ...121..161S} IMF, for the metallicities $0.4Z_{\sun}$  
and $Z_{\sun}$. The spectra are shown in the two narrow wavelength regions 
covered by the \citet{1999ApJ...513..224V} model.}
\label{fig_specomp}
\end{minipage}
\end{figure*}

The recent model of \citet{1999ApJ...513..224V} offers another element of
comparison, as it predicts the spectra of simple stellar populations of 
various ages and metallicities at a resolution of $\sim1.8$~{\AA}. Thus, the 
strengths of H$\delta_A$, Fe5270 and Mg$_2$ can be measured in these spectra in 
the same way as in our model, without having to transform predictions to the 
Lick/IDS system. In the right-hand panels of Fig.~\ref{fig_idxcomp}, we compare
the strengths of H$\delta_A$, Fe5270 and Mg$_2$ in our model with those measured
in the \citet{1999ApJ...513..224V} spectra broadened to a resolution of
3~{\AA} FWHM, for an SSP with solar composition and a \citet{1955ApJ...121..161S} 
IMF. The predictions of both models agree to within the typical theoretical errors
listed in Table~\ref{idx-errors}. As a further check, we compare in 
Fig.~\ref{fig_specomp} the spectra predicted by both models for 10~Gyr-old SSPs
with scaled-solar abundances, for the metallicities $0.4Z_{\sun}$ and $Z_{\sun}$.
The spectra are shown in the two narrow wavelength regions covered by the 
\citet{1999ApJ...513..224V} model, 3820--4500~{\AA} and 4780--5460~{\AA}. The
overall agreement between the two models is excellent at both metallicities. We
conclude that our model agrees reasonably well with previous models of spectral
indices of galaxies.

\section{Summary and conclusions}

We have presented a new model for computing the spectral evolution of stellar 
populations of different metallicities at ages between $1\times10^5\,$yr and
$2\times10^{10}\,$yr at a resolution of 3~{\AA} FWHM across the whole wavelength
range from 3200~{\AA} to 9500~{\AA} (corresponding to a median resolving power 
$\lambda/\Delta \lambda \approx2000$). These predictions are based on a new 
library of observed stellar spectra recently assembled by 
\citet{2003A&A...402..433L}. The spectral evolution can also be computed across
a larger wavelength range, from 91~{\AA} to 160~$\mu$m, at lower resolution.
The model incorporates recent progress in stellar evolution theory and an 
observationally motivated prescription for thermally-pulsing AGB stars, which 
is supported by observations of surface brightness fluctuations in nearby stellar
populations (\citealt{2000ApJ...543..644L}; \citealt{2002ApJ...564..216L}). We
have shown that this model reproduces well the observed optical and near-infrared 
colour-magnitude diagrams of Galactic star clusters of various ages and 
metallicities, and that stochastic fluctuations in the numbers of stars in 
different evolutionary phases can account for the full range of observed 
integrated colours of star clusters in the Magellanic Clouds and in the merger
remnant galaxy NGC~7252.

Our model reproduces in detail typical galaxy spectra extracted from the SDSS 
Early Data Release \citep{2002AJ....123..485S}. We have shown how this type
of spectral fit can constrain physical parameters such as the star formation 
history, metallicity and dust content of galaxies. Our model is also the first
to enable accurate studies of absorption-line strengths in galaxies containing
stars over the full range of ages. We have shown that it can reproduce
simultaneously the observed strengths of those Lick indices that do not depend
strongly on element abundance ratios in 2010 spectra with S/N$_{\rm med} 
\ge 30$, taken from the `main galaxy sample' of the SDSS EDR. This comparison
requires proper accounting for the observed velocity dispersions of the 
galaxies. Indices whose strengths depend strongly on element abundance ratios
cannot always be fitted accurately, because the stars in the model spectral 
library have fixed composition at fixed metallicity. Based on the model with
variable element abundance ratios of \citet{2003MNRAS.339..897T}, we have
identified a few spectral features which depend negligibly on element abundance 
ratios and are well reproduced by our model: \mgfep, \mgofe\ and \mgtfe\
(equations~\ref{mg1fe}--\ref{mg2fe}). These features, when combined with a 
Balmer-line index such as H$\beta$, H$\gamma_A$ or H$\delta_A$, should be 
particularly useful for constraining the star formation histories and 
metallicities of galaxies. Several other popular indices, such as the \caii\
triplet index near 8500~{\AA} \citep{1989MNRAS.239..325D}, can also be measured
directly from the model spectra. Most interestingly, our model offers the 
possibility to explore new indices over the full wavelength range from 3200~{\AA}
to 9500~{\AA}.

It is worth mentioning that, for applications to studies of star-forming 
galaxies, the influence of the interstellar medium on the stellar radiation 
predicted by our model must be accounted for. The emission-line spectrum of
the \hii\ regions and the diffuse gas ionized by young stars can be computed
by combining our model with a standard photoionization code (see for example 
\citealt{2001MNRAS.323..887C}). To account for the attenuation of starlight by 
dust, the simple but realistic prescription of \citet{2000ApJ...539..718C} is
particularly well suited to population synthesis studies. In this prescription,
the attenuation of starlight by dust may be accounted for by inserting a factor 
$\exp[-\hat{\tau}_\lambda(t')]$ in the integrand on the right-hand side of 
equation~(\ref{convol}), where $\hat{\tau}_\lambda(t')$ is the `effective 
absorption' curve describing the attenuation of photons emitted in all 
directions by stars of age $t'$ in a galaxy. This is given by the simple formula
\begin{eqnarray}
\hat{\tau}_\lambda(t')=\cases{
\hskip0.22cm\hat{\tau}_V\left(\lambda/{5500\,\rmn{\AA}}\right)^{-0.7}\,,
&for $t'\leq 10^7$ yr,\cr
{{\mu\hat{\tau}_V}}\left(\lambda/{5500\,\rmn{\AA}}\right)^{-0.7}\,,
&for $t'>10^7$ yr,\cr}
\label{taueff}
\end{eqnarray}
where $\hat{\tau}_V$ is the total effective $V$-band optical depth seen by 
young stars. The characteristic age $10^7\,$yr corresponds to the typical
lifetime of a giant molecular cloud. The adjustable parameter $\mu$ defines the
fraction of the total dust absorption optical depth of the galaxy contributed 
by the diffuse interstellar medium ($\mu \approx 1/3$ on average, with 
substantial scatter).  Note that equation~(\ref{taueff}) neglects the 
absorption of ionizing photons by dust in the \hii\ regions, that should be
accounted for to study line luminosities (see \citealt{2002MNRAS.330..876C}).

The high-resolution population synthesis model presented in this paper enables
more refined spectral analyses of galaxies than could be achieved using previous
low-resolution models. In particular, in recent studies of our own, it has 
become clear that the ability to resolve stellar absorption features in galaxy
spectra demonstrates a need in many galaxies to account for the stochastic 
nature of star formation (e.g., \citealt{2003MNRAS.341...33K}; see also 
Section~4.3 above). Traditional models with continuous star formation 
histories tend to smooth away valuable spectral signatures of stochastic 
starbursts. Our preliminary results also suggest that the new high-resolution
model makes it possible to break, in many cases, the age-metallicity degeneracy
which has impaired most previous population synthesis studies of the star 
formation and enrichment histories of galaxies. We hope that this model will 
contribute to the refinement of such studies in the future. Our model is 
intended for use by the general astronomical community and is available from 
http://www.cida.ve/$\sim$bruzual/bc2003 and http://www.iap.fr/$\sim$charlot/bc2003.

\section*{Acknowledgments}

We are grateful to J.~Brinchmann and H.~Mathis for their help in producing the
results presented in Sections~4.2 and 4.3 of this paper and to T.~Le~Bertre,
R.~Loidl, T.~Rauch and M.~Schultheis for providing data in advance of publication.
Special thanks to C.~Tremonti for her help regarding the implementation of the
STELIB library into our model. G.~Kauffmann and S.~White provided useful advice on
the analysis of SDSS galaxy spectra. We thank the referee, A.~Vazdekis, and 
F.~Schweizer for helpful comments on the original manuscript. We also thank the
many colleagues who, over the last decade, have helped us improve our population
synthesis model through the feedback they provided.

G.B.A. acknowledges generous financial support from the European Commission
(under ALAMED contract No. CIL-CT93-0328VE), the Landessternwarte 
Heidelberg-K\"onigstuhl (under grant No. SFB-328) and the Max-Planck Institut
f\"ur Astrophysik, Germany, the Universitat de Barcelona, Spain, the Swiss 
National Science Foundation (under grant No.  20-40654.94), FAPESP, Brazil,
the Observatoire Midi-Pyr\'en\'ees and the Institut d'Astrophysique de Paris, 
CNRS, France and the Consejo Nacional de Investigaciones Cient{\'\i}ficas y 
Tecnol\'ogicas of Venezuela (under grant No. F-155). S.C. thanks the Alexander
von Humboldt Foundation, the Federal Ministry of Education and Research, and
the Programme for Investment in the Future (ZIP) of the German Government for
their support through a Sofja Kovalevskaja award. This research was supported
in part by the National Science Foundation under Grant No. PHY94-07194 to the
Institute for Theoretical Physics, Santa Barbara.

Funding for the creation and distribution of the SDSS Archive has been provided
by the Alfred P. Sloan Foundation, the Participating Institutions, the National
Aeronautics and Space Administration, the National Science Foundation, the US
Department of Energy, the Japanese Monbukagakusho, and the Max Planck Society.
The SDSS Web site is http://www.sdss.org/. The Participating Institutions are
the University of Chicago, Fermilab, the Institute for Advanced Study, the
Japan Participation Group, the Johns Hopkins University, the Max Planck
Institute for Astronomy (MPIA), the Max Planck Institute for Astrophysics 
(MPA), New Mexico State University, Princeton University, the United States
Naval Observatory, and the University of Washington.

\appendix

\section{Mapping of theoretical isochrones with stellar spectra}

\begin{figure*}
\centering
\begin{minipage}{140mm}
\includegraphics[width=140mm]{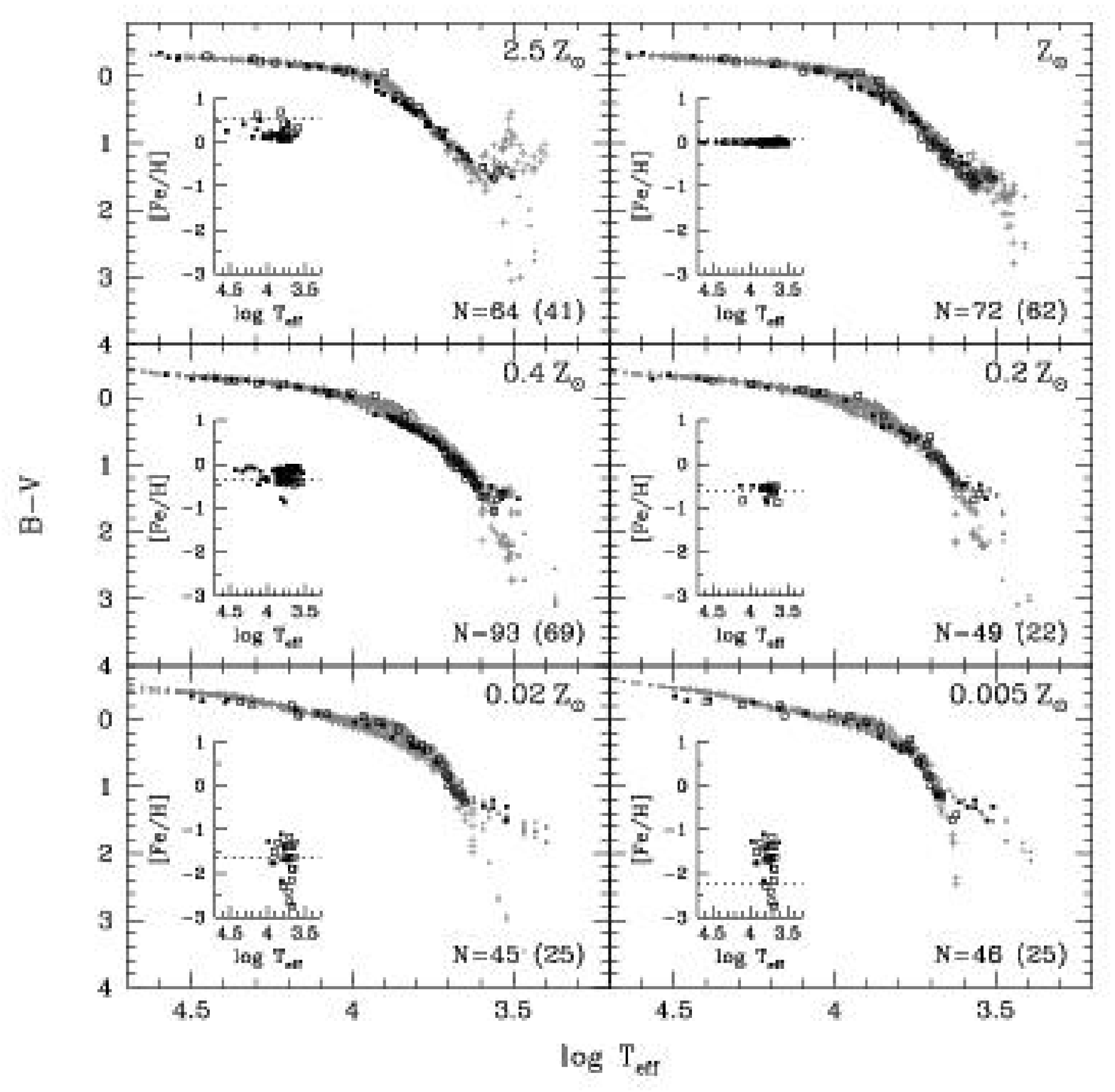}
\caption{\bv\ colour as a function of effective temperature for those
stars of the BaSeL~3.1 (grey dots: dwarfs; grey crosses: giants) and
STELIB/BaSeL~3.1 (black filled squares: dwarfs; black open squares: giants) 
libraries that were selected to map the theoretical isochrones computed using
the Padova 1994 evolutionary tracks, as described in Appendix A. Each panel
corresponds to a different metallicity of the evolutionary tracks, as indicated.
For each metallicity, the inset panel shows \feh\ as a function of $\log T_{\rm
eff}$ for the subset of STELIB stars included in the metallicity bin whose 
abundances are compatible with the \feh\ value of the evolutionary tracks 
(indicated by a dotted line). The number of these stars is given in parentheses 
at the bottom right of each panel. Also indicated is the {\em total} number N of 
STELIB stars included in the metallicity bin and shown on the colour-temperature
relation. This number includes, in addition to the stars shown in the inset panel,
cool K- and M-dwarf stars from the SDSS EDR (whose colours are taken to be the
same at all metallicities) and hot solar-metallicity stars (see text for detail).}
\label{fig_bvcalib}
\end{minipage}
\end{figure*}

\begin{figure*}
\centering
\begin{minipage}{140mm}
\includegraphics[width=140mm]{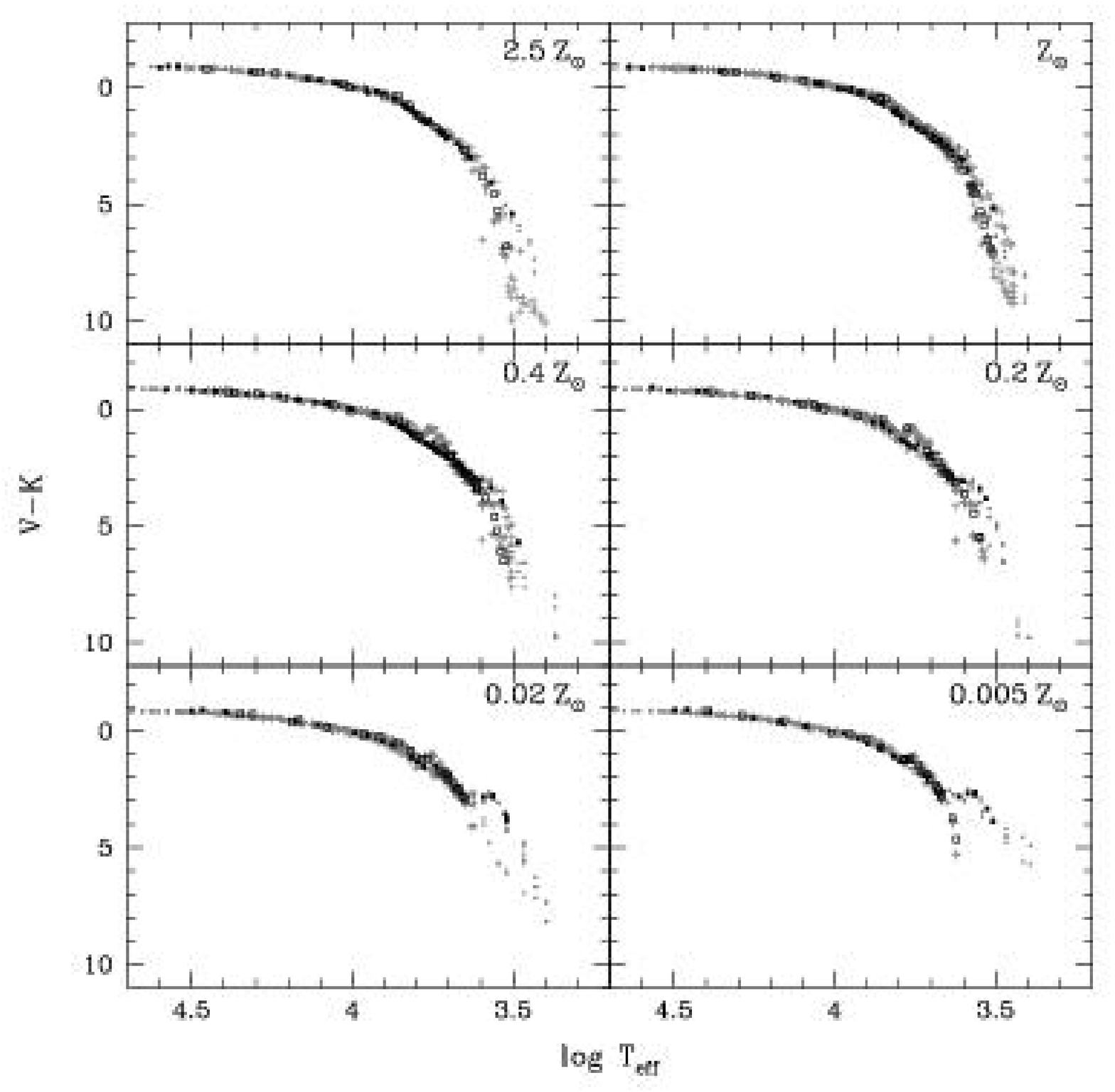}
\caption{\vk\ colour as a function of effective temperature for those
stars of the BaSeL~3.1 (grey dots: dwarfs; grey crosses: giants) and
STELIB/BaSeL~3.1 (black filled squares: dwarfs; black open squares: giants)
libraries that were selected to map the theoretical isochrones computed using
the Padova 1994 evolutionary tracks, as described in Appendix A. Each panel
corresponds to a different metallicity of the evolutionary tracks, as indicated.}
\label{fig_vkcalib}
\end{minipage}
\end{figure*}

We illustrate here in a graphical way the mapping of theoretical isochrones with
stellar spectra in our model. At fixed metallicity $Z$ (or \feh), an isochrone 
interpolated from a set of stellar evolutionary tracks in the HR diagram is 
defined by a sequence of evolutionary phases corresponding to different effective 
temperatures $\log T_{\rm eff}$ and gravities $\log g$. To transform this
theoretical isochrone into an observational one, we must assign a stellar 
spectrum to each of the evolutionary phases. This is straightforward in the case
where the spectra are taken from a library of theoretical model atmospheres (i.e.,
BaSeL~1.0, BaSeL~2.2, BaSeL~3.1; see Tables~\ref{spectra-options} and 
\ref{calib-options}), since in general such models are parametrized in terms of
$Z$, $\log T_{\rm eff}$ and $\log g$. The only practical complication in this 
case is that model atmospheres are available only for discrete values of these
parameters. Thus model spectra must be interpolated at the values of $Z$, $\log
T_{\rm eff}$ and $\log g$ corresponding to the isochrone. The relation between
colours and effective temperature for stars of fixed metallicity and gravity is
then tied to the adopted spectral library (Section 2.2.1).

We now turn to the case where the STELIB library of observed stellar spectra
is used to map the theoretical isochrones in the HR diagram. In this case, we
still rely on the colour-temperature scale of one of the BaSeL libraries in 
Table~\ref{calib-options}. The reason for this is that the effective temperatures
published by \citet{2003A&A...402..433L} for the STELIB stars are incomplete and
were not derived in a homogeneous way. Hence these temperatures are not suited
to model calibration. We therefore assign to each evolutionary stage along the
isochrone the STELIB spectrum of the corresponding luminosity class that best 
matches the theoretical BaSeL spectrum assigned to that stage. In this procedure,
the STELIB spectra are first degraded to the resolution of the BaSeL library 
using a gaussian filter of 20~{\AA} FWHM. The selection is thus driven by the
{\em shape} of the continuum spectrum rather than by the strengths of absorption
features (we have checked that this approach is free of systematic biases). We
refer to the three possible implementations of the STELIB library in our model as
the `STELIB/BaSeL~1.0', the `STELIB/BaSeL~2.2' and the `STELIB/BaSeL~3.1' 
libraries (Section 2.2.2).

Figs.~\ref{fig_bvcalib} and \ref{fig_vkcalib} show the \bv\ and \vk\ colours
of those stars of the BaSeL~3.1 and STELIB/BaSeL~3.1 libraries that were selected
as described above to map the theoretical isochrones computed using the Padova
1994 evolutionary tracks, for different metallicities. The colours are
plotted as a function of effective temperature $T_{\rm eff}$, and different
symbols represent dwarf and giant stars in each library. In the case of the
STELIB/BaSeL~3.1 library, the \bv\ colours rely on the STELIB spectra 
alone, while the \vk\ colours rely on extensions of these spectra at near-infrared
wavelengths using BaSeL~3.1 spectra (Section 2.2.2). 

By construction, at fixed metallicity, the colour-temperature scale of the
STELIB/BaSeL~3.1 library is always similar to that of the BaSeL~3.1 library. 
The number of STELIB stars with metallicities close to that of the evolutionary
tracks is indicated in parentheses at the bottom right of each panel in 
Fig.~\ref{fig_bvcalib}.  The inset panels in Fig.~\ref{fig_bvcalib} show [Fe/H] 
as a function of $\log T_{\rm eff}$ for these stars. Because of the scarcity of 
stars hotter than about 7000~K at non-solar metallicities, we include hot 
solar-metallicity stars to sample the colour-$T_{\rm eff}$ relations at all 
metallicities. As mentioned in Section~2.2.1, the spectra of these stars should
be representative of hot stars at all but the most extreme metallicities. For 
completeness, we also include in the STELIB library a few SDSS-EDR spectra of 
cool K- and M-dwarf stars at all metallicities (we thank C.~Tremonti for kindly
providing us with these spectra). For standard IMFs, these stars never contribute 
significantly to the integrated light of model stellar populations. The total 
number of STELIB stars used to sample the colour-temperature relation at each 
metallicity (including hot solar-metallicity stars and SDSS-EDR cool dwarf 
stars) is indicated at the bottom right of each panel in Fig.~\ref{fig_bvcalib}.

\begin{table*}
 \centering
 \begin{minipage}{140mm}
  \caption{Qualitative assessment of the spectral predictions of the model for
simple stellar populations of various ages and metallicities computed using the
Padova~1994 evolutionary tracks and different spectral libraries. For each 
entry, the expected reliability of the predictions is indicated separately for
young ($\la1\,$Gyr) and old ($\gg1\,$Gyr) stellar populations (listed as
young/old).}
  \label{x-accuracy}
  \begin{tabular}{@{}rcccc@{}}
  \hline
Metallicity&\multicolumn{2}{c}{STELIB/BaSeL~3.1}& BaSeL~3.1 & Pickles\\
(Padova 1994)  
& optical colours & line strengths & UV--NIR colours & UV--NIR colours\\
 \hline
$2.5\,Z_\odot$ & good/good & fair/fair$^a$  & fair/poor & \ldots \\
$Z_\odot$ & very good/very good 
                                       & very good/very good
                                       & very good/good & very good/very good\\
$0.4\,Z_\odot$ & good/good & good/very good & good/fair& \ldots  \\
$0.2\,Z_\odot$ & fair/good & fair/good      & good/fair& \ldots  \\
$0.02\,Z_\odot$      & poor/fair & poor/fair      & fair/poor& \ldots  \\
$0.005\,Z_\odot$           & poor/fair & poor/fair      & poor/poor& \ldots  \\
\hline
\end{tabular}
$^a\,$The STELIB spectra used to map the theoretical isochrones have 
$\feh\approx+0.25$ on average, i.e., slightly lower than the metallicity of the 
evolutionary tracks (Fig.~\ref{fig_bvcalib}). This affects line strengths but
not colour predictions, which are tied to the colour-temperature scale of the
BaSeL~3.1 library for $Z=2.5\,Z_\odot$.
\end{minipage}
\end{table*}

Figs.~\ref{fig_bvcalib} and \ref{fig_vkcalib} allow us to draw the following
conclusions. First, the STELIB spectra of dwarf and giant stars used to
sample the colour-$T_{\rm eff}$ relation in each metallicity bin 
provide reasonable coverage of the HR diagram. In practice, in regions
where the sampling is scarcer (for example, around $\bv\approx0.1$ and 
$\log T_{\rm eff}\approx3.9$ at the metallicity $0.4\,Z_{\sun}$ in 
Fig.~\ref{fig_bvcalib}), we improve it by interpolating between nearby STELIB
spectra of the appropriate luminosity class. We do not perform such
interpolations in the temperature range 3750--5000~K that is critical for bright
giant stars and is always well sampled. Second, the homogeneity in \feh\ of 
STELIB stars included in each metallicity bin varies from bin to bin. It
is reasonable for $Z\geq0.4Z_{\sun}$, although we note that for $Z=2.5Z_{\sun}$
the STELIB stars have $\feh\approx+0.25$ on average, i.e., slightly lower 
metallicity than the tracks. There are no stars at all hotter than 7000~K at 
metallicities $Z\leq0.2Z_{\sun}$. At these metallicities, the model relies 
heavily on our adoption of solar-metallicity spectra at high temperatures.
Third, both Figs.~\ref{fig_bvcalib} and \ref{fig_vkcalib} show that the STELIB
library does not include giant stars as red as the reddest stars selected from
the BaSeL~3.1 library to map the theoretical isochrones, even at solar 
metallicity. These stars, however, do not contribute critically to the integrated
light of model stellar populations for standard IMFs. This is illustrated by the
close agreement between the dotted (BaSeL~3.1) and solid (STELIB/BaSeL~3.1) lines
in Fig.~\ref{fig_calib}. It is worth recalling that, for the brightest 
asymptotic-giant-branch stars, we adopt in all libraries the prescription outlined
in Section~2.2.4 (these stars have redder colours off the scales of 
Figs.~\ref{fig_bvcalib} and \ref{fig_vkcalib}).

We have performed further extensive tests of the $UBVRIJHKL$ colour-temperature
scales, colour-colour relations and bolometric corrections of the different 
spectral libraries used in our model (Table~\ref{spectra-options}). We find that
our procedure to assign spectra from these libraries to stars in the HR diagram
does not introduce any systematic bias in the predicted photometric evolution of
stellar populations. The libraries themselves, however, include their own 
uncertainties. In particular, the ultraviolet and near-infrared colours of the 
spectra, even in the most recent BaSeL~3.1 library, remain significantly more
uncertain than the optical colours because in part of the lack of comparison 
standards at non-solar metallicities \citep{2002A&A...381..524W}. The situation 
should improve as more observed spectra become available to extend the optical 
spectra at ultraviolet and near-infrared wavelengths, as is the case already for
solar metallicity (Pickles library). We further conclude that, despite the 
limitations outlined above, the STELIB library provides reasonable coverage of 
the HR diagram for population synthesis purposes, especially at metallicities 
greater than $0.2Z_\odot$. Based on these arguments, we can assess the expected 
accuracy of the spectral predictions of our model for simple stellar populations
of various ages and metallicities computed using the Padova~1994 evolutionary 
tracks and different spectral libraries. This qualitative assessment is summarized
in Table~\ref{x-accuracy}.

\bsp

\label{lastpage}

\end{document}